\documentclass{aastex63}
\usepackage{amsmath}
\usepackage{lineno}

\received{...}
\revised{...}
\accepted{...}

\submitjournal{ApJ}

\shorttitle{Theoretical blazar variability comparison}
\shortauthors{Thiersen, Zacharias \& B\"ottcher.}
\graphicspath{{./}{figures/}}

\begin{document}

\title{Simulations of Stochastic Long-Term Variability in Leptonic Models for External-Compton and Synchrotron Self-Compton Dominated Blazars}

\correspondingauthor{Hannes Thiersen}
\email{hannesthiersen@gmail.com}

\author[0000-0002-4024-3280]{Hannes Thiersen}
\affiliation{Centre for Space Research, North-West University, Potchefstroom, 2520, South Africa}

\author[0000-0001-5801-3945]{Michael Zacharias}
\affiliation{Laboratoire Univers et Théories, Observatoire de Paris, Université PSL, CNRS, Université de Paris, 92190 Meudon, France}
\affiliation{Centre for Space Research, North-West University, Potchefstroom, 2520, South Africa}

\author[0000-0002-8434-5692]{Markus B\"ottcher}
\affiliation{Centre for Space Research, North-West University, Potchefstroom, 2520, South Africa}

\begin{abstract}

    In this work we investigate the nature of multi-wavelength variability of
    blazars from a purely numerical approach.  We use a time-dependent one-zone
    leptonic blazar emission model to simulate multi-wavelength variability by
    introducing stochastic parameter variations in the emission region.  These
    stochastic parameter variations are generated by Monte Carlo methods and
    have a characteristic power law index of $\alpha=-2$ in their power spectral
    densities. We include representative blazar test cases for a flat
    spectrum radio quasar and a high synchrotron peaked BL Lacertae
    object for which the high energy component of the Spectral Energy
    Distribution is dominated by external Compton and synchrotron
    self-Compton emission, respectively.  The simulated variability is
    analyzed in order to characterise the distinctions between the two blazar
    cases and the physical parameters driving the variability.  We show that
    the variability's power spectrum is closely related to underlying
    stochastic parameter variations for both cases.  Distinct differences
    between the different progenitor variations are present in the
    multi-wavelength cross-correlation functions.

\end{abstract}


\section{Introduction} \label{sec:intro}

Blazars are a type of active galactic nuclei (AGN) that possess a relativistic jet of particles moving on an axis aligned close to our line-of-sight \citep{1995PASP..107..803U}.
They have a characteristic non-thermal spectral energy distribution (SED) consisting of a low-energy emission component ranging from radio frequencies up to infrared or X-rays and a high-energy component that spans from X-rays up to $\gamma$-rays, in some cases extending into the Very High-Energy (VHE, $E > 100$~GeV) regime.
The low energy emission component is produced through synchrotron radiation of relativistic electrons (and positrons).
The dominant mechanism producing the high energy component on the other hand is still under debate, but is typically explained as inverse-Compton (IC) scattering of ambient photons by the jet electrons within leptonic models.
Alternatively, in hadronic models, the high-energy emission is produced through relativistic proton synchrotron and radiation related to proton-photon interactions \citep{2017SSRv..207....5R}.
Blazars are subdivided into two subclasses based on optical emission-line features: flat-spectrum radio quasars (FSRQs) possessing broad optical emission lines and BL Lac objects for which these features are effectively absent.
Another criterion for further classification of BL Lac objects is the value of their SED synchrotron peak frequency for which there are three types: low-synchrotron peaked (LBL; $\nu_\text{sync, peak} < 10^{14}$ Hz), intermediate-synchrotron peaked (IBL; $ 10^{14} \; \text{Hz} \le \nu_\text{sync, peak} < 10^{15}$ Hz) and high-synchrotron peaked (HBL; $\nu_\text{sync, peak} > 10^{15}$ Hz) BL Lac objects \citep{2010ApJ...716...30A}.

Broadband variability on a vast range of timescales is a ubiquitous characteristic of AGN and, in particular, blazars \citep{1995ARA&A..33..163W,2010ApJ...722..520A,2007ApJ...669..862A,2017A&A...598A..39H}.
However, in spite of a large number of observational multi-wavelength campaigns on many blazars and extensive theoretical studies of blazar variability, the causes of the variability are still unclear.
Most research on blazar variability focuses on some significant short-term events, typically flares \citep[e.g.,][]{Aharonian_2008,2013ApJ...762...92A,2018MNRAS.474.1296B}.
These flares may be caused either by changes intrinsic to the jet, such as shocks and/or magnetic reconnection \citep{2017MNRAS.464.4875B,2015MNRAS.450..183S,2019MNRAS.482...65C}, or external to the jet.
Examples of the latter are the ablation of a dust cloud \citep{2017ApJ...851...72Z} or interactions of the jet with a stellar wind \citep{2010A&A...522A..97A,2012ApJ...749..119B}.
However, such flaring events might not be representative of the underlying long-term variability which is still present even in the absence of flares.

Extensive multi-wavelength campaigns on blazars allowed for investigations into the long-term variability in blazars providing in-depth analyses and characterization of observed variability behaviour \citep[e.g., detections of quasi-periodic oscillations and coloured noise characteristics of multi-wavelength variability; see][]{2016ApJ...832...47B,2017ApJ...837..127G,2018ApJ...863..175G,2020ApJ...891..120B}.
However, in this work, we study the underlying long-term blazar variability from a purely theoretical/numerical perspective.
We simulate the multi-wavelength variability by assuming a stochastic process inducing time-dependent changes to some emission region variables.
These parameter changes are fed into a time-dependent leptonic one-zone code in order to obtain the variable light curves in various energy bands.
Lastly, we derive the power density spectra for the individual light curves and the cross-correlation functions between the different energy bands.
Two representative test cases of blazars are considered: (a) an FSRQ, dominated by inverse-Compton scattering of external photon fields (external Compton; EC) in its high-energy SED component and (b) an HBL, dominated by synchrotron self-Compton (SSC) emission.

A similar study was done by \cite{2013MNRAS.434.2684M} who investigated variability in X-rays and $\gamma$-rays for one-zone lepto-hadronic emission scenarios in which the high-energy SED component is produced by synchrotron radiation of protons or secondary particles from photopair/photopion production, and/or the decay of neutral pions.
Random walks were used to generate variations of injection luminosity and maximum particle energy introduced into their emission model to induce variability.
Here, we focus solely on leptonic emission processes, consider wavelengths from optical up to TeV $\gamma$-rays, and generate independent variations for the electron injection luminosity, the magnetic field strength, the electron injection spectral index, and the maximum electron energy by means of Fourier transforms.

An overview of the blazar emission model, the representative blazar test cases and the generation of the stochastic variations is given in Section \ref{sec:model}.
Results are presented and discussed in Section \ref{sec:results_discuss}.

\section{Model} \label{sec:model}

\subsection{Model Setup}

This work employs the time-dependent homogenous one-zone leptonic blazar emission model created by \cite{2014JHEAp...1...63D,2016ApJ...826...54D} and further developed by \cite{2017ApJ...851...72Z}.
We refer to these papers for the details of the code.
It assumes that the multi-wavelength emission is produced by a homogenous region/zone in which a rapid acceleration process (e.g., shocks or magnetic reconnection) injects relativistic electrons with a power-law distribution in energy.
The emission region travels along the jet-axis at relativistic speed and is pervaded by a tangled magnetic field of uniform strength.

The model uses Fokker-Planck and radiation transfer equations to keep track of the electron and photon populations within the emission region over time.
This allows for the extraction of snap-shot SEDs and photon-energy-dependent light curves.
The relativistic electrons in the emission region produce synchrotron and IC radiation which then, due to the emission region's relativistic motion, produce Doppler boosted emission in the observer's frame of reference.

\begin{figure}[ht]
    \gridline{
        \fig{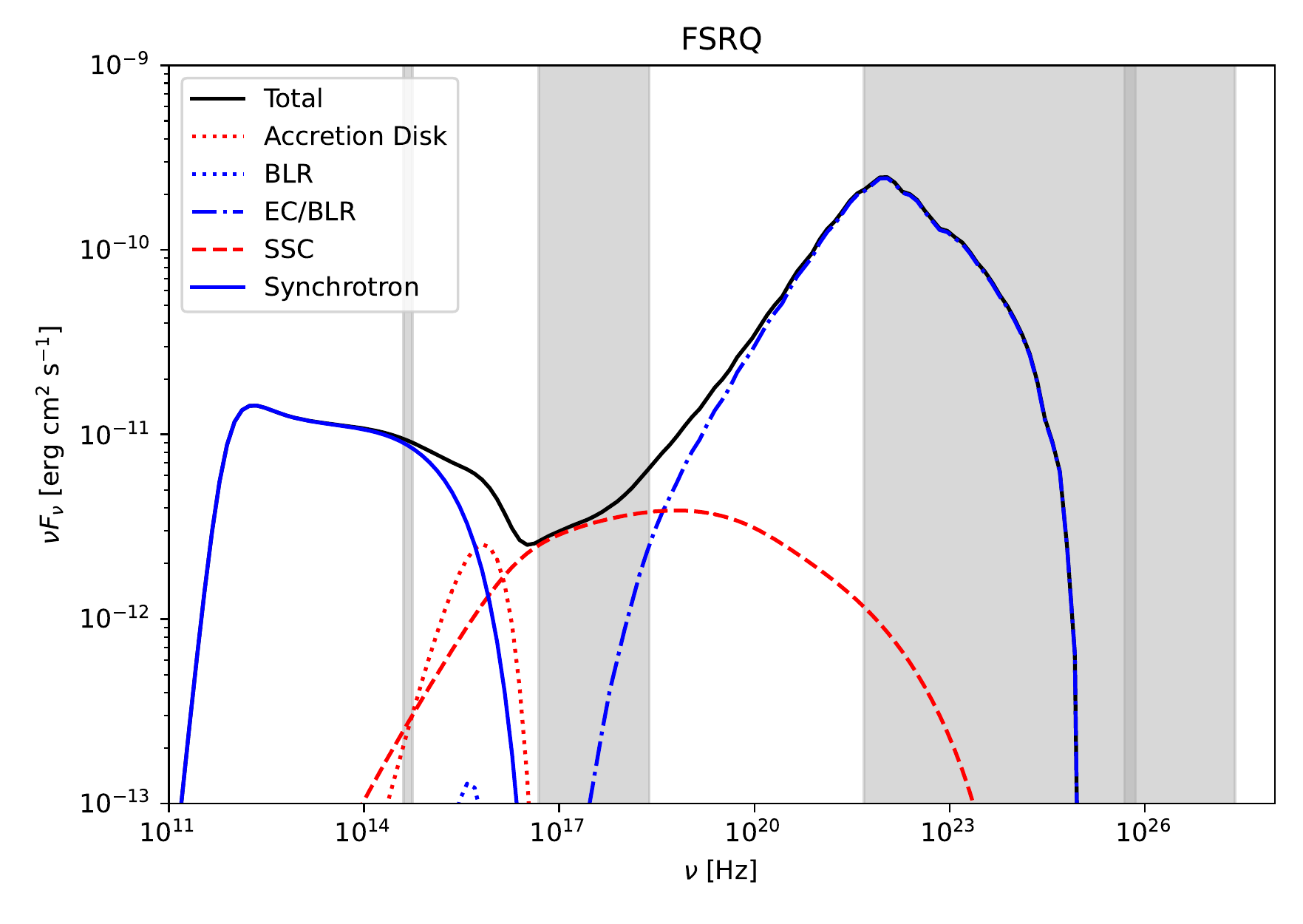}{.5\textwidth}{}
        \fig{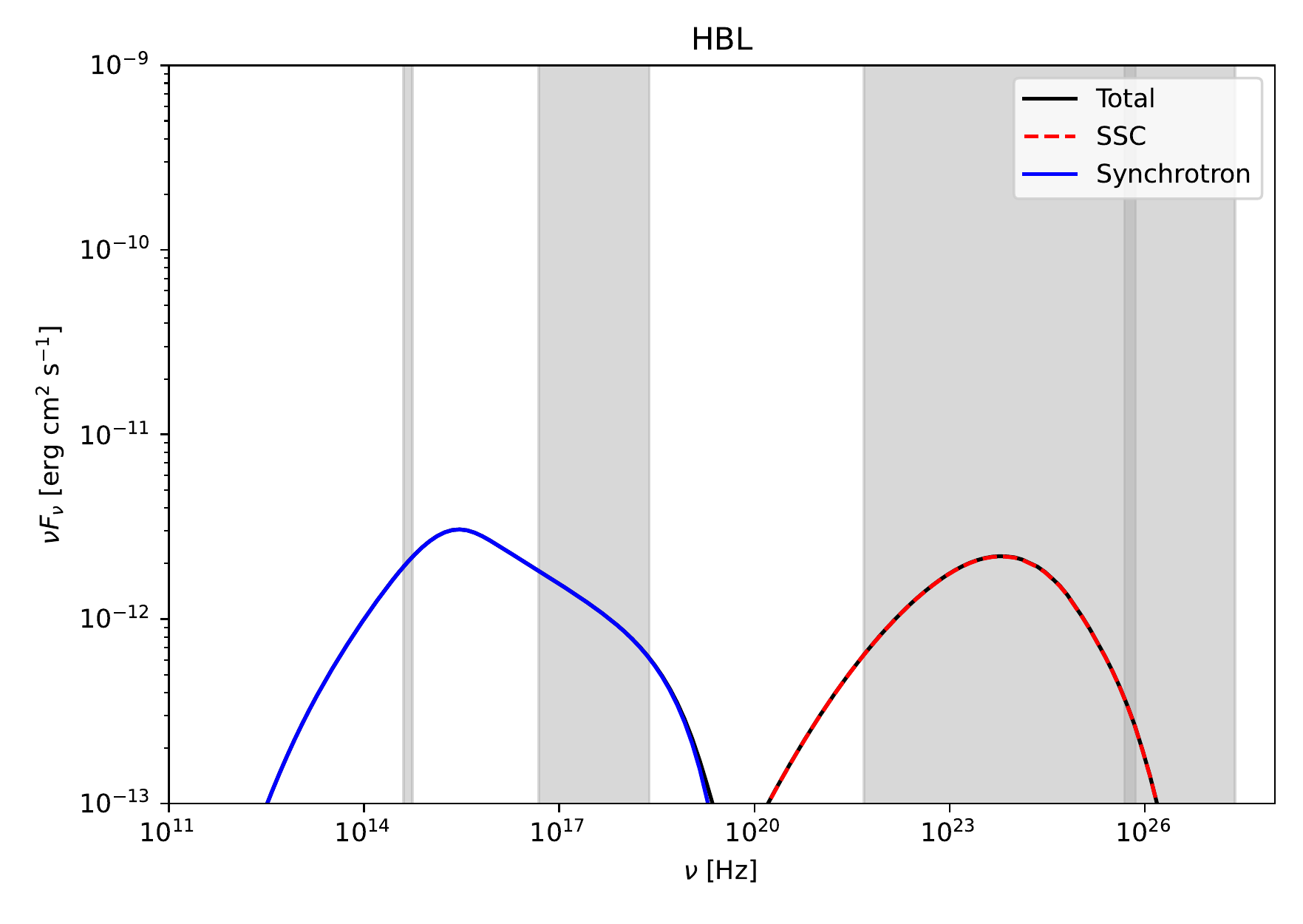}{.5\textwidth}{}
    }
    \caption{
        Broad band spectral energy distributions for the representative FSRQ
        (left) and HBL (right) blazar cases. Spectral components are as
        indicated in the legend. The shaded areas indicate the frequency ranges
        for which the optical, X-ray, HE $\gamma$-ray and VHE $\gamma$-ray
        light curves have been derived. We note that there is an overlap in the
        frequency ranges of the HE and VHE
        $\gamma$-rays.  }
    \label{fig:SEDs}
\end{figure}

Figure~\ref{fig:SEDs} shows the SEDs of the two different representative blazar test cases.
The parameter sets used to generate those SEDs are defined and listed in Table~\ref{tab:blazar_params}.
The low-energy components in both cases are produced by synchrotron radiation (in the FSRQ case from radio up to optical frequencies; in the HBL case from radio up to X-rays).
In the FSRQ case X-ray emission (low-energy end of the high-energy component) is dominated by SSC while the $\gamma$-rays are dominated by the IC scattering of the broad-line region (BLR) photon field.
In the HBL case the low-energy synchrotron photons serve as target photons for IC scattering producing the high-energy component through SSC (soft $\gamma$-rays up to VHE $\gamma$-rays).

The shaded areas in Figure~\ref{fig:SEDs} indicate the frequency ranges over which fluxes have been integrated to obtain light curves.
The selected frequency ranges are motivated by the ranges of filters and instruments.
This also implies the existence of archival light curves from extensive monitoring campaigns on many blazars in these bands.
The ranges are: R~band in the optical regime, 0.2 - 10 keV in X-rays (as typically observed with e.g. Swift-XRT), 20 MeV - 300 GeV in high-energy (HE) $\gamma$-rays (the sensitivity range of the \textit{Fermi}-LAT), and 100 GeV - 10 TeV in very high-energy (VHE) $\gamma$-rays (as observed with e.g. H.E.S.S, MAGIC, and VERITAS).
Although detailed light curves also exist in radio  wavelengths ($\sim$~GHz), we do not consider them in this work since the emission region considered here is optically thick at those frequencies, and most of the radio emission is produced in the larger-scale jet structures.

\begin{deluxetable*}{llcc}
    \tablecaption{
        Initial model parameters for the representative FSRQ and HBL blazar
        cases. These parameters remain fixed unless they represent the
        varying parameter of the respective simulation realization (see
        Section~\ref{sec:variations})
    }
    \label{tab:blazar_params}
    \tablewidth{0pt}
    \tablehead{
    \colhead{Definition} & \colhead{Symbol} & \colhead{FSRQ} & \colhead{HBL}
    }
    \startdata
        Magnetic field  & $B_0$ & 1.70 G & 0.40 G \\
        Blob radius & $R$ & $ 3.0 \times 10^{16}$  cm & $3.0 \times 10^{16}$ cm \\
        Ratio of the acceleration to escape time scales & $\eta$ &  1.00 & 1.00  \\
        Escape time scale & $t_{\text{esc}}$  &  $10.0\ R/c$ & $10.0\ R/c$ \\
        Redshift to the source & $z$ &  1.0 &  1.0 \\
        Minimum Lorentz factor of the electron injection spectrum & $\gamma_{\text{min}}$ &  $1.0 \times 10^2$ & $1.0 \times 10^4$\\
        Maximum Lorentz factor of the electron injection spectrum & $\gamma_{\text{max}, 0}$ & $1.0 \times 10^4$ & $1.0 \times 10^6$\\
        Bulk Lorentz factor & $\Gamma$ &  20.0 & 20.0 \\
        Observing angle relative to the axis of the BH jet & $\theta_{\text{obs}}$  &  $5.0 \times 10^{-2}$ rad & $5.0 \times 10^{-2}$ rad\\
        Doppler factor & $\delta$ &  20.0 & 20.0 \\
        Electron injection index & $q_0$ &  2.8 & 2.5\\
        Co-moving injection luminosity  of the electron spectrum  & $L_{\text{inj}, 0}$ &  $5.0 \times 10^{43}$ erg/s & $1.0 \times 10^{42}$ erg/s\\
        Mass of the super massive black hole & $M_{\text{BH}}$ &  $8.5 \times 10^8 M_{\odot}$  & $8.5 \times 10^8 M_{\odot}$ \\
        Eddington ratio & $l_{\text{edd}}$ &  $1.0 \times 10^{-1}$ & $1.0 \times 10^{-4}$\\
        Initial location of the blob along jet axis  & $d$ &  $6.5 \times 10^{17}$ cm & $6.5 \times 10^{17}$ cm\\
        Radius of the BLR  & $R_{\text{BLR}}$ &  $6.7 \times 10^{17}$ cm & - \\
        Effective temperature of the BLR  & $T_{\text{eff}}$ & $ 5.0 \times 10^{4}$  K & - \\
        Effective luminosity of the BLR  & $L_{\text{BLR}}$ &  $1.0 \times 10^{45}$ erg/s & - \\
    \enddata
\end{deluxetable*}

\subsection{Stochastic Variations} \label{sec:variations}

In order to produce the light curves, we introduce a stochastic process that results in the variation of one of the emission region parameters.
This work does not attempt to explain the stochastic process itself.
It is merely assumed that a process exists that results in the stochastic variations of individual source parameters considered here.
One may speculate that the stochastic variation of the jet parameters is related to the stochastic variations observed in accretion disks \citep{2018ApJ...859L..21C}.

The stochastic variations are generated with the algorithm developed by \cite{1995A&A...300..707T}.
In the generation procedure an underlying spectrum --- specifically, a pure power law in this work --- is assumed for the stochastic process' power spectral density (PSD) in temporal frequency.
Using the underlying PSD spectrum as a baseline a unique periodogram\footnote{Also typically known as the Fourier transform of a time series that contains power and phase information for each temporal frequency.} is produced by means of a Monte Carlo scheme in which complex numbers are generated for the periodogram given the power of the underlying PSD at a given temporal frequency.
This effectively produces a noisy version of the underlying spectrum and, in the case of an underlying pure power law spectrum, conserves the power law index.
A complex valued time series is then obtained from the inverse Fourier transform of the generated periodogram whose real values are considered to create parameter variations.

The observed light curve variability of accreting systems such as AGN and X-ray binaries has shown evidence of colored noise \citep[e.g.][]{2014ApJ...785...76P,2018A&A...620A.185N,Goyal_2021}.
Therefore, in this work all variations were generated for an underlying PSD being a pure power law with index of $\alpha=-2$ such that $P(f) \propto f^{\alpha}$ where $P(f)$ represents the PSD as function of temporal frequency $f$.

For each generated time series the parameter variations are obtained for the maximum electron Lorentz factor, $\gamma_{\text{max}}$, the electron injection luminosity, $L_{\text{inj}}$, the magnetic field strength, $B$, and the electron injection spectral index, $q$, and introduced into the emission region in independent realizations of the simulation.
These parameters are naturally expected to be subject to changes over time making them prime candidates for causing variability.
Induced variability from the underlying accretion flow will most likely impact jet power and, thus, injection luminosity.
Magnetic fields are significantly affected by shock-compression and/or magnetic reconnection which in turn will lead to variability.
Particle spectral indices and electron Lorentz factors are characteristic of the considered acceleration mechanisms and can even vary for the same mechanisms depending on the state of the environment \citep[e.g., shock obliquity in the case of mildly relativistic shocks; see][]{SB12}.

\begin{figure}[ht]
    \centering
    \includegraphics[width=.8\textwidth]{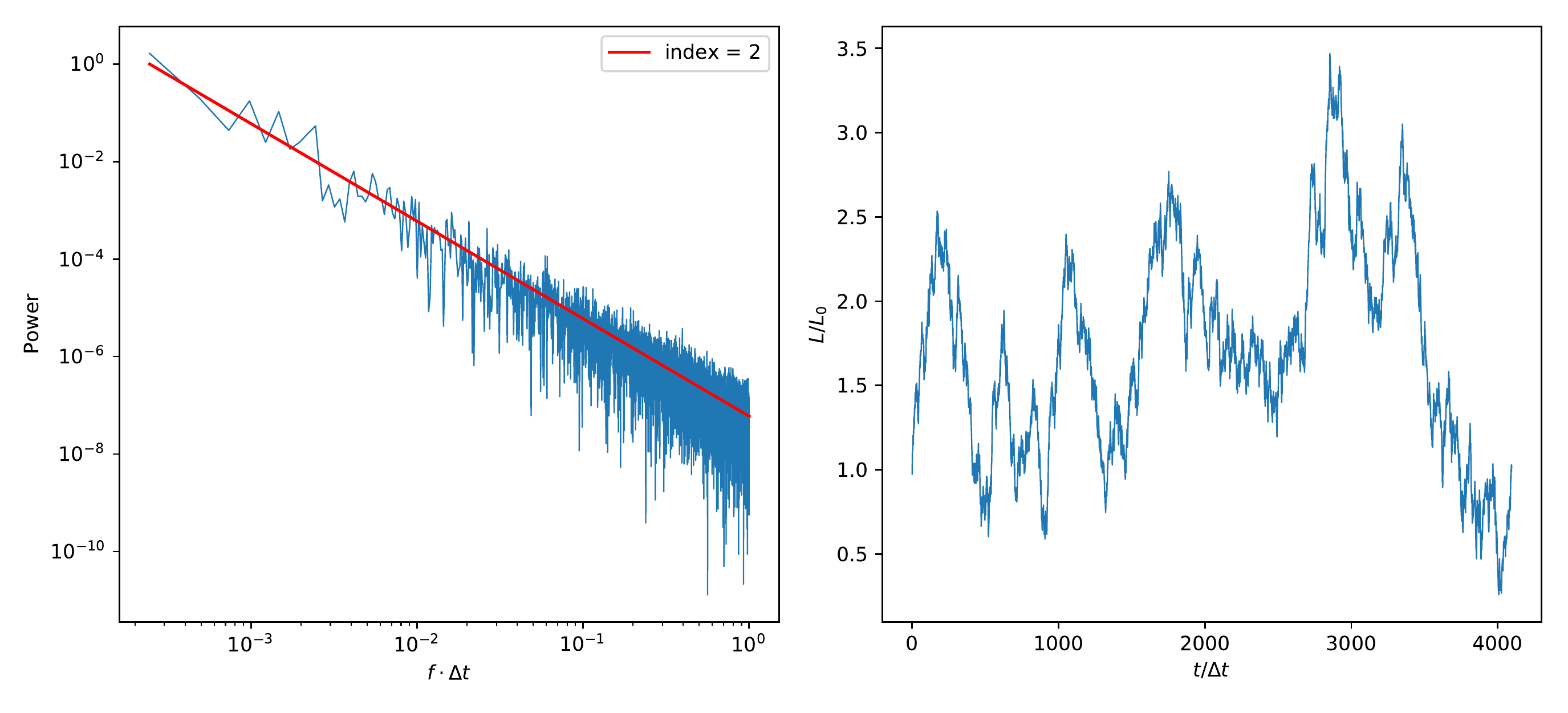}
    \caption{
        Example of a generated stochastic variation. Left: The PSD obtained
        from the produced periodogram (blue) follows the underlying pure power
        law (red). Right: The variation obtained from the inverse FFT of the
        periodogram exibiting red noise. Specifically, this represents
        the value of $L_\text{inj}$ compared to the initial value at each time
        step
    }
        \label{fig:example_variation}
\end{figure}

The parameter variations are generated by assuming that the time series represents the scale factor of the initial parameter value at each time step, except for the spectral index variations for which the time series represents the constant added to the initial spectral index value instead.
The time series is further manipulated by adding a single constant value throughout to produce maximum Lorentz factor, injection luminosity, and magnetic field variations such that the occurence of non-physical parameter values is prevented (ensuring that $\gamma_\text{max} > \gamma_\text{min}$ and $L_{inj}, B > 0$).
In the case of spectral index variations the time series is normalized within the range [-1,1] for the same reason.
The overall minimum and maximum values of the aforementioned parameters during the variations are as follows: $\gamma_\text{max}/\gamma_{\text{max},0} \in [0.1, 5.6]$; $L_\text{inj} / L_{\text{inj},0} \in [0, 5.5]$; $B / B_0 \in [0.1, 5.6]$ where $X_0$ represents the respective parameter's initial value in Table \ref{tab:blazar_params}.
The minimum and maximum spectral index values are in the range $q \in [q_0 -1, q_0 +1]$.
Note that the aforementioned modifications to time series do not change the PSD power law index of the underlying parameter variations.
An example of a variation generated with this method is shown in Figure~\ref{fig:example_variation}.

The time and frequency intervals of the variations produced by the algorithm may be chosen arbitrarily.
In the simulations a total of 4000 steps have been used for a fixed time interval of 2 hours in the emission region's rest frame for both blazar cases, corresponding to 12 minute intervals in the observer's frame.
A total of 100 time series with different random seeds were generated and used to simulate 100 individual realizations for each permutation of blazar case and varied parameter (8 permutations in total).

\section{Results and Discussions} \label{sec:results_discuss}

\subsection{Light Curves}

\begin{figure}[ht]
    \gridline{
        \centering
        \fig{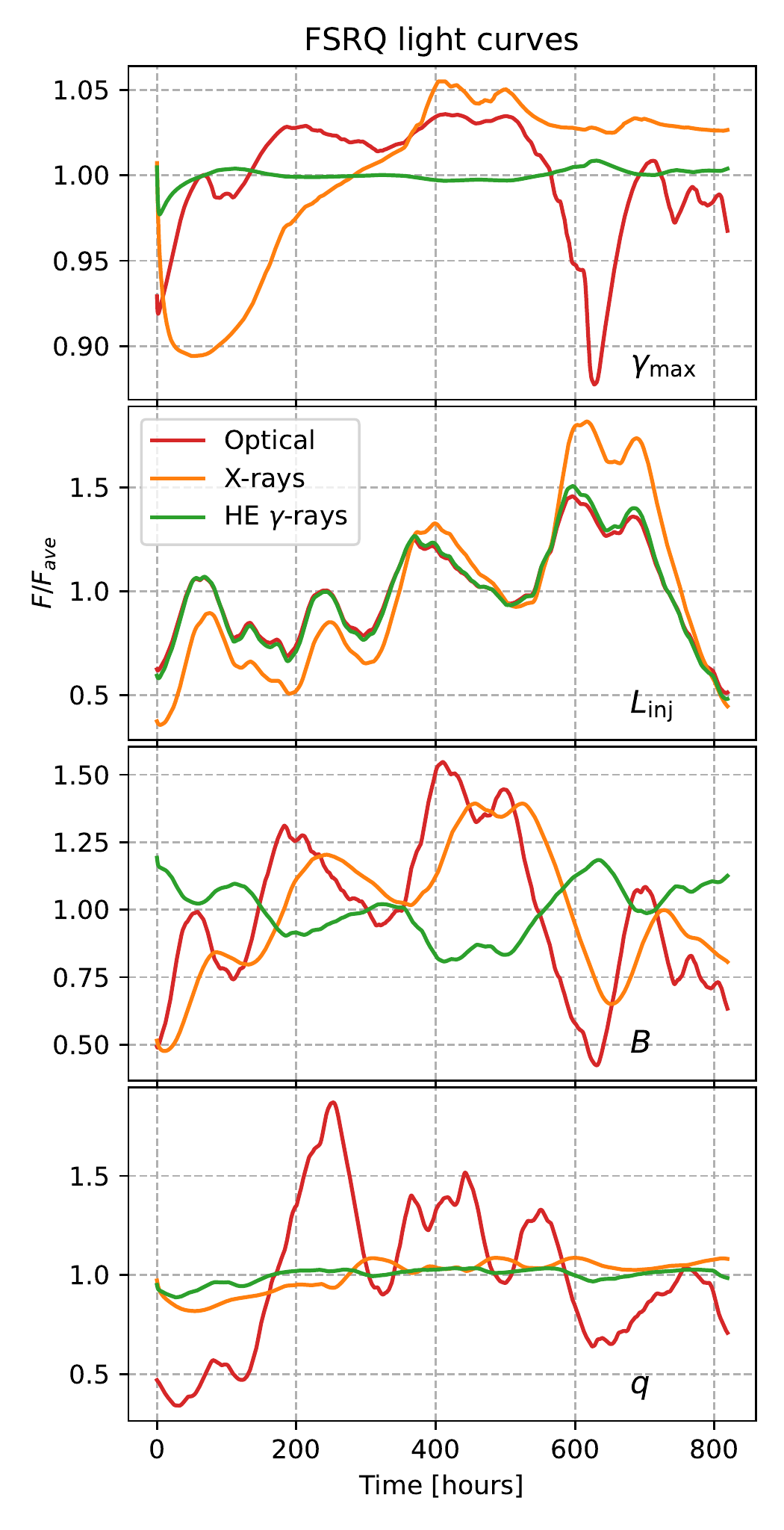}{.45\textwidth}{}
        \fig{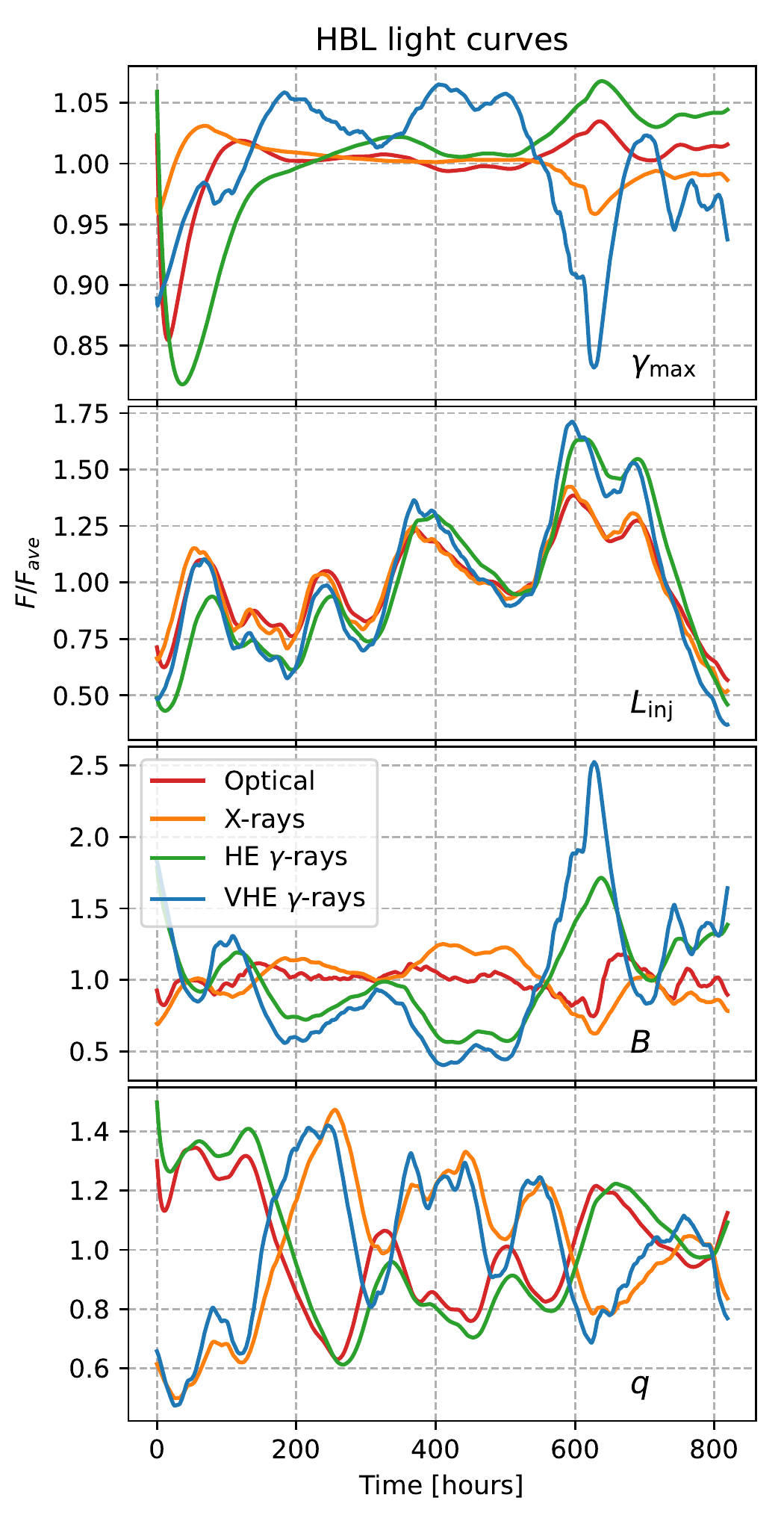}{.45\textwidth}{}
    }
    \caption{
        Simulated light curves exhibiting variability for the same stochastic
        variation applied to different emission region parameters as indicated.
        Refer to symbols in Table \ref{tab:blazar_params}. Note that the
        optical and HE $\gamma$-ray light curves are practically identical for
        the FSRQ case and injection luminosity variation. }
    \label{fig:light_curves}
\end{figure}

Figure~\ref{fig:light_curves} shows the resulting multi-wavelength light curves for one representative realization for each of the 8 cases (FSRQ vs. HBL; $\gamma_\text{max}$, $L_\text{inj}$, $B$, $q$ variation).
The figure illustrates that variations in injection luminosity, magnetic field, and spectral index variations produce significant flux variability across all photon frequencies with flux changes of $\pm \sim 50\%$ compared to the average flux.
Qualitatively, the variability produced by the different parameter variations is also distinct from one another.
Note that the VHE $\gamma$-ray light curve is omitted from the FSRQ results due to its insignificant flux (see Figure~\ref{fig:SEDs}).

Flux changes due to $\gamma_\text{max}$ variations are less volatile with only about $\pm 5-10\%$.
However, FSRQ optical and HBL VHE $\gamma$-ray light curves show the most significant drops in flux at the point where $\gamma_\text{max}$ is at a minimum.
This sensitivity indicates that these respective wavelengths are produced by the highest-energy electrons in the electron population.

Correlated variability for injection luminosity variations can be seen for both the FSRQ and HBL case evident from the synchronous flux increases and decreases across all wavelengths due to uniform changes in the number of radiating electrons across the whole electron spectrum.
Note that the optical and HE $\gamma$-ray light curves in the FSRQ case are practically identical which is a consequence of being produced by electrons of similar energies.

In the case of magnetic field variations for the FSRQ the optical and X-rays show high sensitivity to such changes, whereas the reaction in HE $\gamma$-rays is weaker.
This is expected since the HE $\gamma$-rays are dominated by EC processes that depend only indirectly on the magnetic field through increased synchrotron cooling of the radiating electrons.
The HBL case shows comparable variability in all wavelengths, however, in this specific run the SSC radiation is much more sensitive to the magnetic field changes than the synchrotron.

The prominent variability patterns produced by spectral index variations in both blazars are similar to the variability exhibited from variations in injection luminosity since the normalization of the electron distribution depends on both the injection luminosity and the spectral index (see Eq.~\ref{eq:inj_norm} in appendix \ref{app:lum_norm}).
However, unlike for the injection luminosity the spectral index variations induce an energy dependent variation of the electron distribution resulting in different amplitudes of the variability in the light curves.
Only the EC radiation components in the FSRQ show little variability over the course of the parameter variation.
This is because the $\gamma$-ray spectrum pivots around a central energy in the HE regime and the X-rays are produced by SSC radiation of low-energy electrons.
A lowering (hardening) of the spectral index enhances the synchrotron target photon field for SSC emission but depletes low-energy electrons, so that the two effects almost cancel each other out.

\begin{deluxetable*}{lcc}
    \tablecaption{
        FSRQ cooling time-scales in the observer's frame. }
    \label{tab:fsrq_cooling}
    \tablewidth{0pt}
    \tablehead{
        \colhead{Waveband / energy} & \colhead{Time-scale (s)} & \colhead{Temporal Frequency (Hz)}
    }
    \startdata
        R Band (658 nm)             &   $4.42 \times 10^{2}$   &   $2.26 \times 10^{-3}$ \\ 
        X-ray (0.2 keV)             &   $2.75 \times 10^{5}$   &   $3.64 \times 10^{-6}$ \\ 
        X-ray (10 keV)              &   $3.89 \times 10^{4}$   &   $2.57 \times 10^{-5}$ \\ 
        HE $\gamma$-ray (100 MeV)   &   $6.35 \times 10^{3}$   &   $1.57 \times 10^{-4}$ \\ 
        HE $\gamma$-ray (30 GeV)    &   $3.67 \times 10^{2}$   &   $2.73 \times 10^{-3}$ \\ 
    \enddata
\end{deluxetable*}

\begin{deluxetable*}{lcc}
    \tablecaption{
        HBL cooling time-scales in the observer's frame. }
    \label{tab:hbl_cooling}
    \tablewidth{0pt}
    \tablehead{
        \colhead{Waveband / energy} & \colhead{Time-scale (s)} & \colhead{Temporal Frequency (Hz)}
    }
    \startdata
        R Band (658 nm)             &  $5.57 \times 10^{4}$   &   $1.79 \times 10^{-5}$ \\ 
        X-ray (0.2 keV)             &  $5.41 \times 10^{3}$   &   $1.85 \times 10^{-4}$ \\ 
        X-ray (10 keV)              &  $7.65 \times 10^{2}$   &   $1.31 \times 10^{-3}$ \\ 
        HE $\gamma$-ray (100 MeV)   &  $1.02 \times 10^{5}$   &   $9.78 \times 10^{-6}$ \\ 
        HE $\gamma$-ray (30 GeV)    &  $5.90 \times 10^{3}$   &   $1.69 \times 10^{-4}$ \\ 
        VHE $\gamma$-ray (200 GeV)  &  $7.41 \times 10^{3}$   &   $1.35 \times 10^{-4}$ \\ 
        VHE $\gamma$-ray (10 TeV)   &  $1.48 \times 10^{2}$   &   $6.74 \times 10^{-3}$ \\ 
    \enddata
\end{deluxetable*}

\subsection{Power Spectral Densities} \label{sec:psd_results}

\begin{figure}[ht]
    \gridline{
        \fig{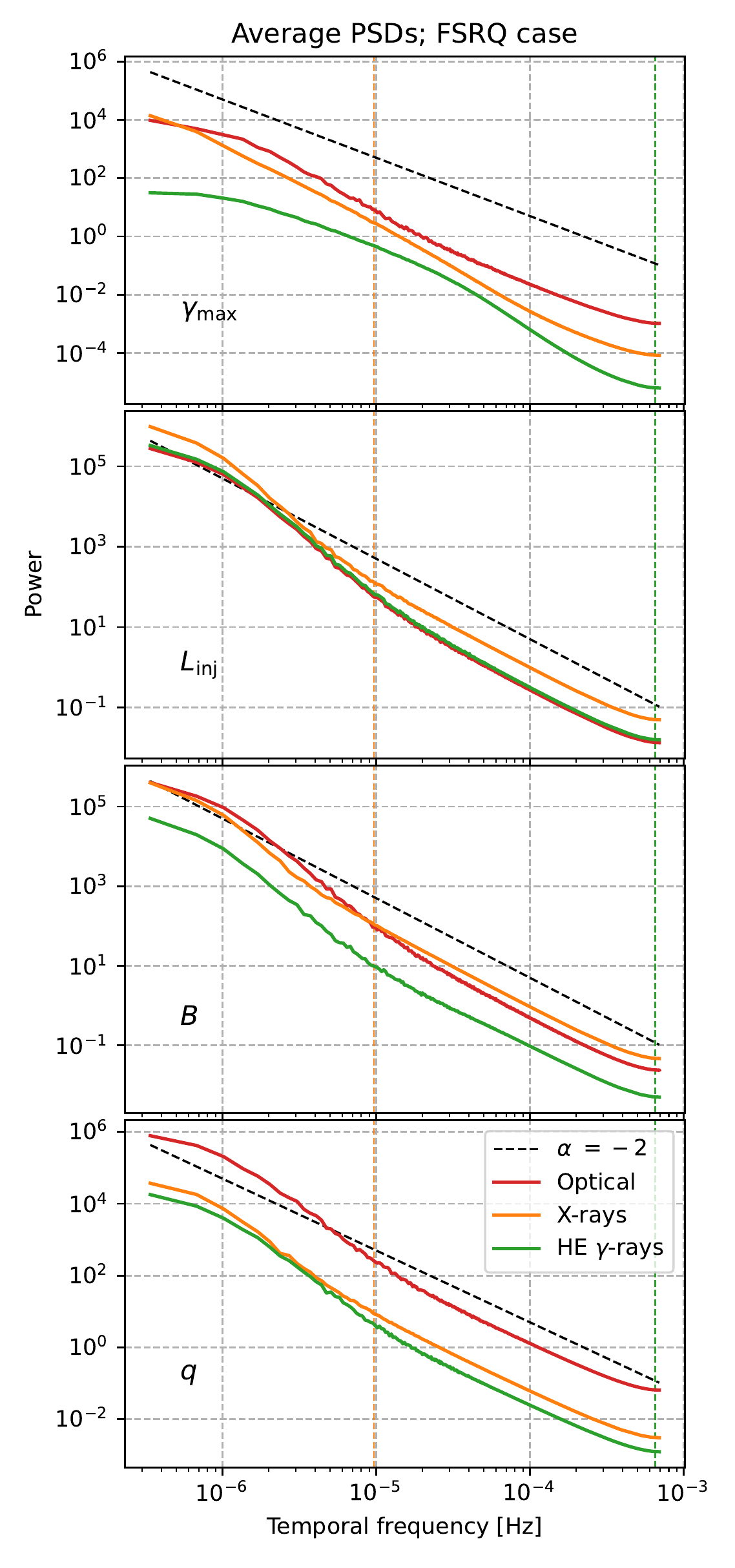}{.4\textwidth}{}
        \fig{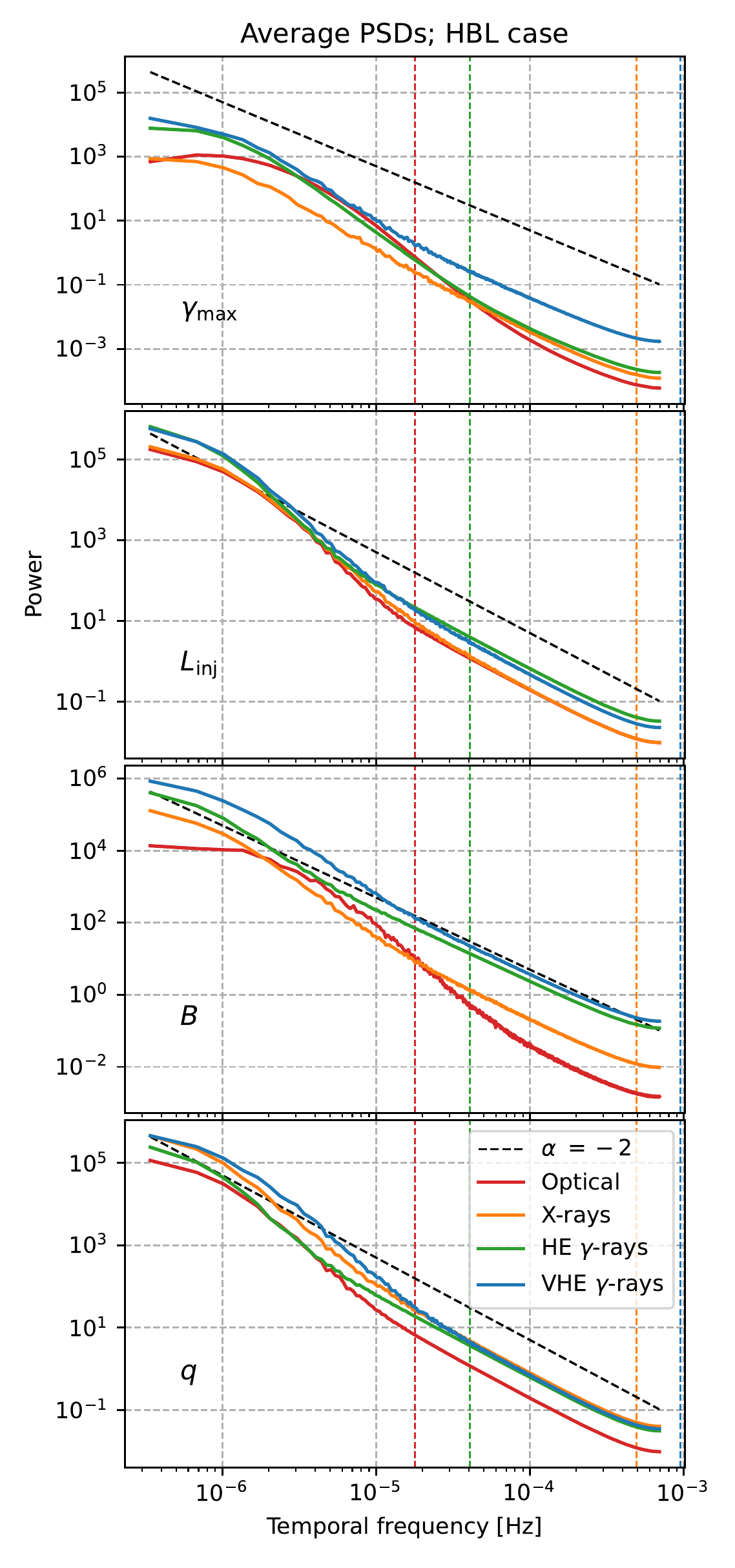}{.4\textwidth}{}
    }
    \caption{
        Average light curve PSDs for the FSRQ (left) and HBL (right) cases for
        the different parameter variations. Vertical dotted lines indicate
        average characteristic cooling frequencies (cf.
        Tables~\ref{tab:fsrq_cooling} and \ref{tab:hbl_cooling}) corresponding
        to the waveband indicated by color in the legend.}
    \label{fig:PSD_ave}
\end{figure}

The average results for the light curve PSDs of the 100 different realizations for each case are presented in Figure~\ref{fig:PSD_ave}.
The individual simulation realizations in comparison to the averages can be found in Figures~\ref{fig:EC-PSD-ave_vs_indiv} and \ref{fig:SSC-PSD-ave_vs_indiv} in Appendix~\ref{app:indiv_results}, respectively.
The average PSDs follow closely the underlying power law of the parameter variation and do not show significant differences in PSD power law index for the different wavelengths.
This result is similar to what \cite{2014ApJ...791...21F,2015ApJ...809...85F} found in their analytical analysis of variability in Fourier space with respect to matching PSD power law indices for different wavelengths.
They also predict spectral breaks between $\sim 10^{-6}$ and $\sim 10^{-4}$Hz that arise from frequencies related to the light crossing and escape time-scales of electrons and photons in the emission region.
However, these clear breaks do not appear in our results because these characteristic frequencies, $t_\text{esc} \approx 10^{7}$~s, corresponding to $f_\text{esc} \approx 10^{-7}$~Hz, fall outside the bounds of frequencies considered in this work for the chosen model parameters (see  Table~\ref{tab:blazar_params}).

Similar PSD power law indices across the different wavelengths supplement the argument that one-zone models lack the complexity to accurately produce observed variability patterns of blazars, which have been found to exhibit differently coloured noise between the low-energy synchrotron and high-energy emission components \cite[see][and references therein]{2020MNRAS.494.3432G}.
The lack of distinctive differences in the PSDs also makes it unlikely to be an effective diagnostic tool for identifying progenitor and/or cooling mechanisms.
However, since they do follow the underlying variation's power law closely they can be a useful tool to probe the stochastic behaviour of variations that cause variability.

Notably, several individual light curve PSDs smooth out at the higher frequencies which highly contrasts the Monte Carlo generated variations (cf. Figure~\ref{fig:example_variation} with Figures~\ref{fig:EC-PSD-ave_vs_indiv} and \ref{fig:SSC-PSD-ave_vs_indiv}).
This is a consequence of the delayed response of the emission processes that cannot keep up with the changes induced by the variations on the shortest timescales.
This is especially evident for the emission processes with longer cooling time-scales; see, for example, the X-ray PSD of the FSRQ case (Figure~\ref{fig:EC-PSD-ave_vs_indiv}) and practically all PSDs of the HBL case (Figure~\ref{fig:SSC-PSD-ave_vs_indiv}).

\subsection{Cross-correlation functions}

\begin{figure}[ht]
    \gridline{
        \fig{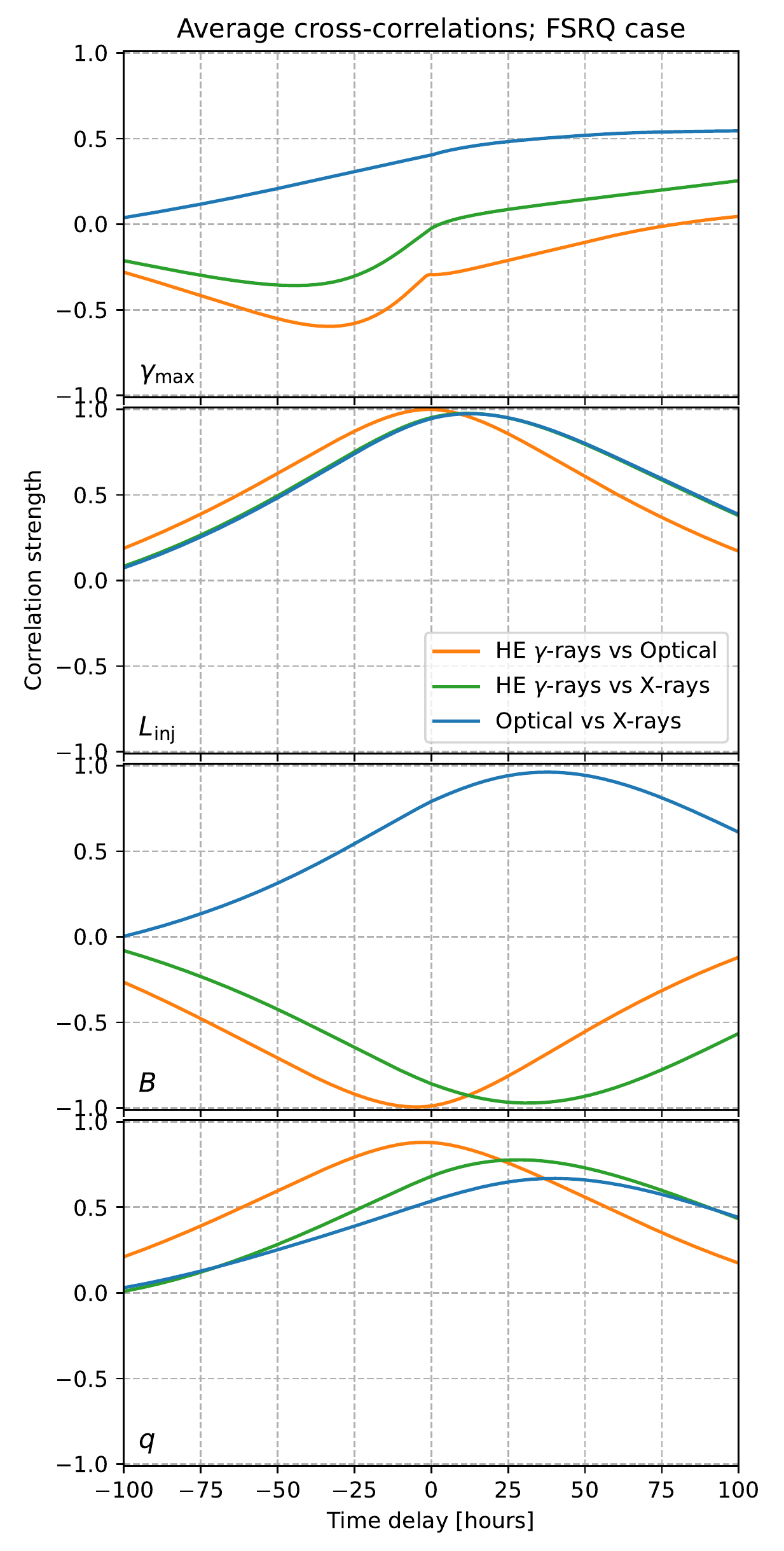}{.4\textwidth}{}
        \fig{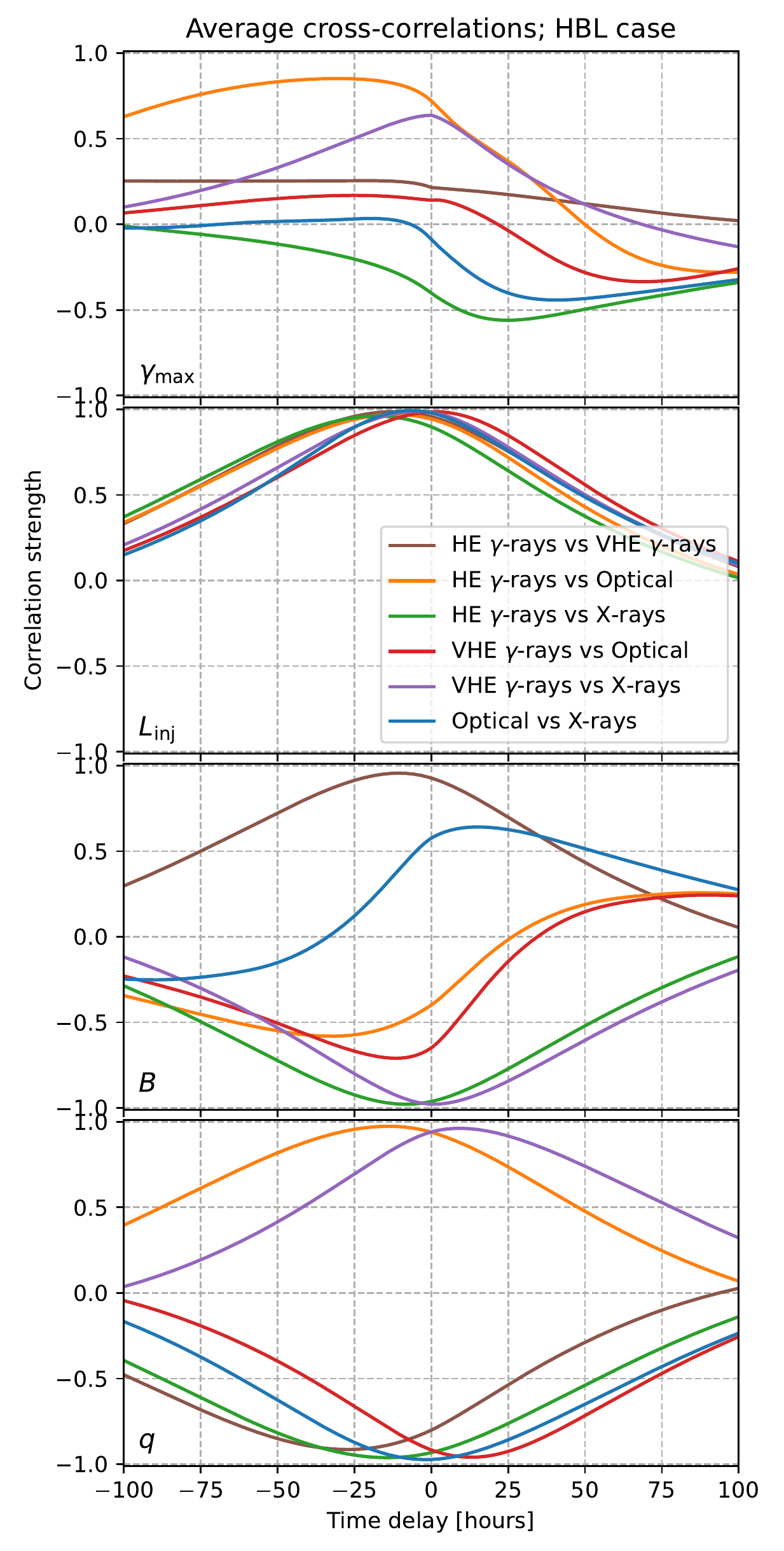}{.4\textwidth}{}
    }
    \caption{
        Average multi-wavelength cross-correlations for the EC (left) and SSC
        (right) cases for the different parameter variations.}
    \label{fig:Corr_ave}
\end{figure}

The average multi-wavelength cross-correlations of the 100 simulation realizations for each case are presented in Figure~\ref{fig:Corr_ave}.
The individual realizations in comparison to the averages can be found in Figures~\ref{fig:EC-Corr-ave_vs_ind}, \ref{fig:SSC-Corr-inj-ave_vs_indiv}, \ref{fig:SSC-Corr-mag-ave_vs_indiv}, and \ref{fig:SSC-Corr-spc-ave_vs_indiv} in Appendix \ref{app:indiv_results}, respectively.
Furthermore the concentrations of cross-correlation peaks of all realizations are presented in Appendix \ref{app:cc_peaks} with the delay times and correlation strengths of the average peaks listed in Tables \ref{tab:fsrq-corr_peaks} and \ref{tab:hbl-corr_peaks}.
The results show strong distinctive features across the different scenarios.

The maximum Lorentz factor variation shows the least consistent behavior between all simulations in both blazars evident from the weak correlations on average and the large deviations between individual realizations.
The only exception in this regard is the HBL HE $\gamma$-ray vs optical cross-correlation function which does present a consistent feature peaking at $-30 \pm 23$ hours with correlation strength $0.8 \pm 0.3$.
This is due to the HE $\gamma$-rays and optical photons being produced by the low-energy electrons in the HBL that are similarly, weakly affected by changes in $\gamma_\text{max}$ leading to a positive correlation.
The fact that the optical precedes the HE $\gamma$-rays stems from HE $\gamma$-rays being SSC radiation which depends on the optical synchrotron photons, even though the cooling time-scales of electrons radiating in these wavebands are comparable.
A similar counterpart is absent in the FSRQ case.

The injection luminosity variations produce consistently strong correlations ($\approx 1.0$) which deviate little in time (standard deviations $\le 2$ hours) between all wavebands for both blazars.
The time delays of each peak are indicative of the differences in cooling time scales of electrons radiating in the respective wavebands.
The FSRQ optical and HE $\gamma$-rays produced by high-energy electrons precede the X-rays produced by low-energy electrons since the cooling time-scales of electrons radiating in optical and HE $\gamma$-rays are shorter than those of electrons radiating in X-rays (cf. Table~\ref{tab:fsrq_cooling}).
The time delays presented in the HBL cross-correlations further confirm this, as the X-ray band, with the shortest cooling time-scale, precedes all other wavebands.
Although electrons radiating in VHE $\gamma$-rays have similar cooling time-scales compared to X-rays, the production due to SSC depends on the synchrotron production.
This explains how the optical variability, with a longer electron cooling time-scale, can precede VHE $\gamma$-ray variability.

The cross-correlation functions for magnetic field variations also produce strong correlations/anti-correlations ($\approx \pm 1$).
However, contrary to injection luminosity variations these correlations/anti-correlations show larger deviations in time delays (standard deviations up to 7 hours) among individual realizations.
The exception to these results are HBL cross-correlations involving optical.
In these cases, a consistent behaviour in all simulations is observed, even though the cross-correlation peaks are much weaker ($\approx 0.6$) and more scattered (time delay standard deviations up to 34 hours).
This can be attributed to the fact that the magnetic field changes have their most prominent effect on the optical emission (cf. Equation~\ref{eq:t_cool,sync}) combined with the long cooling time-scale compared to other wavebands leading to weaker correlations/anti-correlations.

Furthermore, both blazars exhibit correlations between wavebands that share a cooling mechanism and anti-correlations when the cooling mechanisms differ.\footnote{That is in the HBL case: ``SSC'' vs ``synchrotron'' (e.g. HE $\gamma$-rays vs optical) $\Rightarrow$ anti-correlation; in the FSRQ case: ``EC'' vs ``synchrotron'' $\Rightarrow$ anti-correlation.}
The optical vs X-ray correlation in the FSRQ is a special case and seems to contrast the results of the HBL case.
However, in the FSRQ case, the X-rays are produced by SSC, and are thus linked to the optical emission -- which is synchrotron -- resulting in a positive correlation.
On the other hand, the HE $\gamma$ rays are EC emission, which react only weakly to magnetic field changes providing an anti-correlation with both the optical and the X-ray emission.

For the case of spectral index variations the FSRQ and HBL results differ.
Correlation functions produced by the FSRQ are noticeably less consistent compared to the HBL leading to weaker average correlations ($\approx 0.8$) over a much wider range of time delays (standard deviations up to 48 hours) while in the HBL case correlations/anti-correlations are strong ($|DCF| > 0.9$) and time delays are found in a much narrower range (standard deviations $\le 6$ hours).
This discrepancy is due the specific electron energy range producing both FSRQ X-rays and HE $\gamma$-rays.
These electrons range from the lowest energies up to intermediate energies and beyond compared to the total electron spectrum.
This would then include electron energies higher than the ``pivot'' electron energy, $\epsilon_\text{pivot}$, below which the number of electrons does not change significantly in spite of changes in electron spectral index.
This explains the small amplitude variability in FSRQ X-rays and HE $\gamma$-rays compared to optical (cf. Figure~\ref{fig:light_curves}).
The HBL results show that correlations between wavebands arise for wavebands produced by equivalent energy electrons and anti-correlations arise otherwise.
That is ``high-energy'' vs ``low-energy'' emissions (e.g. VHE $\gamma$-rays vs optical) exhibit anti-correlations.

\begin{deluxetable*}{llcc}
    \tablecaption{
        FSRQ case cross-correlation peaks. }
    \label{tab:fsrq-corr_peaks}
    \tablewidth{0pt}
    \tablehead{
        \colhead{Variation parameter} & \colhead{Wavelengths} & \colhead{Time delay (hours)} & \colhead{Correlation strength}
    }
    \decimals
    \startdata
        {}                      & HE $\gamma$-rays vs X-rays    & 12 $\pm$ 76       & -0.2 $\pm$ 0.5 \\
        {$\gamma_{\text{max}}$} & HE $\gamma$-rays vs Optical   & -27 $\pm$ 25      & -0.6 $\pm$ 0.4 \\
        {}                      & Optical vs X-rays             & 30 $\pm$ 76       & 0.5 $\pm$ 0.4 \\
        \hline
        {}                      & HE $\gamma$-rays vs X-rays    & 12 $\pm$ 2        & 1.0 $\pm$ $8 \times 10^{-3}$ \\
        {Injection luminosity}  & HE $\gamma$-rays vs Optical   & -0.8 $\pm$ 0.1    & 1.0 $\pm$ $6 \times 10^{-5}$ \\
        {}                      & Optical vs X-rays             & 13 $\pm$ 2        & 1.0 $\pm$ $8 \times 10^{-3}$ \\
        \hline
        {}                      & HE $\gamma$-rays vs X-rays    & 32 $\pm$ 7        & -1.0 $\pm$ $1 \times 10^{-2}$ \\
        {Magnetic field}        & HE $\gamma$-rays vs Optical   & -5 $\pm$ 0.9      & -1.0 $\pm$ $2 \times 10^{-3}$ \\
        {}                      & Optical vs X-rays             & 38 $\pm$ 7        & 1.0 $\pm$ $2 \times 10^{-2}$ \\
        \hline
        {}                      & HE $\gamma$-rays vs X-rays    & 25 $\pm$ 33       & 0.8 $\pm$ 0.3 \\
        {Spectral index}        & HE $\gamma$-rays vs Optical   & -3 $\pm$ 12       & 0.9 $\pm$ $8 \times 10^{-2}$ \\
        {}                      & Optical vs X-rays             & 30 $\pm$ 48       & 0.7 $\pm$ 0.4 \\
    \enddata
\end{deluxetable*}

\begin{deluxetable*}{llCC}
    \tablecaption{
        HBL case cross-correlation peaks. }
    \label{tab:hbl-corr_peaks}
    \tablewidth{0pt}
    \tablehead{
        \colhead{Variation parameter} & \colhead{Wavelength} & \colhead{Time delay (hours)} & \colhead{Correlation strength}
    }
    \decimals
    \startdata
        {}                      & HE $\gamma$-rays vs X-rays            & 27 $\pm$ 29   & -0.5 $\pm$ 0.5 \\
        {}                      & HE $\gamma$-rays vs VHE $\gamma$-rays & -1 $\pm$ 72   & 0.3 $\pm$ 0.5 \\
        {$\gamma_{\text{max}}$}  & HE $\gamma$-rays vs Optical           & -30 $\pm$ 23  & 0.8 $\pm$ 0.3 \\
        {}                      & VHE $\gamma$-rays vs X-rays           & 14 $\pm$ 41   & 0.5 $\pm$ 0.6 \\
        {}                      & VHE $\gamma$-rays vs Optical          & 34 $\pm$ 46   & -0.1 $\pm$ 0.6 \\
        {}                      & Optical vs X-rays                     & 25 $\pm$ 51   & -0.4 $\pm$ 0.5 \\
        \hline
        {}                      & HE $\gamma$-rays vs X-rays            & -18 $\pm$ 2   & 1.0 $\pm$ $ 2 \times 10^{-2} $  \\
        {}                      & HE $\gamma$-rays vs VHE $\gamma$-rays & -12 $\pm$ 2   & 1.0 $\pm$ $ 8 \times 10^{-3} $  \\
        {Injection luminosity}  & HE $\gamma$-rays vs Optical           & -12 $\pm$ 2   & 1.0 $\pm$ $ 9 \times 10^{-3} $  \\
        {}                      & VHE $\gamma$-rays vs X-rays           & -4 $\pm$ 1    & 1.0 $\pm$ $ 4 \times 10^{-3} $  \\
        {}                      & VHE $\gamma$-rays vs Optical          & 1 $\pm$ 1     & 1.0 $\pm$ $ 3 \times 10^{-3} $  \\
        {}                      & Optical vs X-rays                     & -6 $\pm$ 1    & 1.0 $\pm$ $ 3 \times 10^{-3} $  \\
        \hline
        {}                      & HE $\gamma$-rays vs X-rays            &  -9 $\pm$  3  & -1.0 $\pm$ $1 \times 10^{-2}$ \\
        {}                      & HE $\gamma$-rays vs VHE $\gamma$-rays & -11 $\pm$  5  &  1.0 $\pm$ $2 \times 10^{-2}$ \\
        {Magnetic field}        & HE $\gamma$-rays vs Optical           & -28 $\pm$ 34  & -0.5 $\pm$ 0.4 \\
        {}                      & VHE $\gamma$-rays vs X-rays           &   1 $\pm$  2  & -1.0 $\pm$ $9 \times 10^{-3}$ \\
        {}                      & VHE $\gamma$-rays vs Optical          &  -9 $\pm$ 22  & -0.7 $\pm$ 0.4 \\
        {}                      & Optical vs X-rays                     &  12 $\pm$ 27  &  0.6 $\pm$ 0.4 \\
        \hline
        {}                      & HE $\gamma$-rays vs X-rays            & -15 $\pm$ 5   & -1.0 $\pm$ $2 \times 10^{-2}$ \\
        {}                      & HE $\gamma$-rays vs VHE $\gamma$-rays & -28 $\pm$ 6   & -0.9 $\pm$ $3 \times 10^{-2}$ \\
        {Spectral index}        & HE $\gamma$-rays vs Optical           & -15 $\pm$ 4   &  1.0 $\pm$ $9 \times 10^{-3}$ \\
        {}                      & VHE $\gamma$-rays vs X-rays           &  10 $\pm$ 5   &  1.0 $\pm$ $3 \times 10^{-2}$ \\
        {}                      & VHE $\gamma$-rays vs Optical          &  13 $\pm$ 4   & -1.0 $\pm$ $2 \times 10^{-2}$ \\
        {}                      & Optical vs X-rays                     &  -1 $\pm$ 3   & -1.0 $\pm$ $9 \times 10^{-3}$ \\
    \enddata
\end{deluxetable*}

\section{Summary \& Conclusions}

This work presents a method for simulating long-term multi-wavelength variability of blazars by means of varying an emission region parameter in a stochastic manner in the framework of a time dependent one-zone leptonic blazar emission model.
The simulated variability for two representative blazar cases, one dominated in $\gamma$-rays by EC and the other by SSC, was investigated.

It was found that the simulated variability PSDs closely resemble the PSD shape of the progenitor variation, in this case a pure power law.
Hence, according to the model, the produced variability in blazar light curves likely reflects the variability characteristics of the relevant particle acceleration mechanism, which in this case were represented by the stochastic variations.
No significant distinctive differences are present among the different test cases.
This indicates that the one-zone emission model might be incapable of producing light curve variability that posses different PSD indices for different wavebands \citep{2020MNRAS.494.3432G}.

The PSDs resulting from our simulations did not present any conclusive spectral breaks at characteristic frequencies corresponding to cooling, light-crossing, or escape time-scales.
This is likely a consequence of the fact that the probed frequency range in this study was limited and, in most cases, did not include the frequencies where such breaks would be expected \citep[see][]{2014ApJ...791...21F,2015ApJ...809...85F}.
However, the frequencies corresponding to some electron cooling time-scales, were within the probed frequency range.
Hence, the results on the absence of corresponding spectral breaks seem unexpected and are still inconclusive.

The cross-correlation results showed distinctive features for most of the different scenarios (FSRQ vs. HBL, maximum Lorentz factor, injection luminosity, magnetic field and spectral index variations).
As expected, in the case of injection luminosity and spectral index variations, photons of different wavelengths produced by the same electron populations (low-energy vs high-energy) are usually strongly correlated.
The time delays obtained between the different wavelengths are primarily caused by differences in electron cooling time-scales.
This illustrates that cross-correlation analysis is a viable strategy to identify the photon frequencies produced by equivalent energy electrons and indicate relative cooling time-scales when it is known that the acceleration mechanism that drives the variability modifies the electron spectrum in a similar manner.

Overall, the cross-correlation functions presented similar features in both blazars for the respective variations.
Varying the maximum Lorentz factor of the electron spectrum produced inconsistent results across all simulations leading to weak correlations/anti-correlations on average.
HBL HE $\gamma$-rays vs optical are the only cross-correlation functions that presented a consistent feature due to the almost negligible effect of the maximum Lorentz factor variations on these wavelengths.
A similar counterpart is not present in the FSRQ cross-correlation comparisons.
Electron injection luminosity variations produced all-round strong correlations across all wavebands in all cases.
The variations of magnetic field strength showed strong anti-correlations between synchrotron and IC wavebands and strong correlations when wavebands share a common cooling mechanism.
However, comparisons with FSRQ X-rays and HBL optical produce noticeably weaker correlations/anti-correlations.
A varying electron spectral index leads to strong correlations between wavebands produced by equivalent energy electrons and strong anti-correlations otherwise except for FSRQ X-rays.
These distinctions between the different varied parameters suggest the possibility to identify the underlying changing quantities from multi-wavelength variability cross-correlations.

This provides a compelling argument to use cross-correlation results as a diagnostic tool to constrain dominant radiation mechanisms as well as changing quantities in the emission region that are the cause of the variability.
However, we acknowledge that such an analysis would be difficult to implement in practice since observational data are, more often than not irregularly sampled and true simultaneous multi-wavelength data with similar temporal sampling are scarce.
Therefore, the theoretical results presented in this paper are practically challenging to test against currently available observational data.
Strategies to test our results against observations are currently under development, and we defer their presentation to a future publication.

\section*{Acknowledgements}
We thank the anonymous referee for helpful and constructive feedback.
The work of M.B. is supported through the South African Research Chair Initiative of the National Research Foundation\footnote{Any opinion, ﬁnding, and conclusion or recommendation expressed in this material is that of the authors, and the NRF does not accept any liability in this regard.} and the Department of Science and Innovation of South Africa, under SARChI Chair grant No. 64789. M.Z. acknowledges postdoctoral financial support from LUTH, Observatoire de Paris.

\appendix

\section{Characteristic cooling time-scales of radiation mechanisms}\label{app:time-scales}

The total energy loss rate, $P(\gamma)$, of individual electrons with energy $\gamma m_e c^2$ -- where $\gamma$ is the electron Lorentz factor, $m_e$ the electron mass, and $c$ the speed of light -- for synchrotron and IC radiation in the Thomson regime can be evaluated as
\begin{eqnarray}
    P(\gamma)^\text{Total} &=& P(\gamma)^\text{sync} + P(\gamma)^\text{C} \nonumber \\
                &=& -\frac{4}{3} \sigma_T c u_B (1 + k_\text{C}) \gamma^2 \label{electron_energy_loss_rate}
\end{eqnarray}
with the Compton dominance parameter $k_\text{C} = \frac{u_\text{\rm ph}}{u_B} \approx \nu F_{\nu}^{\rm p, IC}/\nu F_{\nu}^{\rm p, sy}$, where $\nu F_{\nu}^{\rm p, sy}$ and $\nu F_{\nu}^{\rm p, IC}$ are the $\nu F_{\nu}$ peak fluxes of the synchrotron and Compton emission components, respectively.
Furthermore, $\sigma_T$ represents the Thomson cross-section and $u_B$ the magnetic field energy density.

The cooling time scale for the electrons in the emission-region rest frame is then
\begin{eqnarray}
    T_\text{cool}^{\rm em}(\gamma) &=& \frac{\gamma m_e c^2}{|P(\gamma)^\text{Total}|} \nonumber \\
                    &=& \left[ \frac{4}{3} \sigma_T c \frac{u_B}{m_e c^2} (1 + k_\text{C}) \gamma \right]^{-1} \label{cooling_time-scale}
\end{eqnarray}
and is related to the observer's frame cooling time scale through
\begin{equation}
T_{\rm cool}^{\rm obs} (\gamma) = \frac{1 + z}{\delta} \, T_{\rm cool}^{\rm em} (\gamma).
\end{equation}
where $\delta = \left( \Gamma \, [1 - \beta_{\Gamma} \cos\theta_{\rm obs}] \right)^{-1}$ is the Doppler factor and $z$ the redshift to the source.
The electron Lorentz factor $\gamma$ may be related to an observed photon frequency/energy, depending on the radiation mechanism through which any observed frequency $\nu_{\rm obs}$ is produced.

For frequencies in the synchrotron SED component, we may set $\nu_{\rm obs}$ equal to the critical synchrotron frequency of an electron with Lorentz factor $\gamma$, such that $\nu_{\rm obs} = \frac{\delta}{1 + z}  \frac{3eB}{4\pi m_e c} \gamma^2$ where $e$ is elementary charge.
With the magnetic field energy density $u_B = \frac{B^2}{8\pi}$, this then yields
\begin{eqnarray}
    T_\text{cool, sync}^{\rm obs} (\nu_{\rm obs}) = \sqrt{ \frac{1 + z}{\delta} } \left[ \frac{1}{3} \sigma_T (1 + k_\text{C}) \sqrt{\frac{B^3 \nu^\text{obs}}{3 \pi e m_e c} } \right]^{-1} \label{eq:t_cool,sync}
\end{eqnarray}

We assume that frequencies in X-rays up to HE $\gamma$-rays in the IC SED component are within the Thomsom regime and approximate the relation between $\gamma$ and $\nu^\text{obs}$ with $\epsilon^\text{em}_s = \gamma^2 \epsilon_\text{t}^\text{em}$ where $\epsilon$ represent the photon energy normalized w.r.t. electron rest energy and leads to $\nu_s^\text{obs} = \frac{\delta}{1+z} \nu_\text{t}^\text{em} \gamma^2$ where the subscripts $s$ and t indicates the scattered and target photons respectively.
The cooling time-scale is then
\begin{eqnarray}
    T_\text{cool, T}^{\rm obs} (\nu_{\rm obs}) = \sqrt{ \frac{1 + z}{\delta} } \left[ \frac{1}{6} \sigma_T \frac{B^2}{\pi m_e c} (1 + k_\text{C}) \sqrt{\frac{\nu_{\rm obs}}{\nu_\text{t}^\text{em}} } \right]^{-1}
\end{eqnarray}

For frequencies in VHE $\gamma$-rays we approximate IC radiation in the Klein-Nishina regime, $\epsilon = \gamma$ leading to the relation $\nu^\text{obs} = \frac{\delta}{1 + z} \frac{m_e c^2}{h} \gamma$ where $h$ is Planck's constant, giving the following result for the cooling time-scale:
\begin{eqnarray}
    T_\text{cool, KN}^{\rm obs} (\nu_{\rm obs}) =  \left[ \frac{1}{6} \sigma_T \frac{B^2}{\pi m_e^2 c^3} (1 + k_\text{C}) h \nu_{\rm obs}  \right]^{-1}
\end{eqnarray}

\section{Injection luminosity within the one-zone model} \label{app:lum_norm}

In the model the emission region is continuously injected with electrons at each time step, $t$, where the rate of electrons injected with Lorentz factor $\gamma$ follow a power-law:
\begin{eqnarray}
    Q(\gamma, t) = Q_0(t) \gamma^{-q(t)} H(\gamma; \gamma_\text{min}, \gamma_\text{max})
\end{eqnarray}
with the normalization factor $Q_0(t)$, and the Heaviside function $H(\gamma; \gamma_\text{min}, \gamma_\text{max})$ which is 1 for $\gamma_\text{min} \le \gamma \le \gamma_\text{max}$ and 0 otherwise, and the electron injection spectral index, $q(t)$.
Normalization with respect to the electron injection luminosity, $L_\text{inj}$, and the size of the emission region, $V_b$,
yields
\begin{eqnarray}
    Q_0(t) =   \begin{cases}
                \frac{L_\text{inj}(t)}{V_b m_e c^2} \frac{2-q(t)}{\gamma_\text{max}^{2-q(t)} - \gamma_\text{min}^{2-q(t)}} & \text{if}\ q(t) \ne 2 \\
                \frac{L_\text{inj}(t)}{V_b m_e c^2 \ln{\frac{\gamma_\text{max}}{\gamma_\text{min}}}} & \text{if}\  q(t) = 2 \\
            \end{cases} \label{eq:inj_norm}
\end{eqnarray}

This shows how changing the injection luminosity and/or the electron injection index can have a large affect on the number of electrons injected into the emission region.
Increasing the electron injection luminosity increases the number of electrons of all energies equally, while a change in the spectral index results in an energy-dependent variation of the electron density.

\section{Individual simulation realization results} \label{app:indiv_results}

\begin{figure}[h]
    \gridline{
        \fig{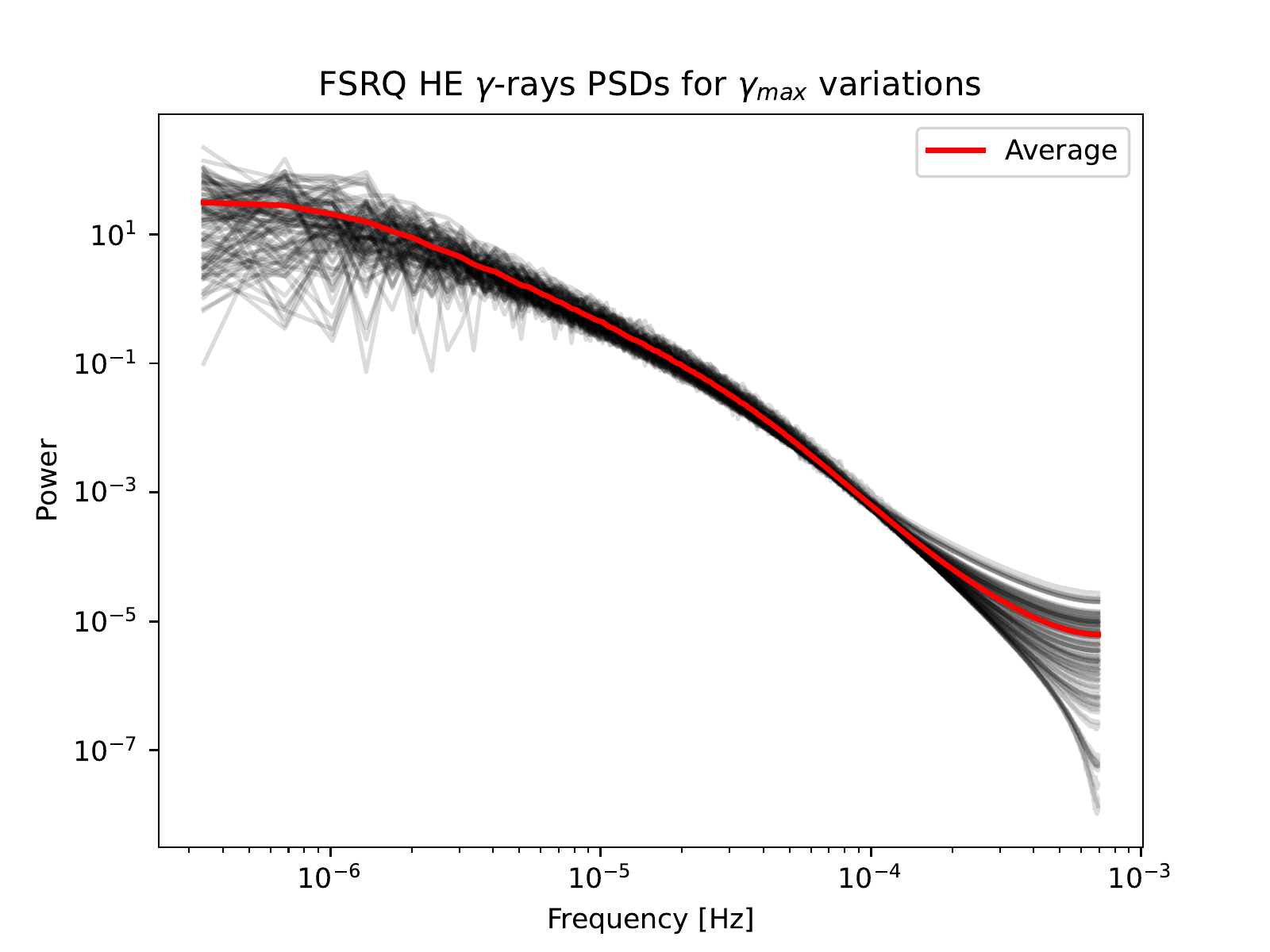}{.25\textwidth}{}
        \fig{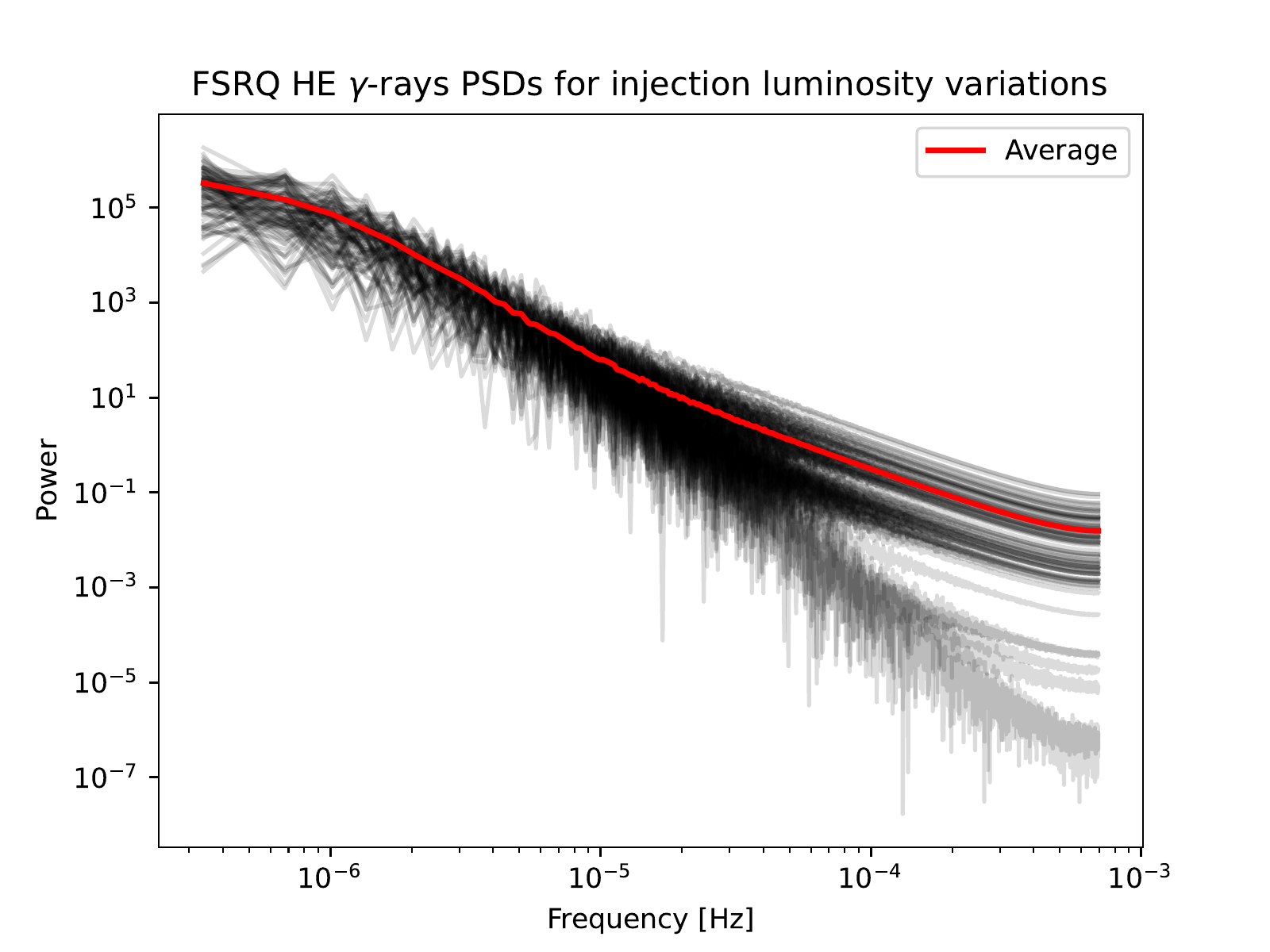}{.25\textwidth}{}
        \fig{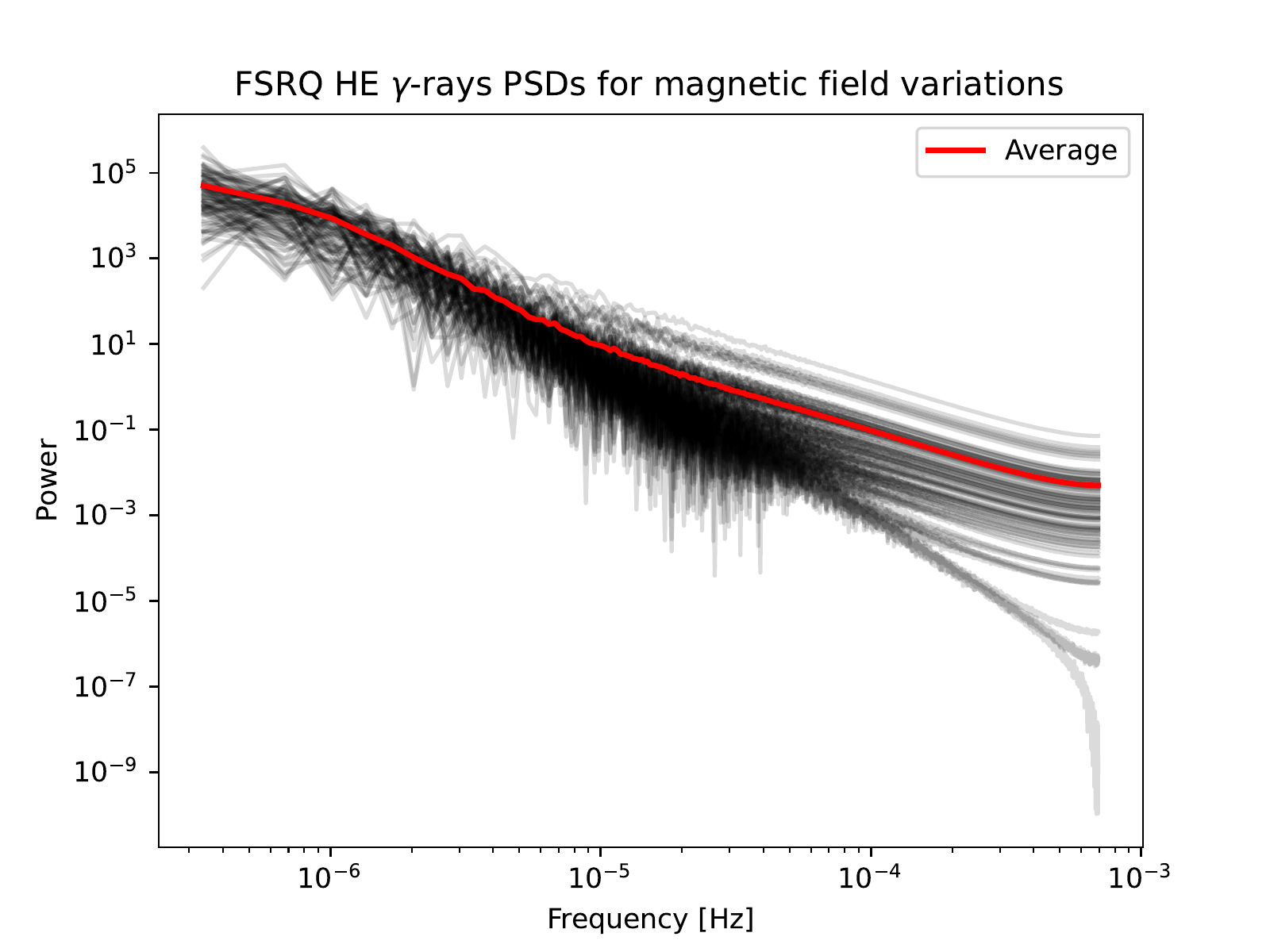}{.25\textwidth}{}
        \fig{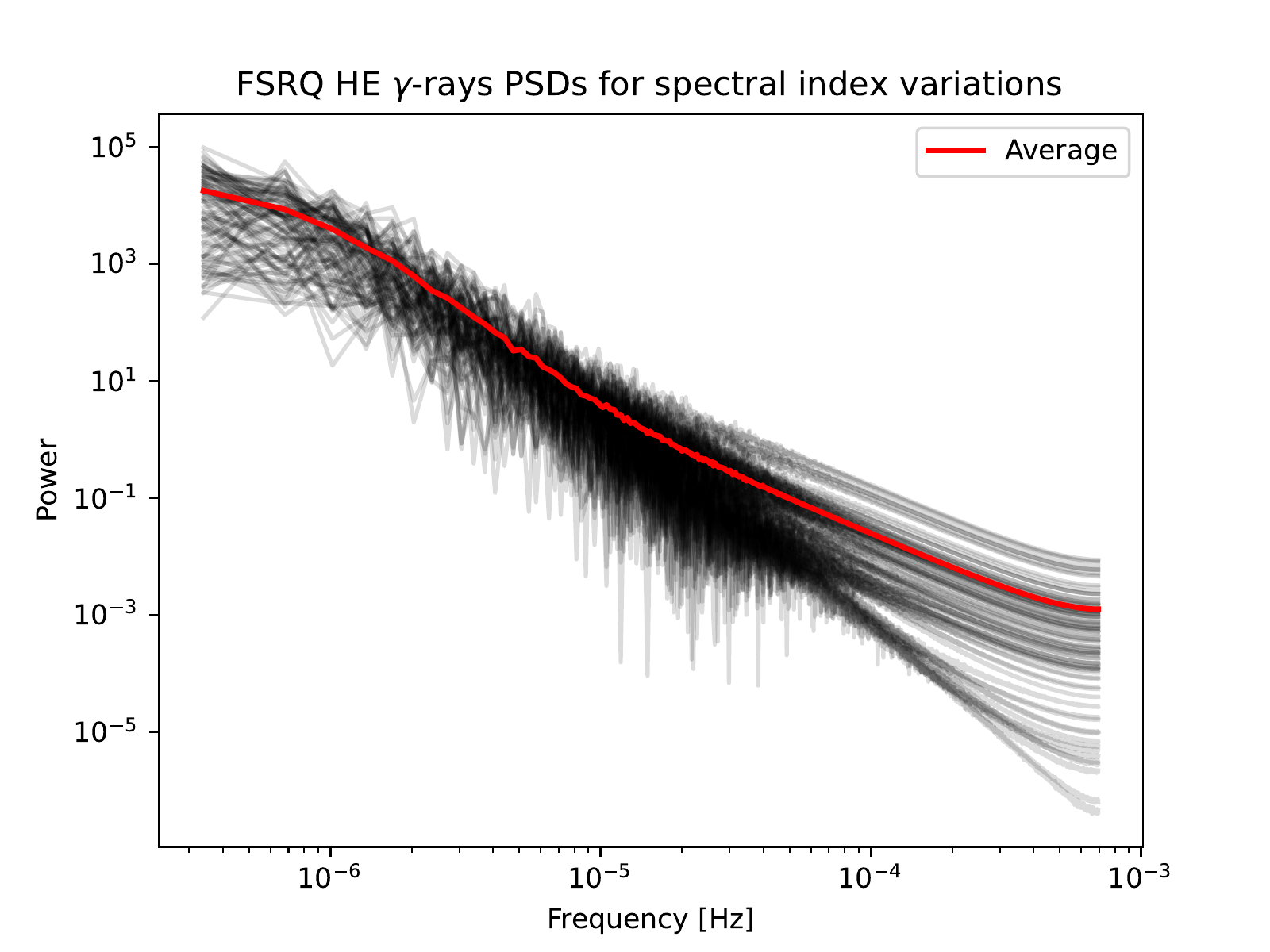}{.25\textwidth}{}
    }
    \gridline{
        \fig{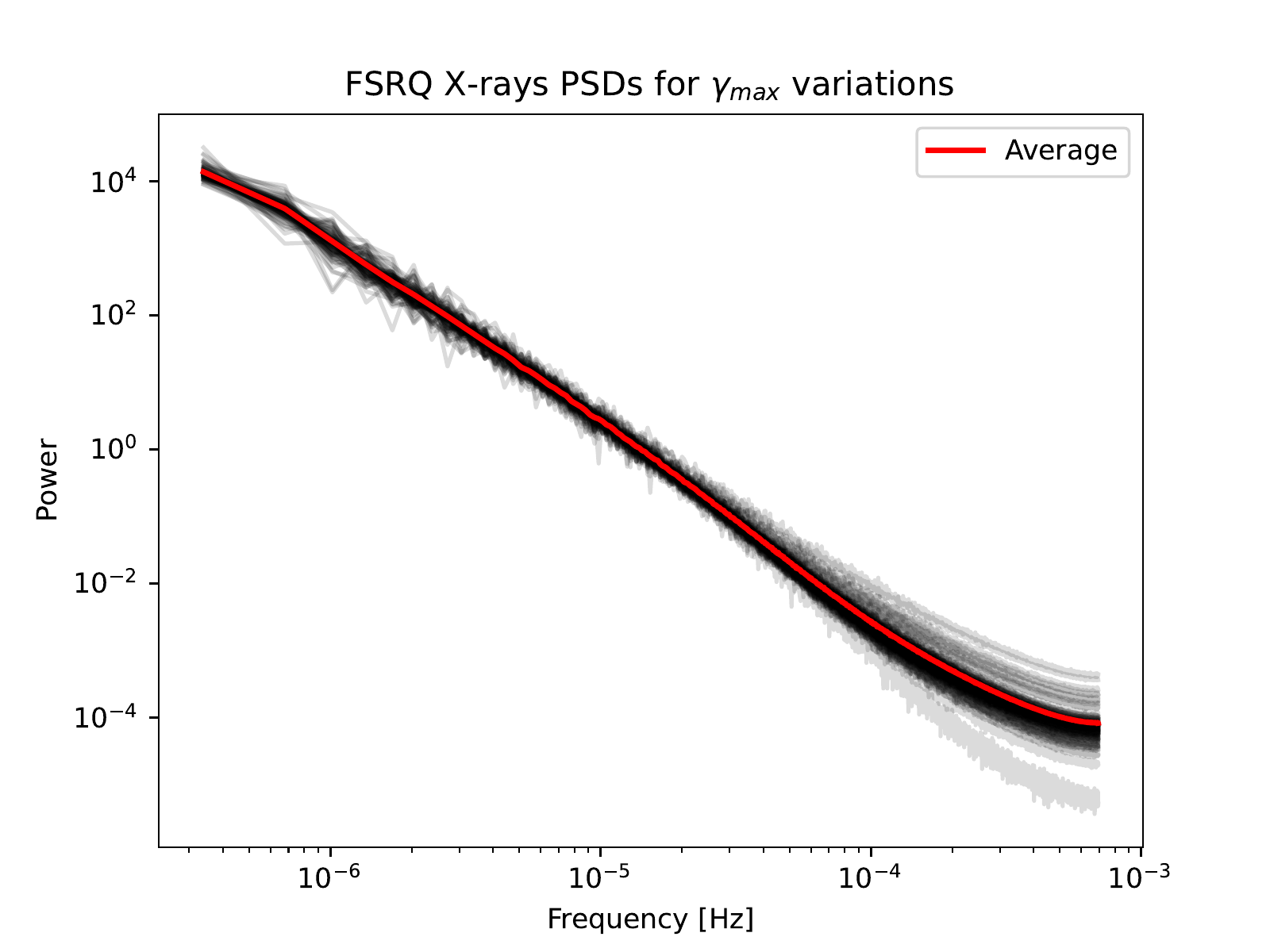}{.25\textwidth}{}
        \fig{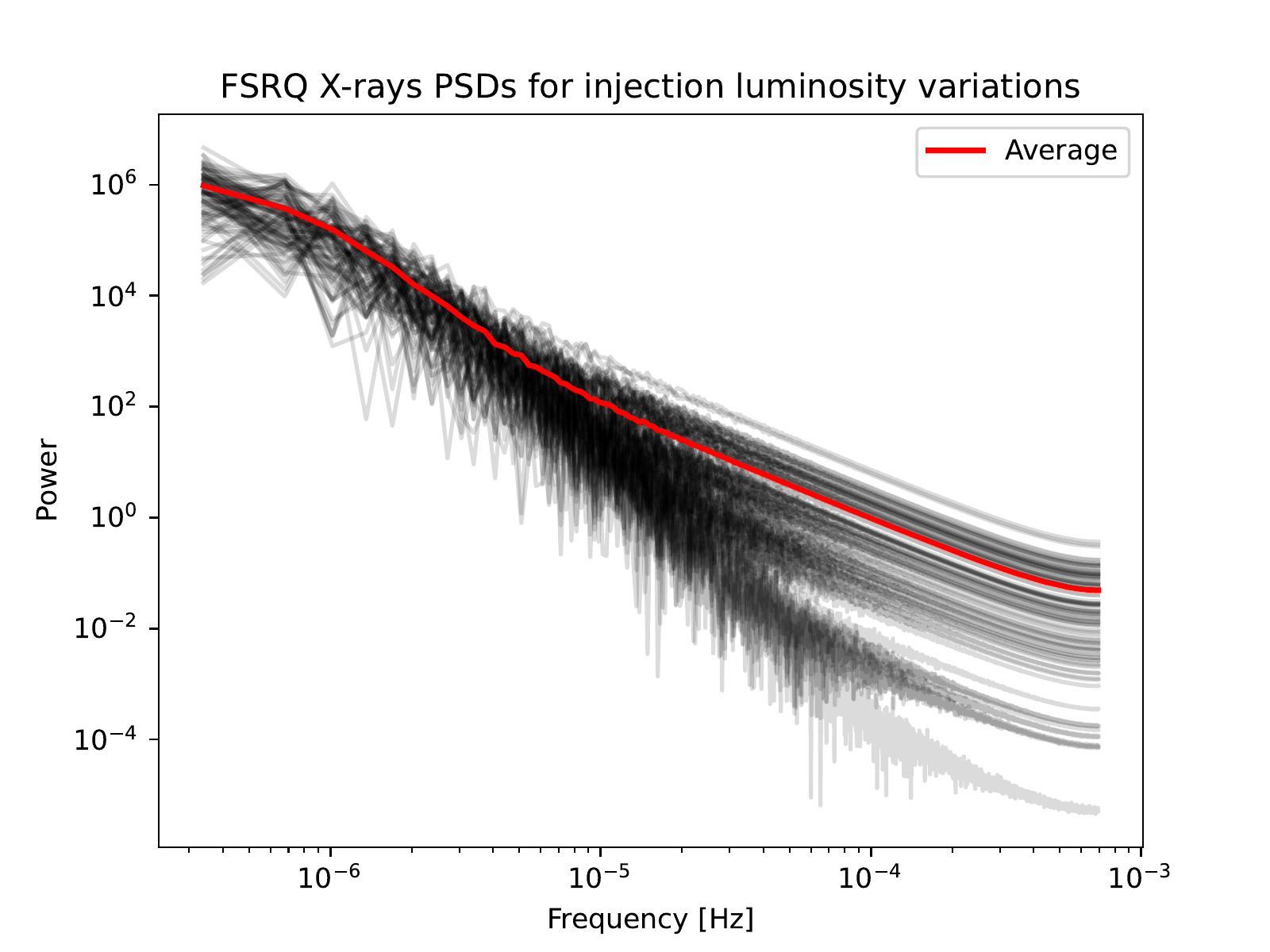}{.25\textwidth}{}
        \fig{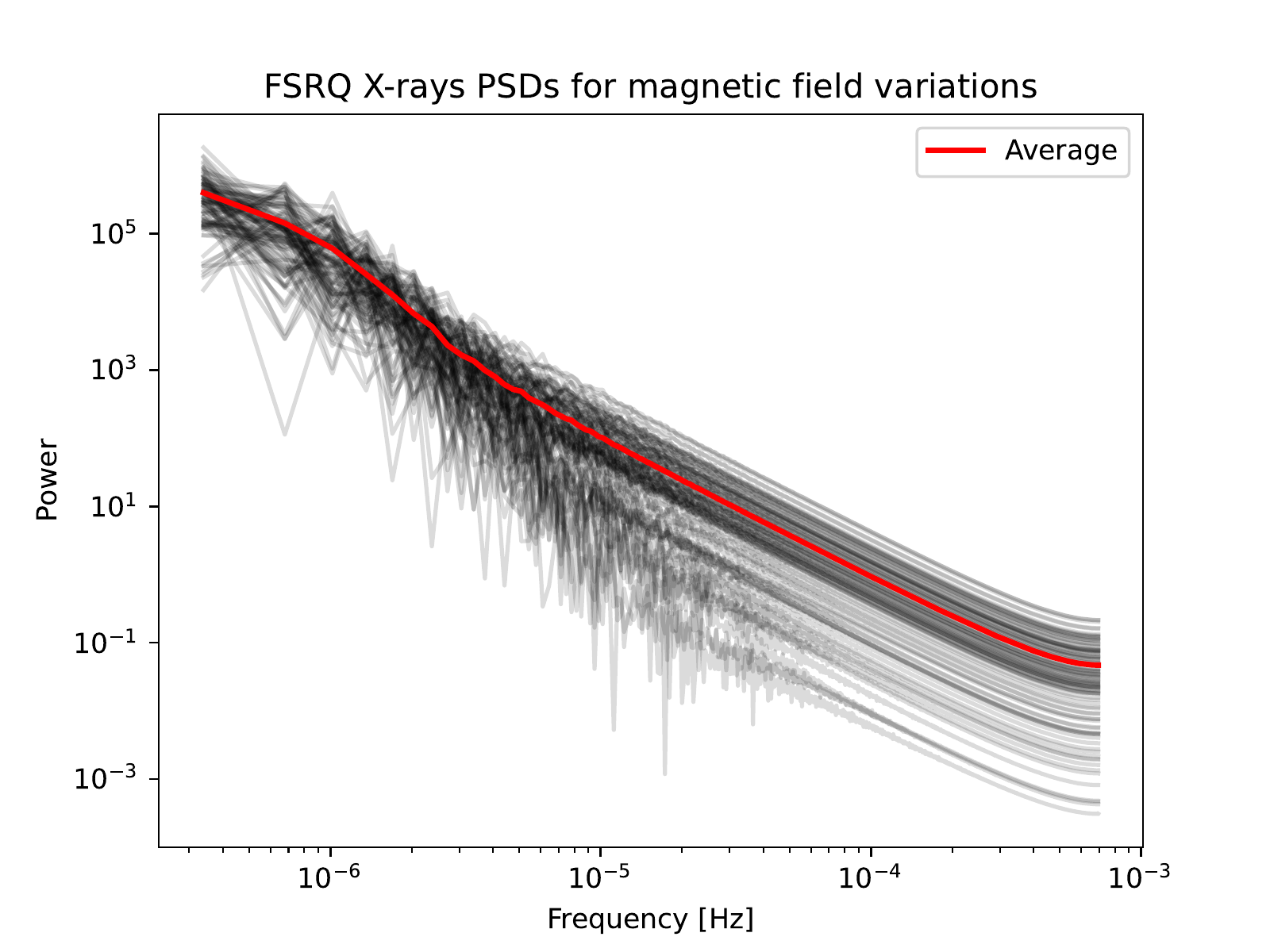}{.25\textwidth}{}
        \fig{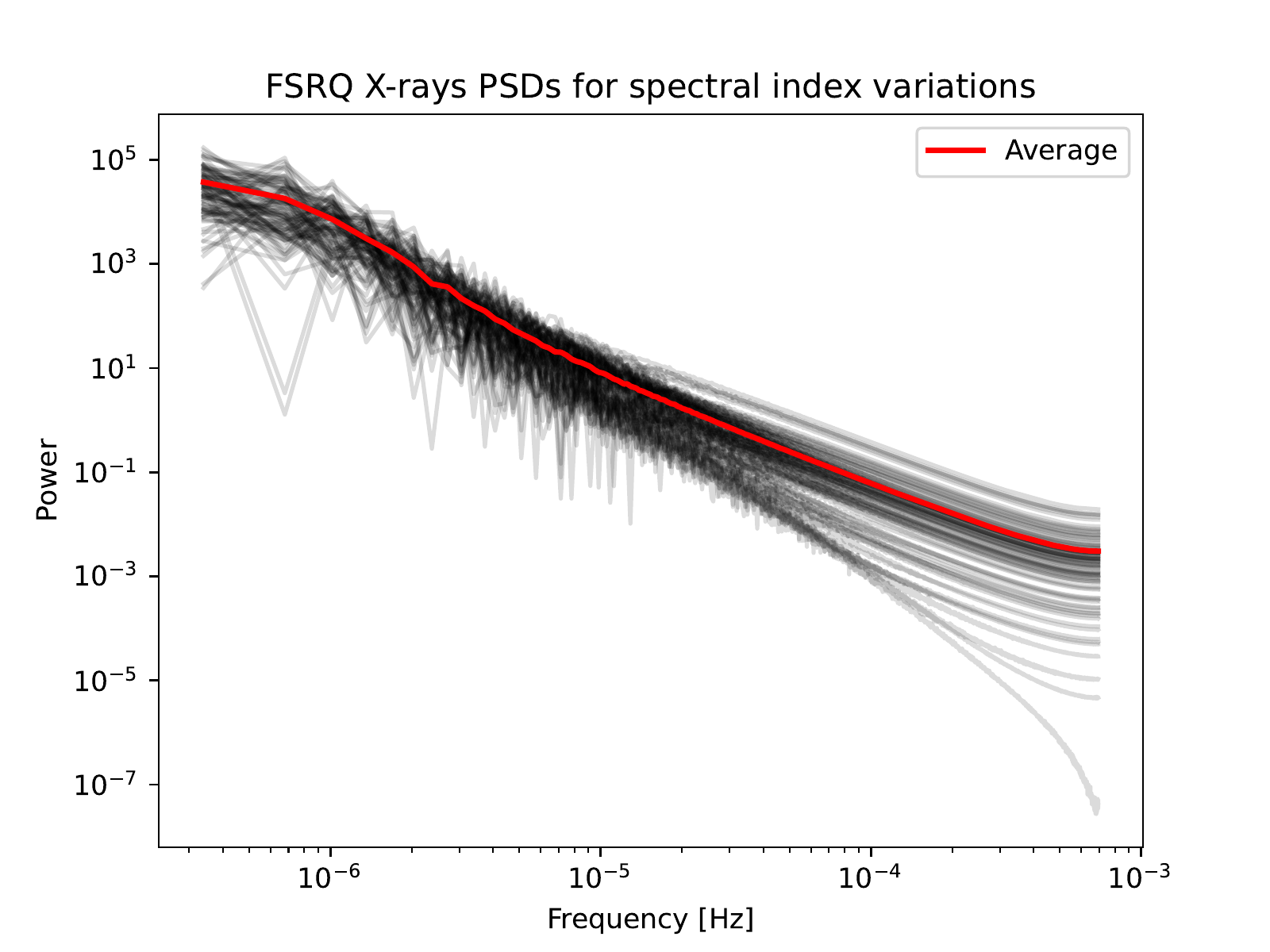}{.25\textwidth}{}
    }
    \gridline{
        \fig{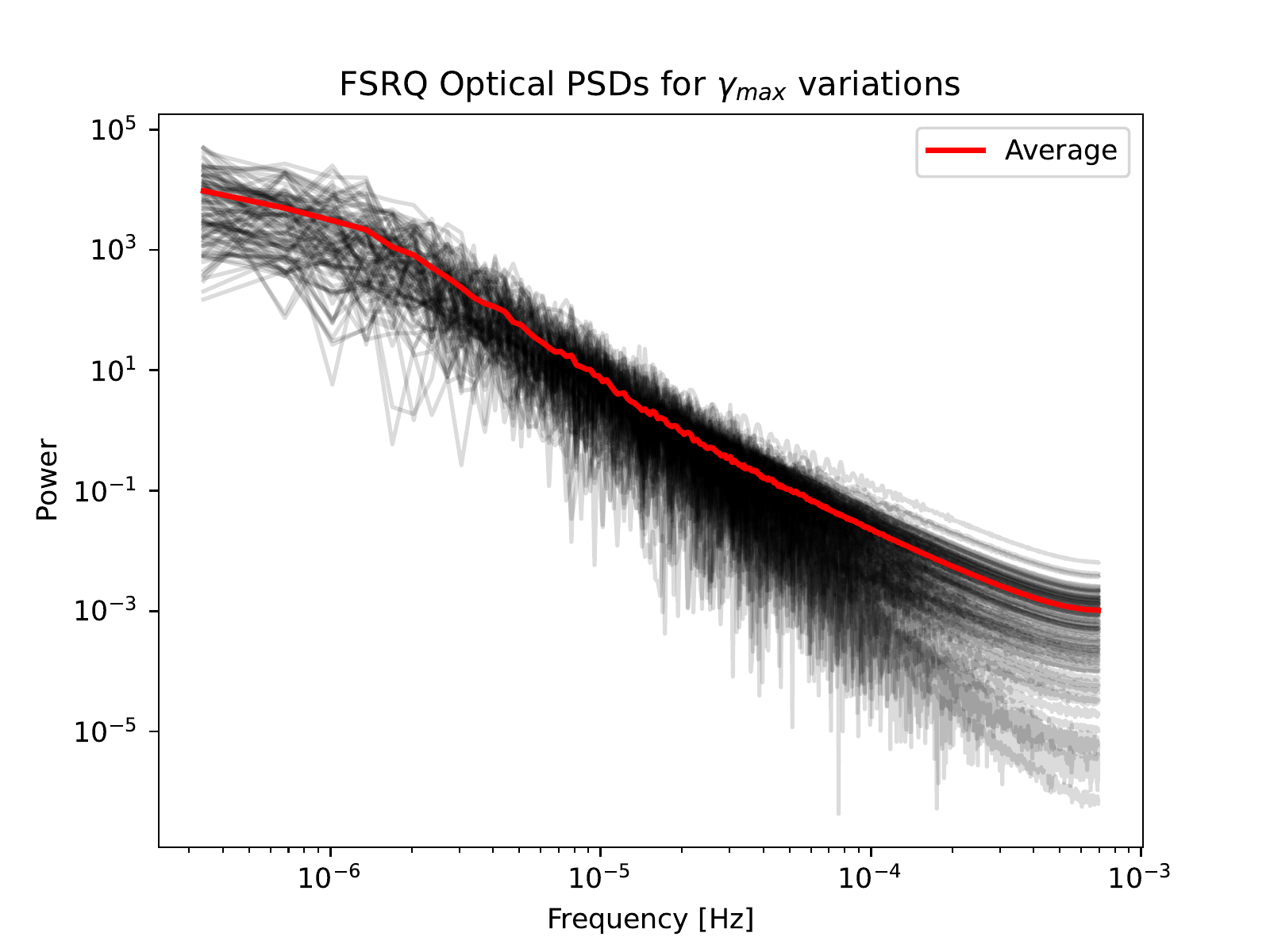}{.25\textwidth}{}
        \fig{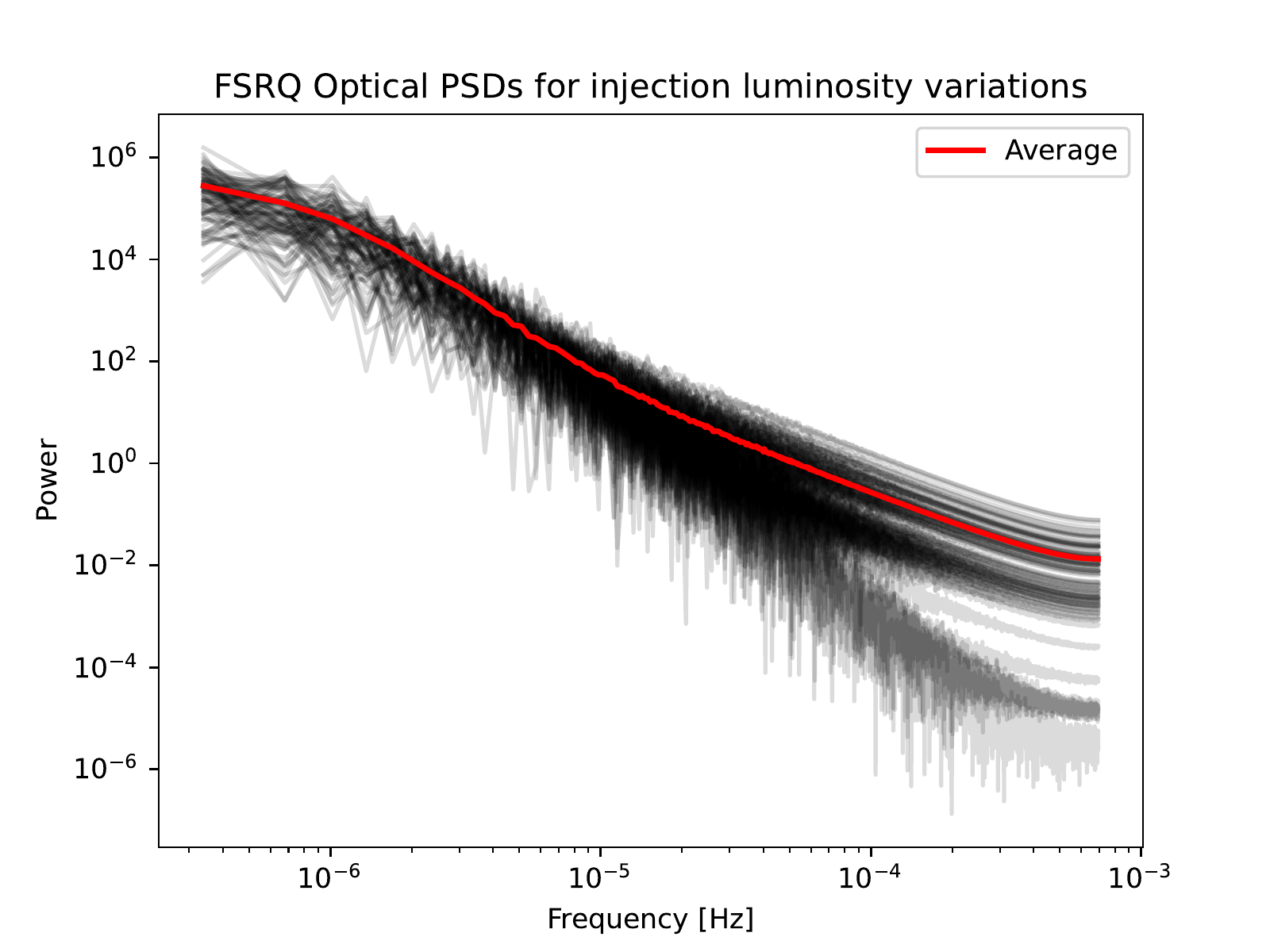}{.25\textwidth}{}
        \fig{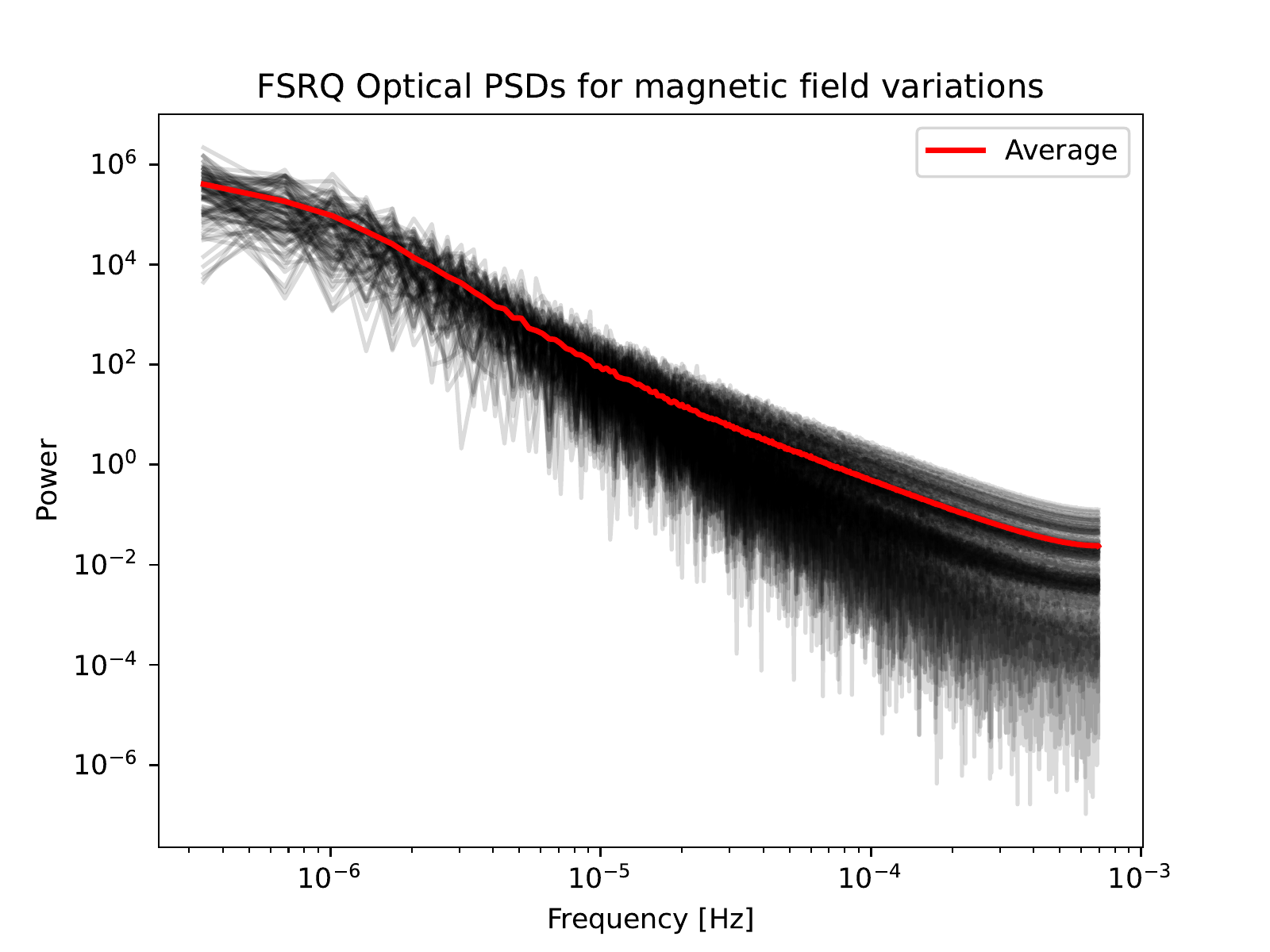}{.25\textwidth}{}
        \fig{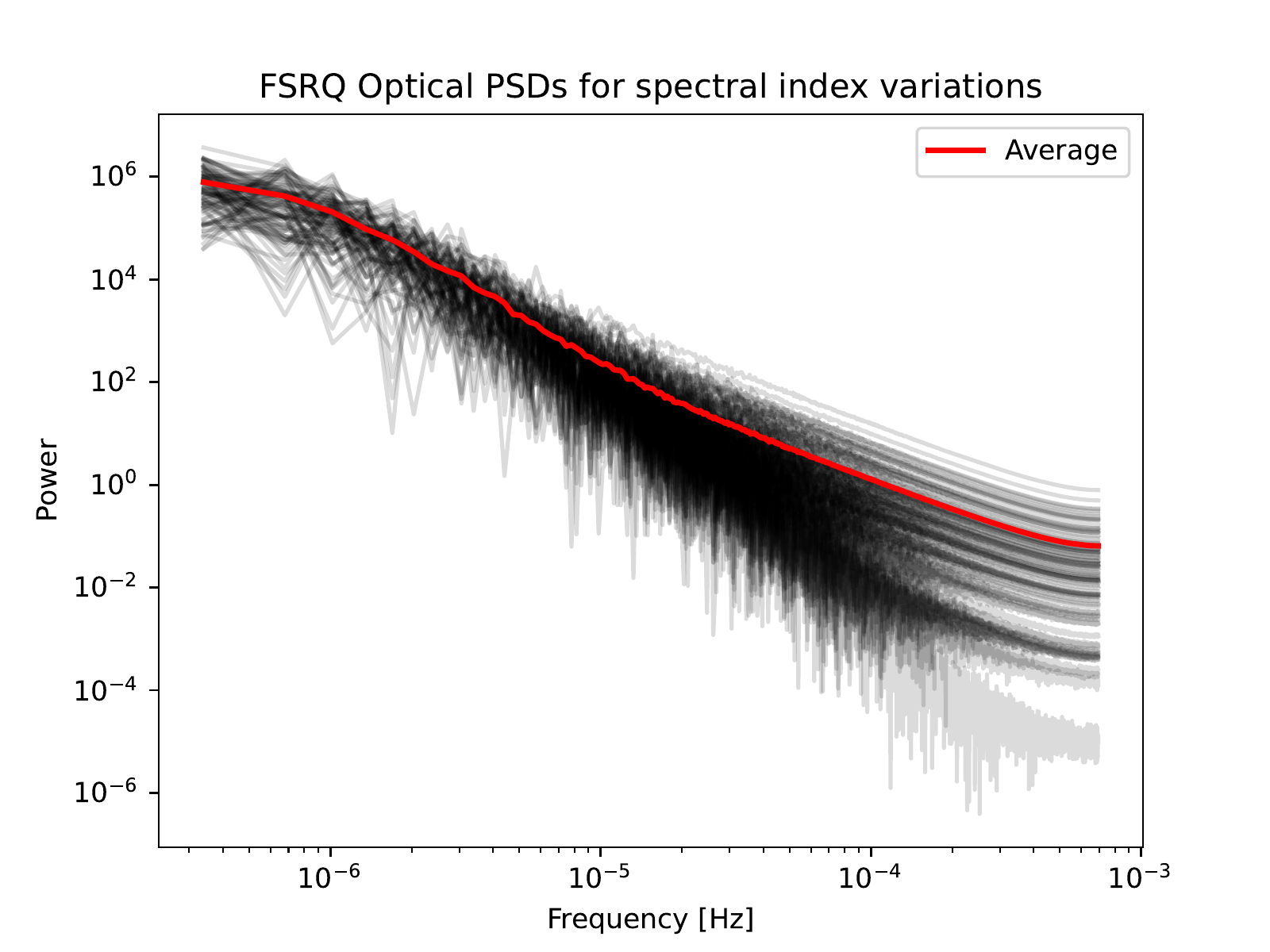}{.25\textwidth}{}
    }
\caption{
    Average and individual PSD comparisons for FSRQ simulation realizations.}
\label{fig:EC-PSD-ave_vs_indiv}
\end{figure}

\begin{figure}[h]
    \gridline{
        \fig{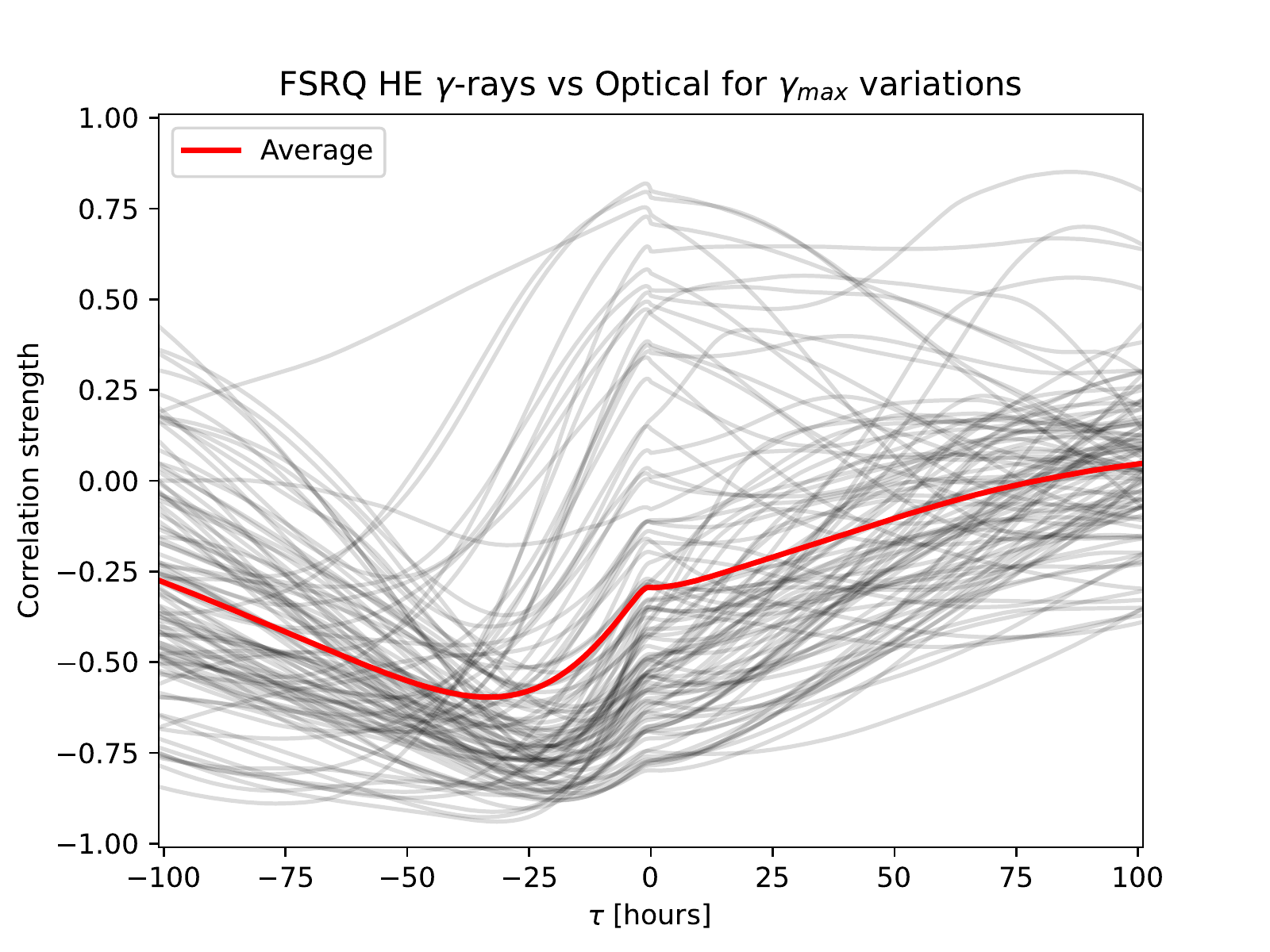}{.25\textwidth}{}
        \fig{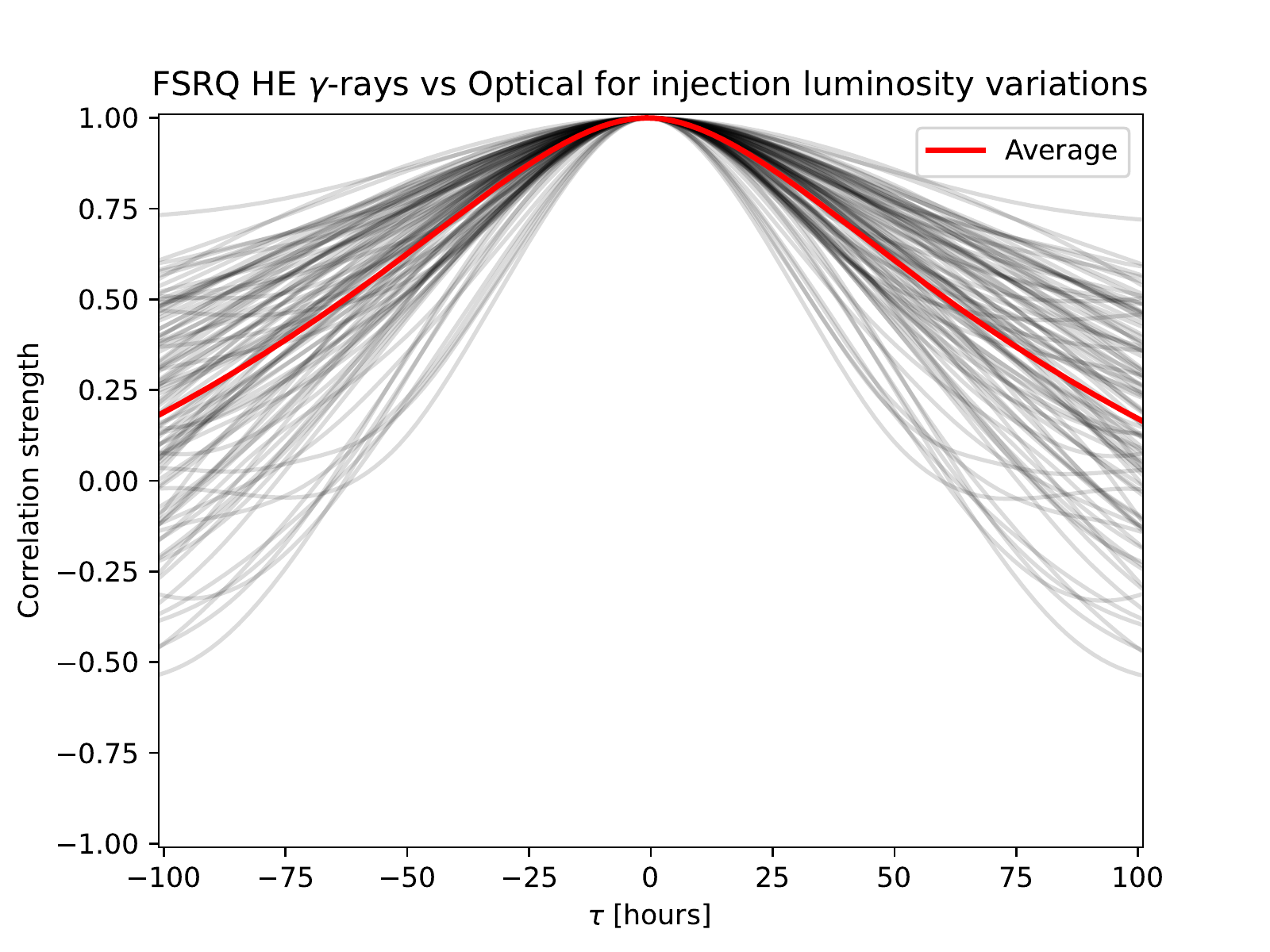}{.25\textwidth}{}
        \fig{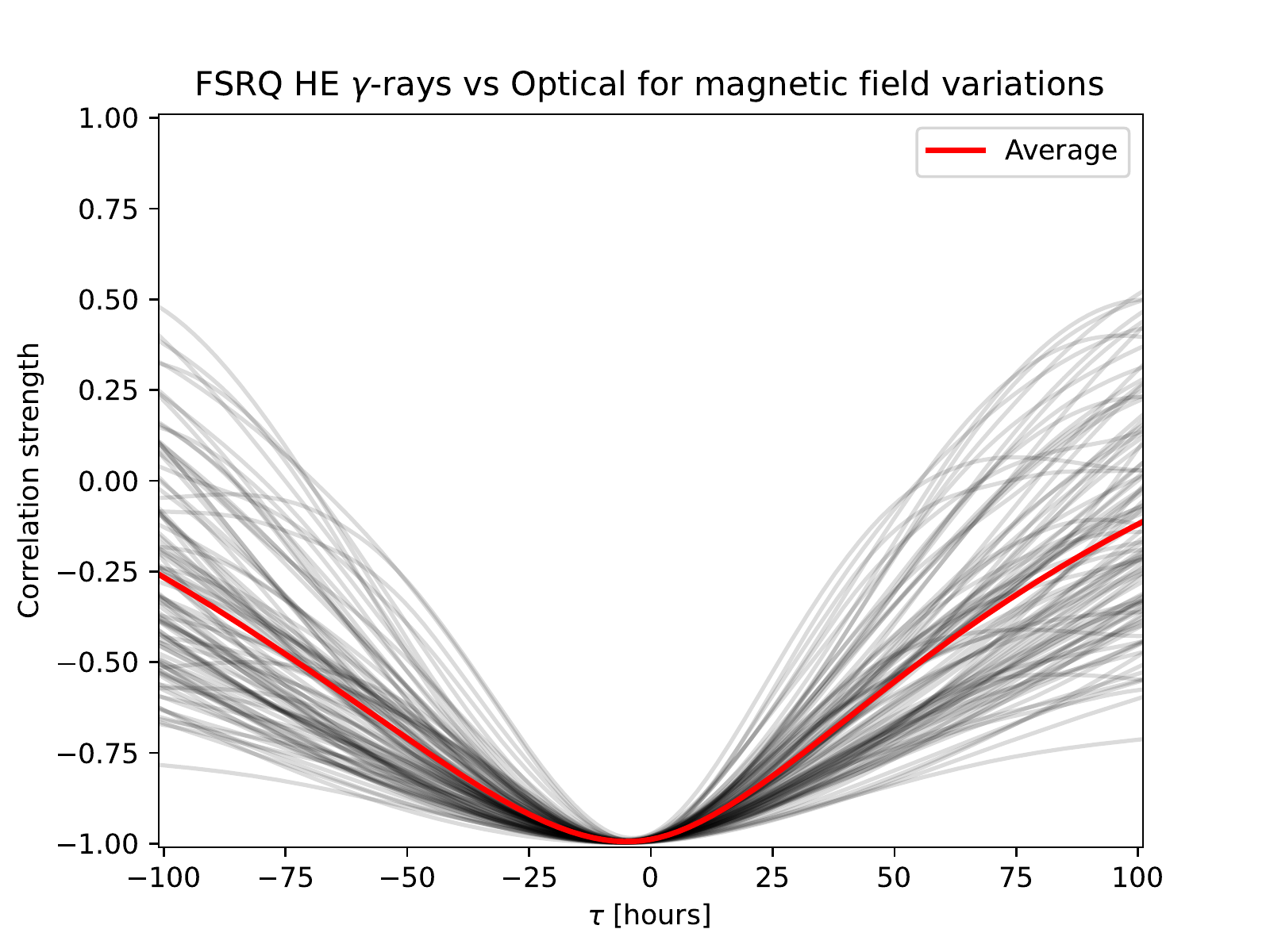}{.25\textwidth}{}
        \fig{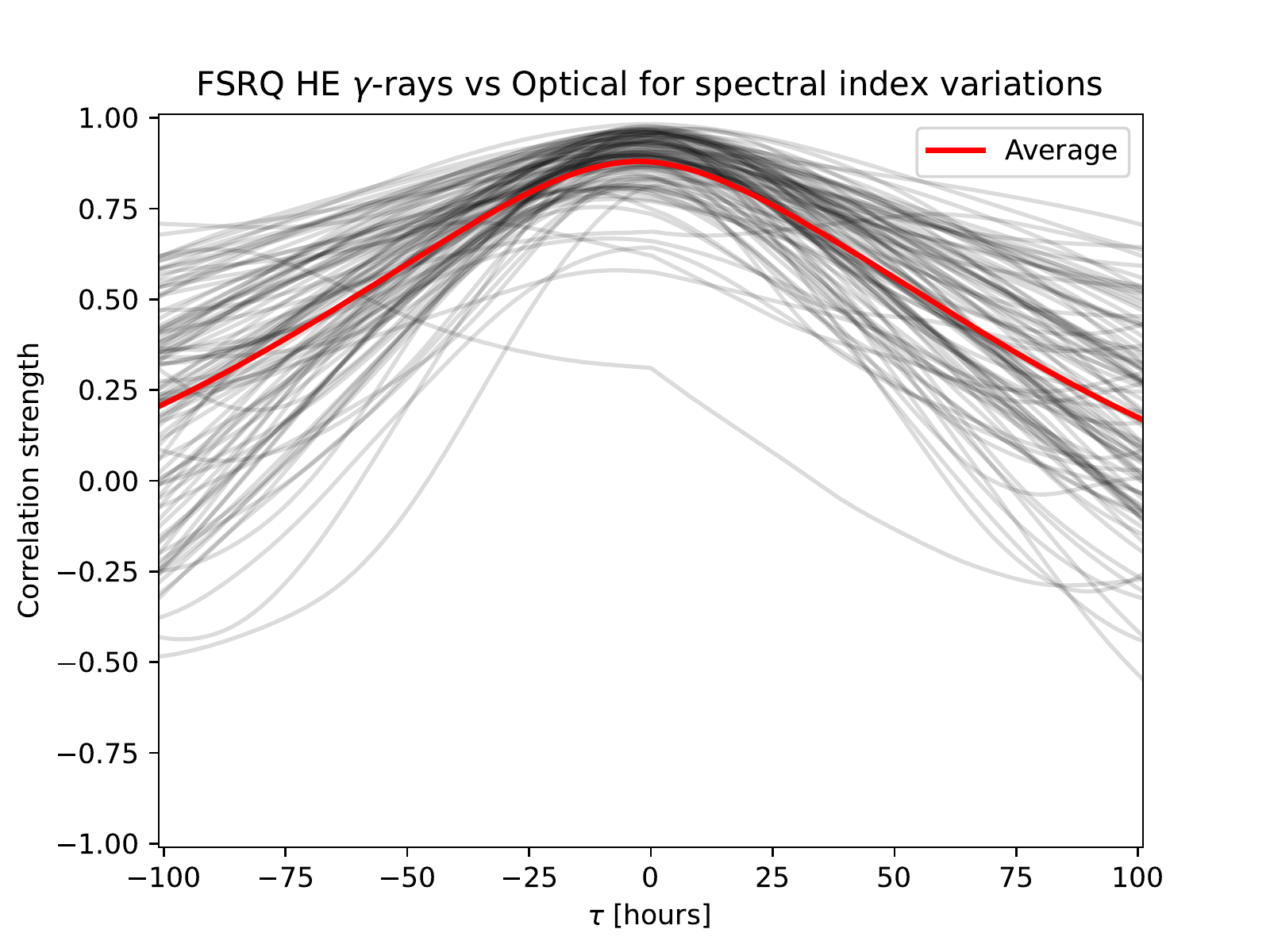}{.25\textwidth}{}
    }
    \gridline{
        \fig{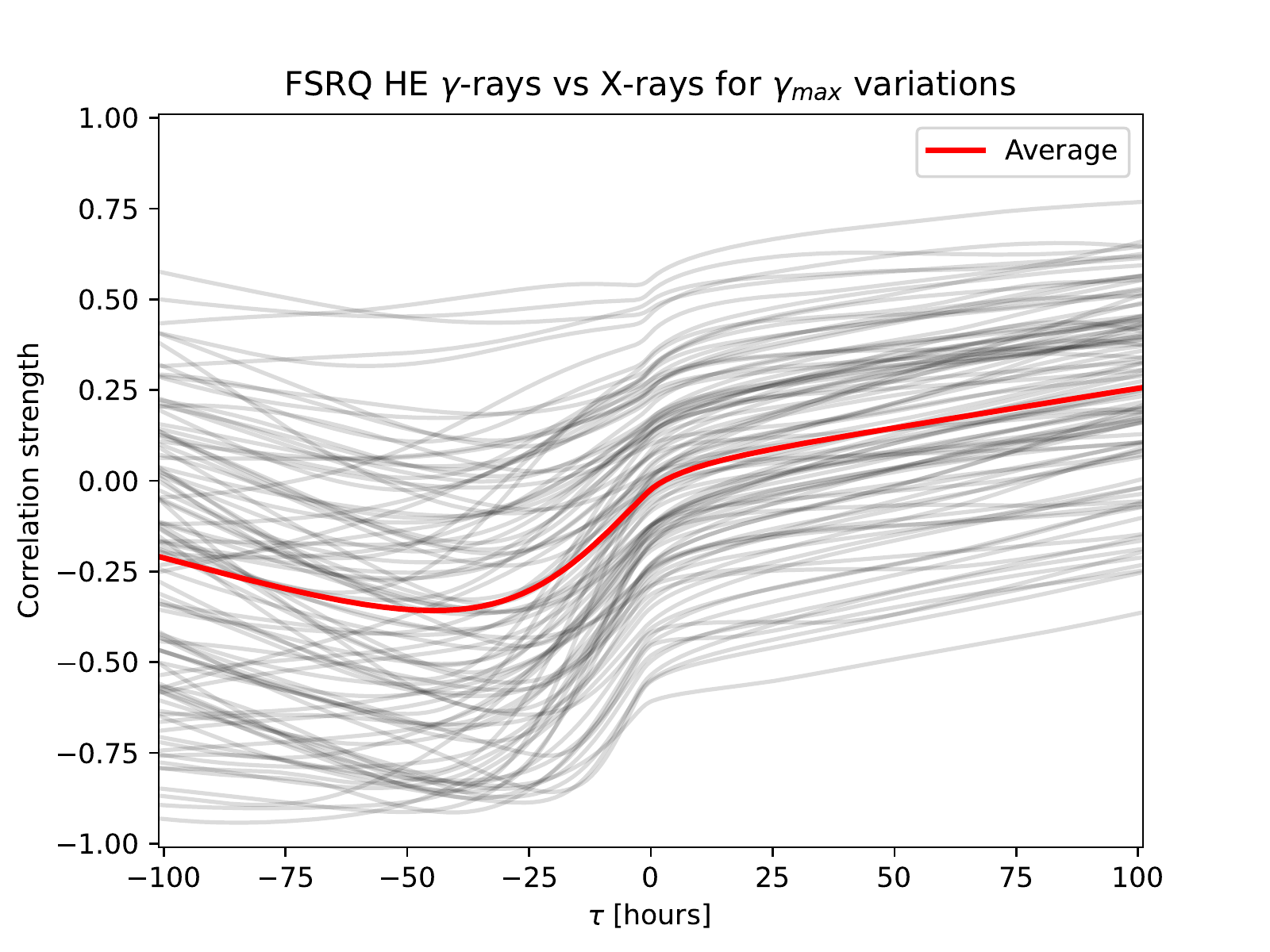}{.25\textwidth}{}
        \fig{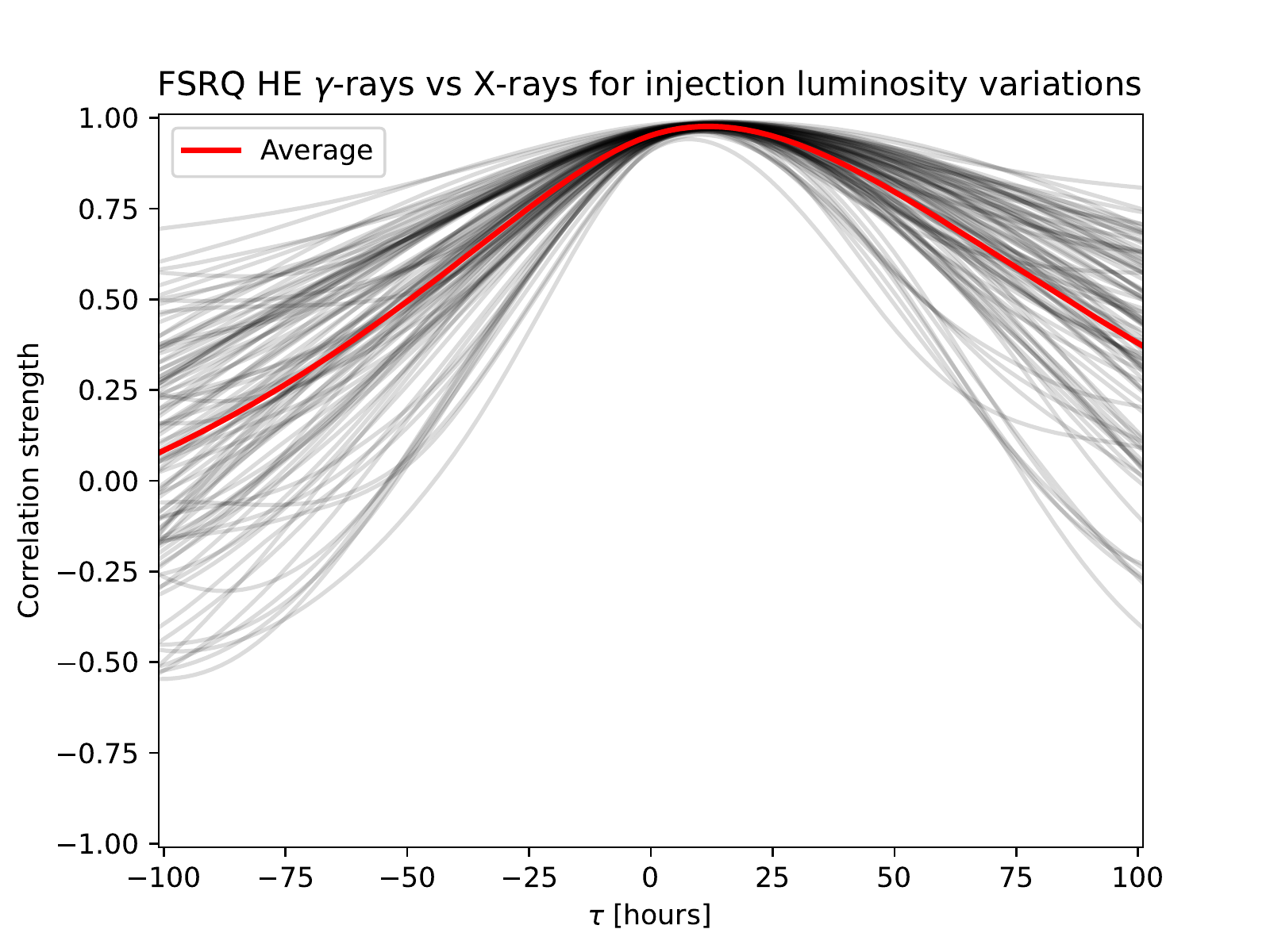}{.25\textwidth}{}
        \fig{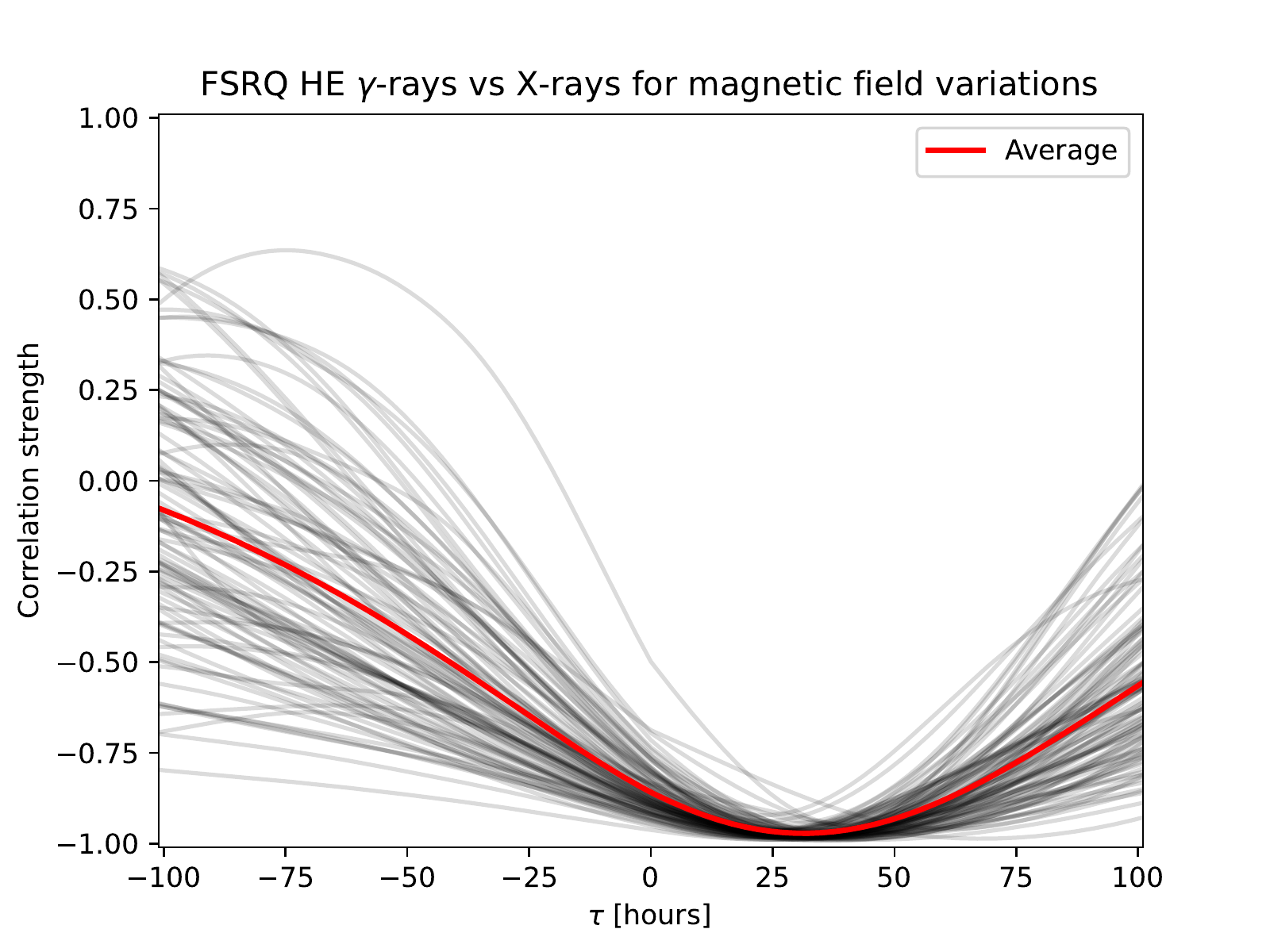}{.25\textwidth}{}
        \fig{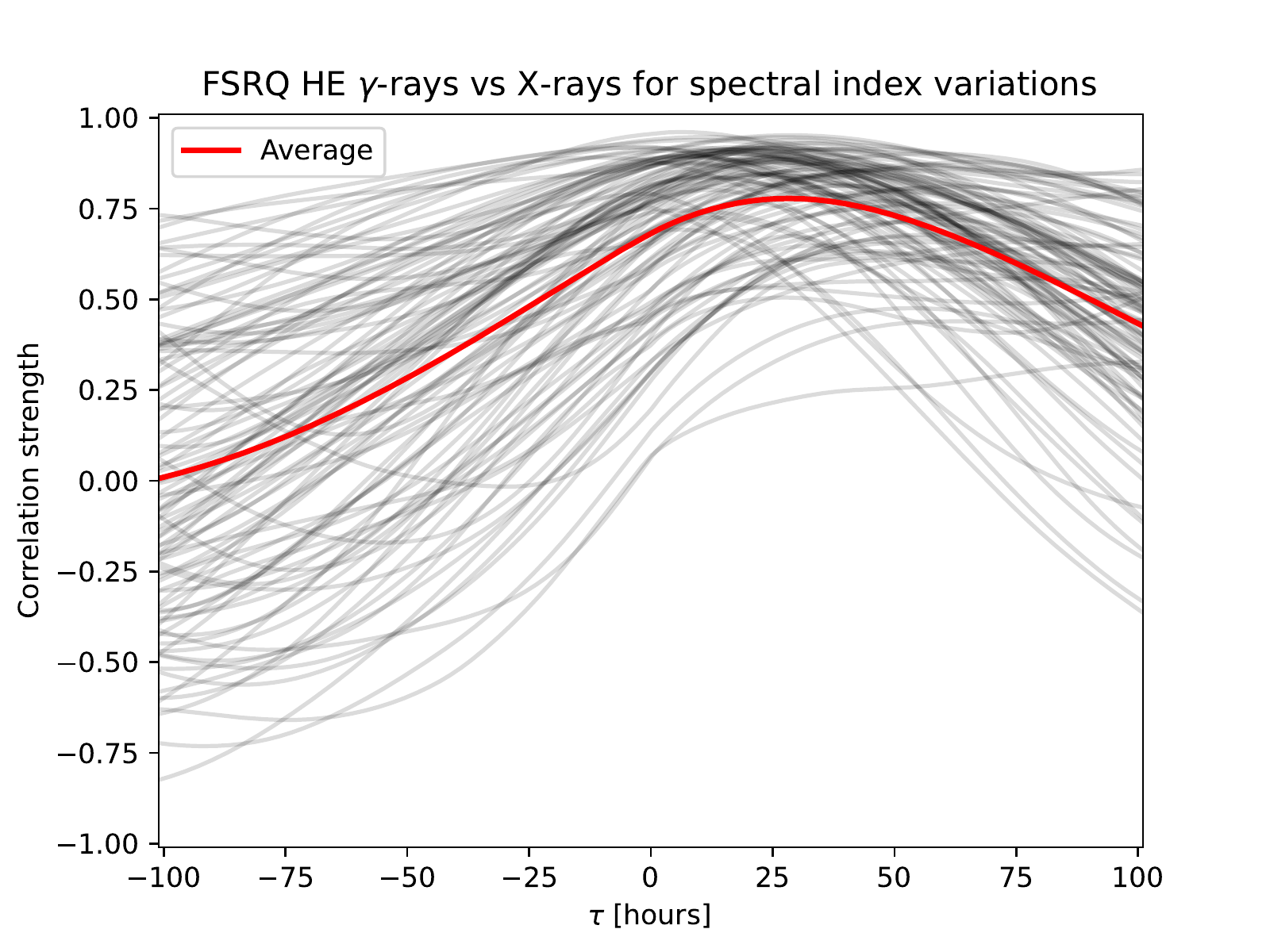}{.25\textwidth}{}
    }
    \gridline{
        \fig{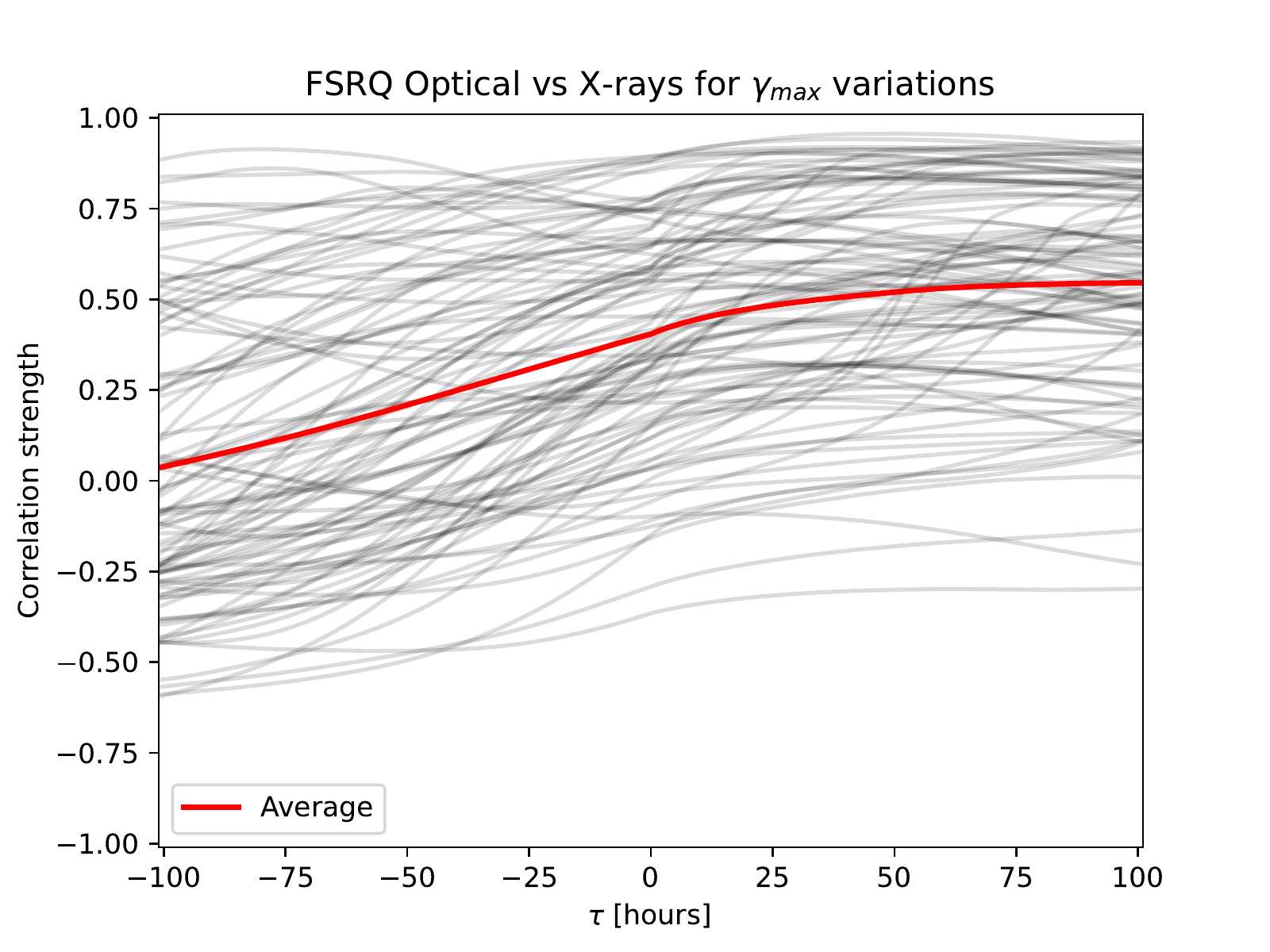}{.25\textwidth}{}
        \fig{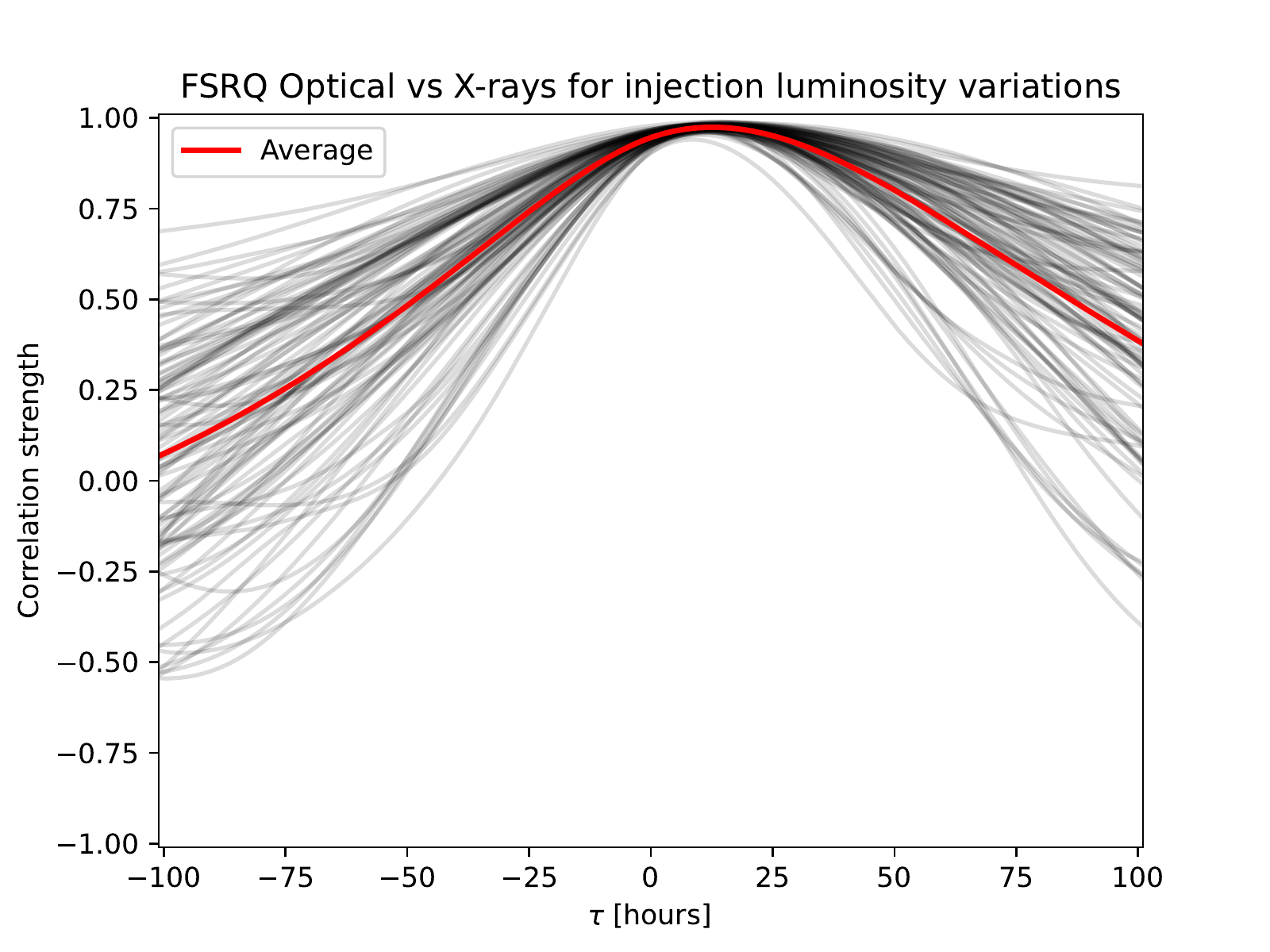}{.25\textwidth}{}
        \fig{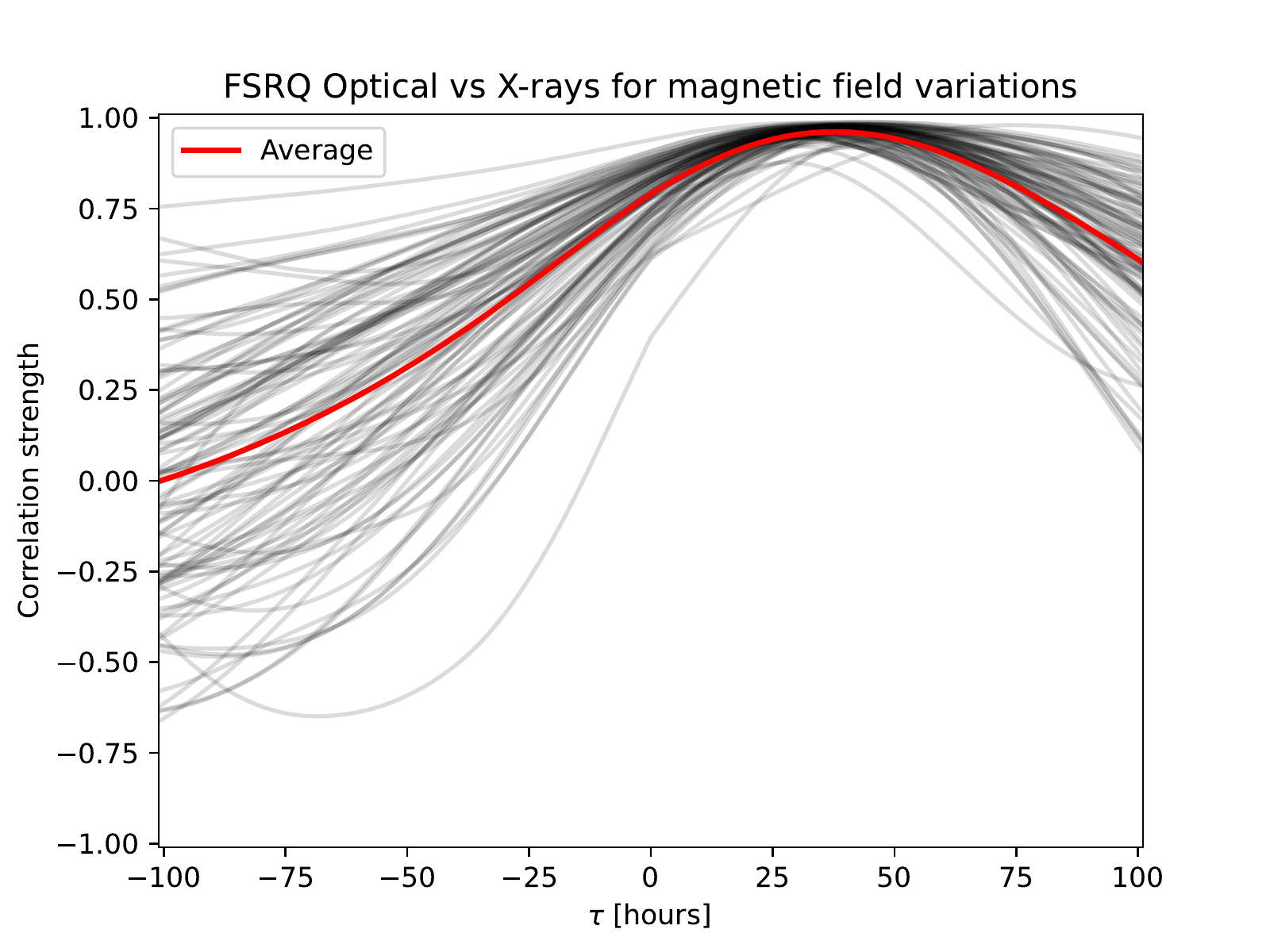}{.25\textwidth}{}
        \fig{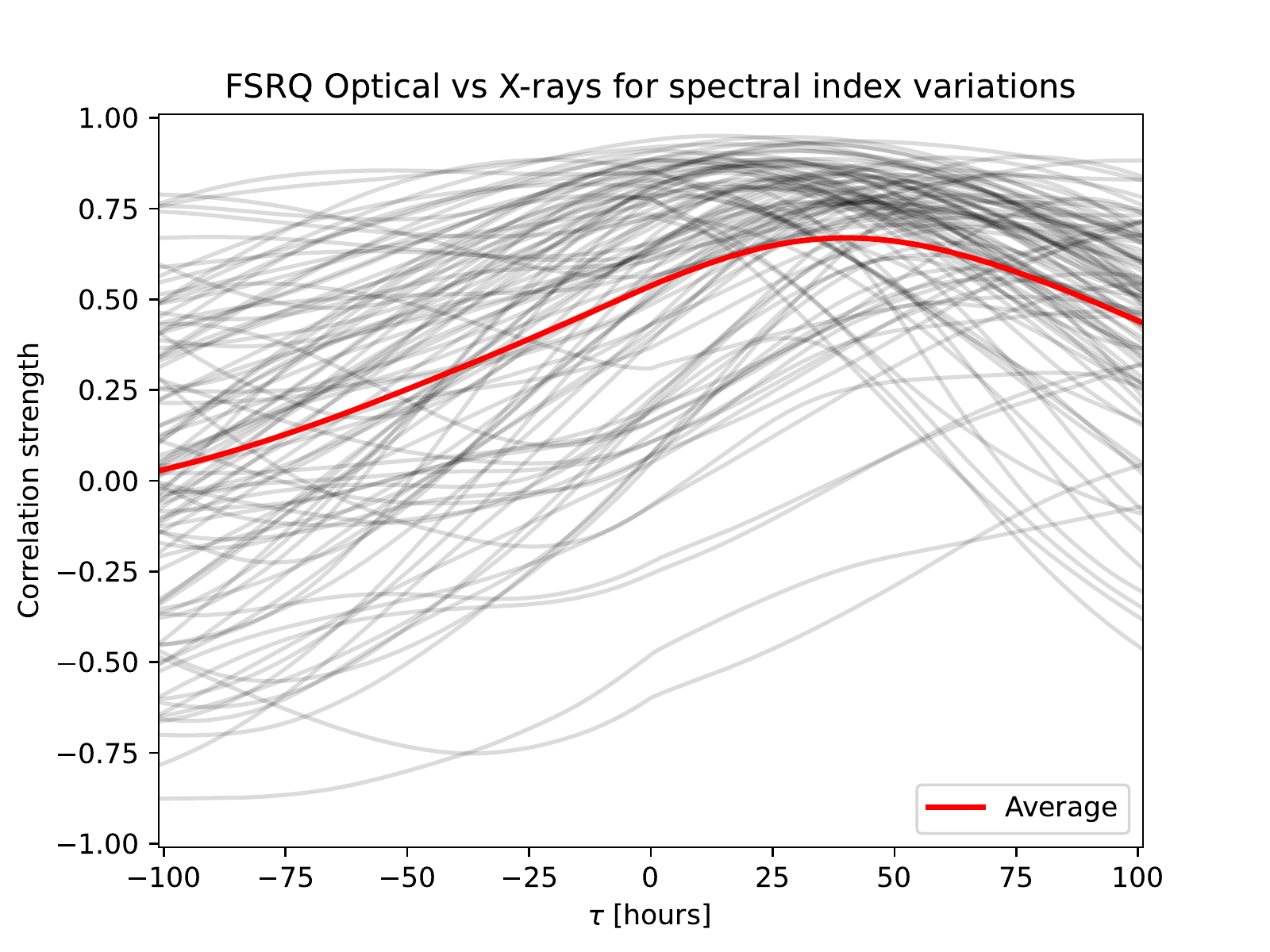}{.25\textwidth}{}
    }
    \caption{
        Average and individual cross-correlation comparisons for FSRQ
        simulation realizations.}
    \label{fig:EC-Corr-ave_vs_ind}
\end{figure}

\begin{figure}[h]
    \gridline{
        \fig{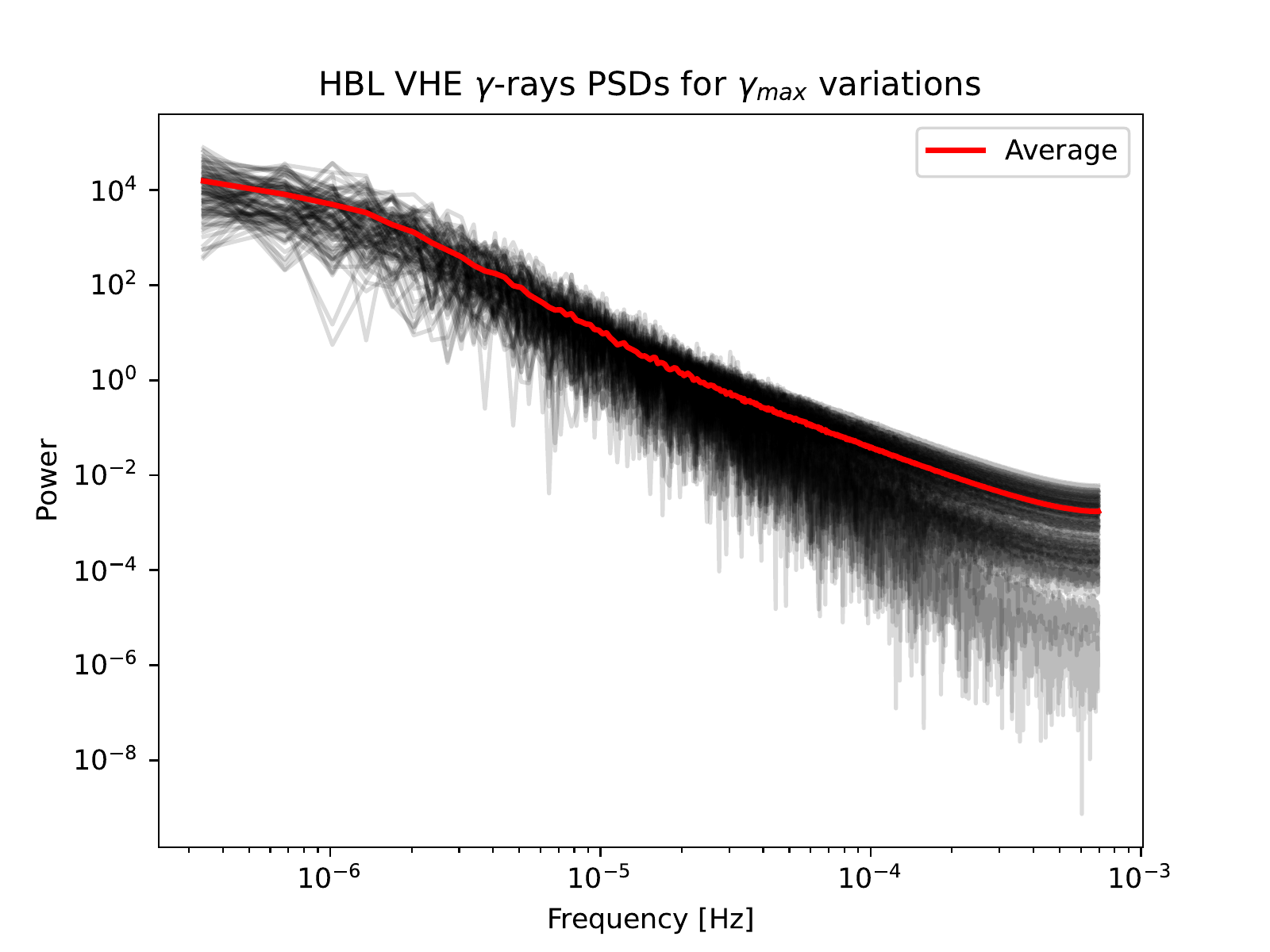}{.25\textwidth}{}
        \fig{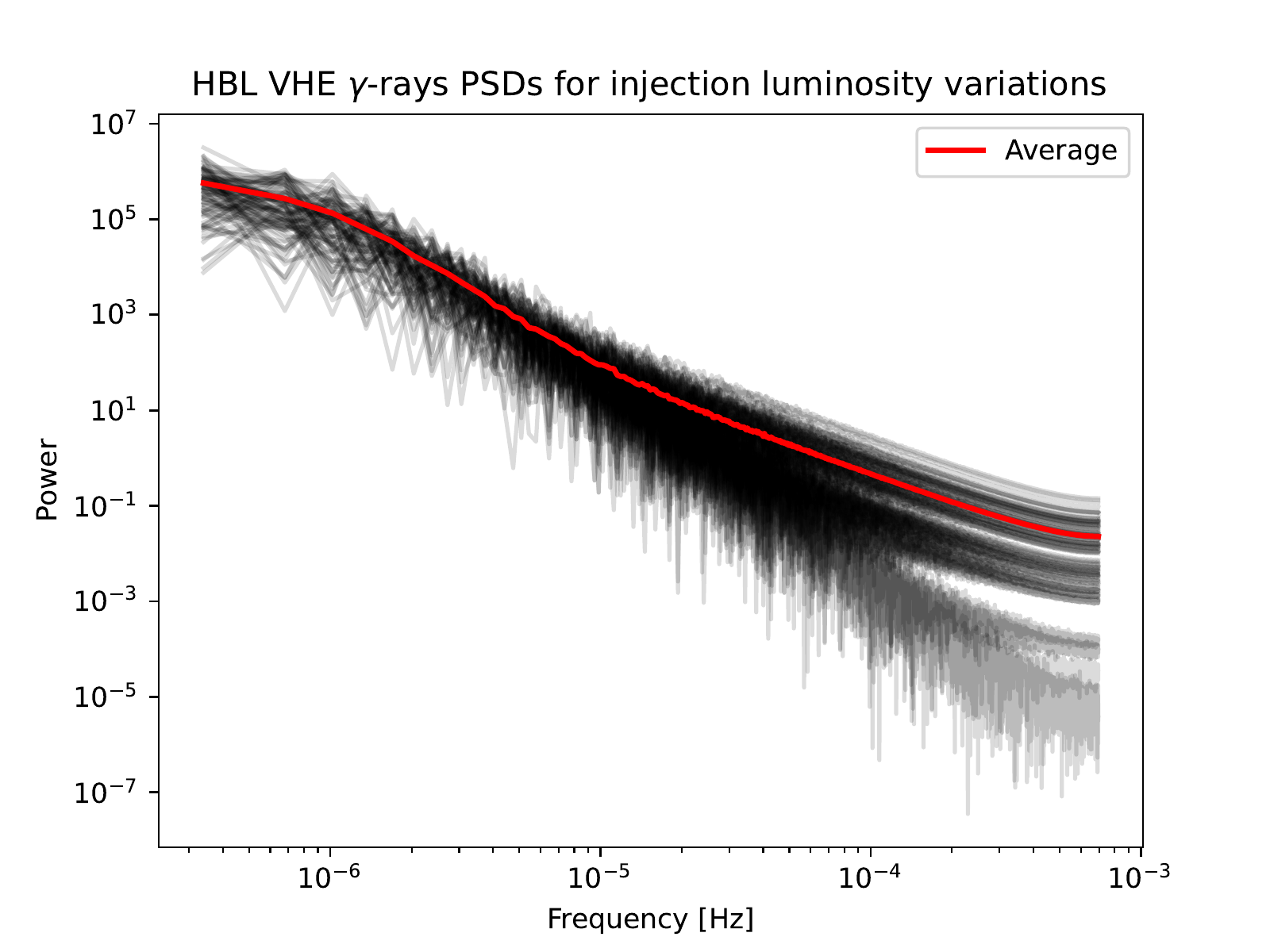}{.25\textwidth}{}
        \fig{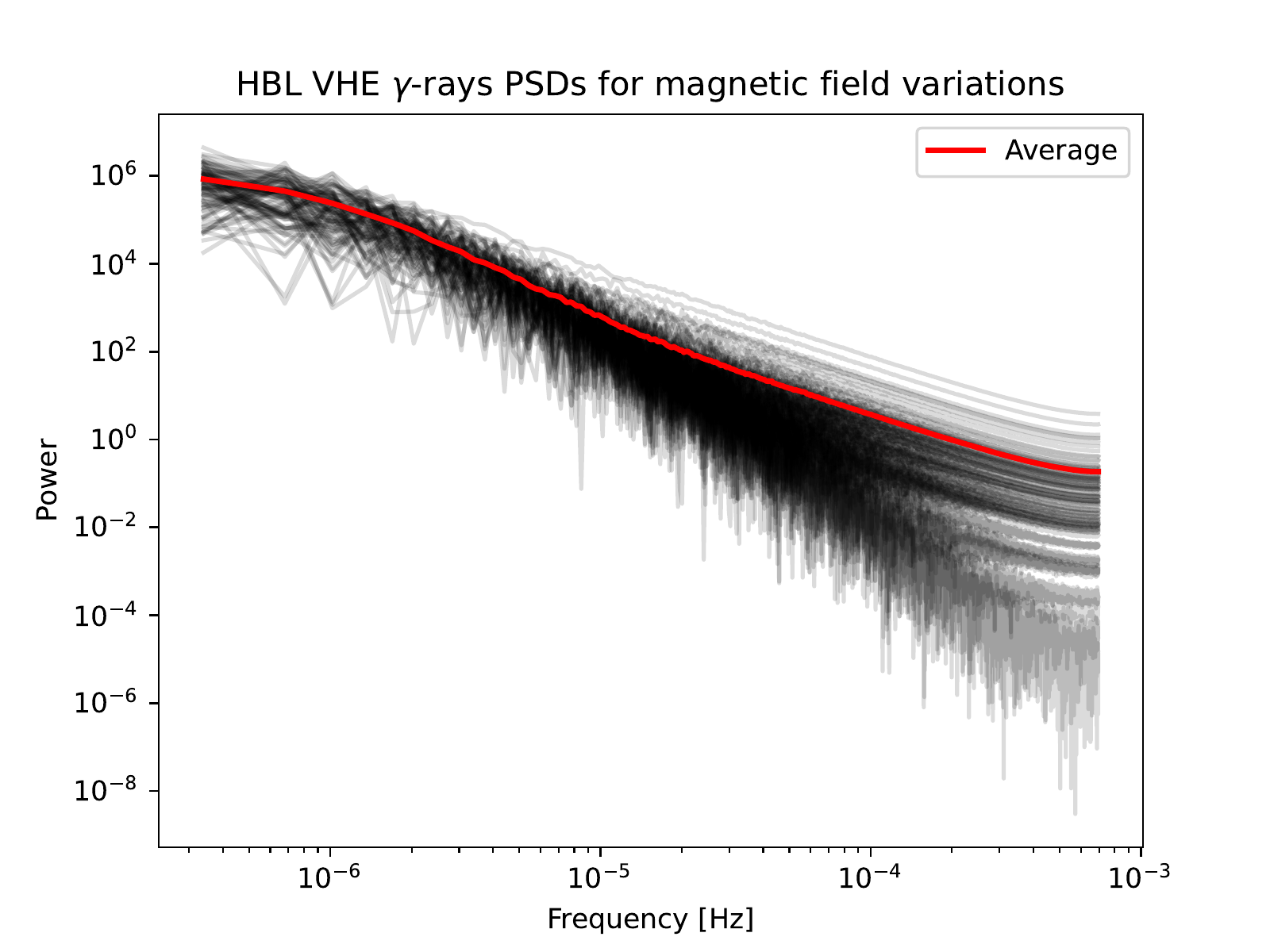}{.25\textwidth}{}
        \fig{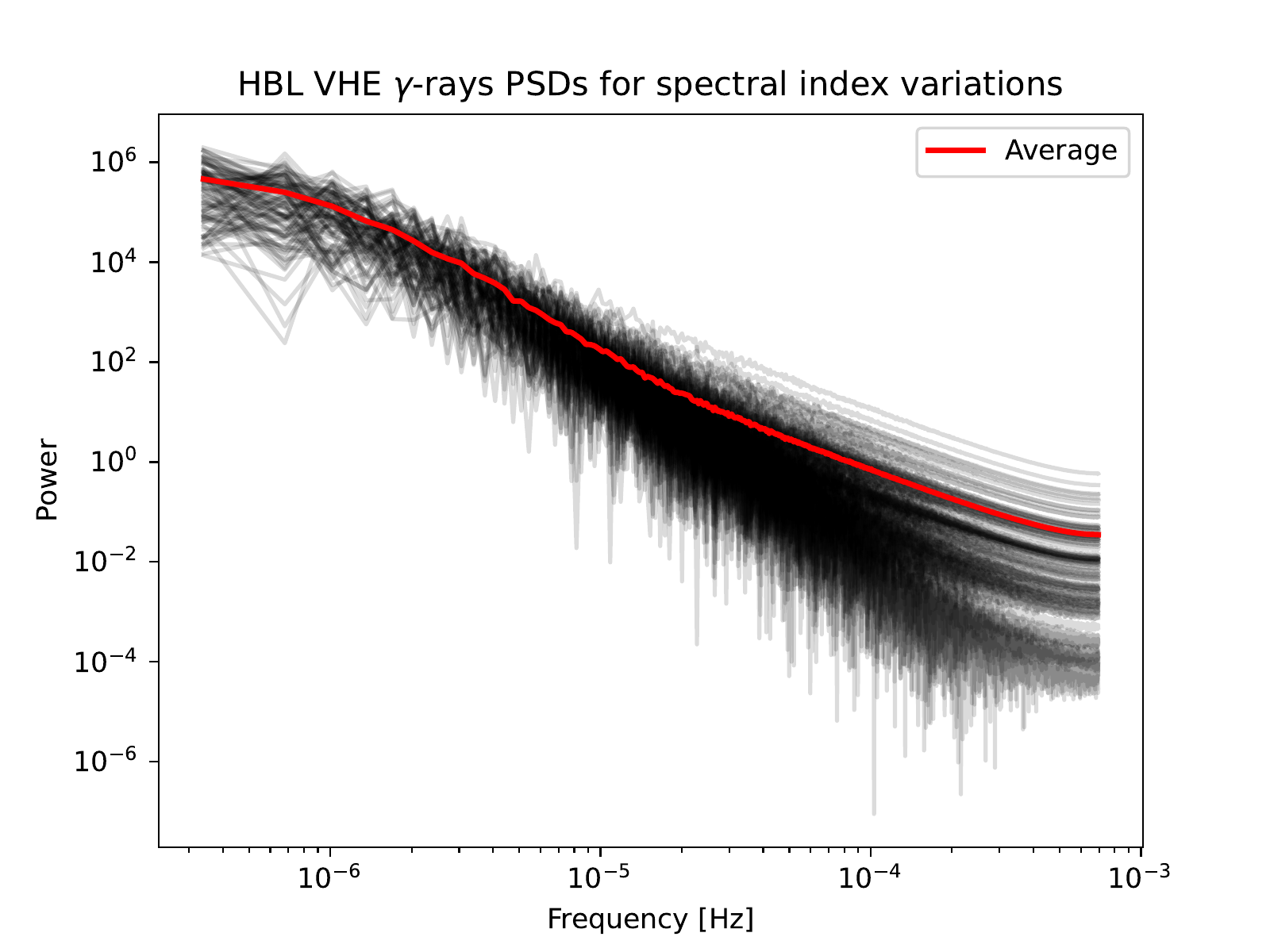}{.25\textwidth}{}
    }
    \gridline{
        \fig{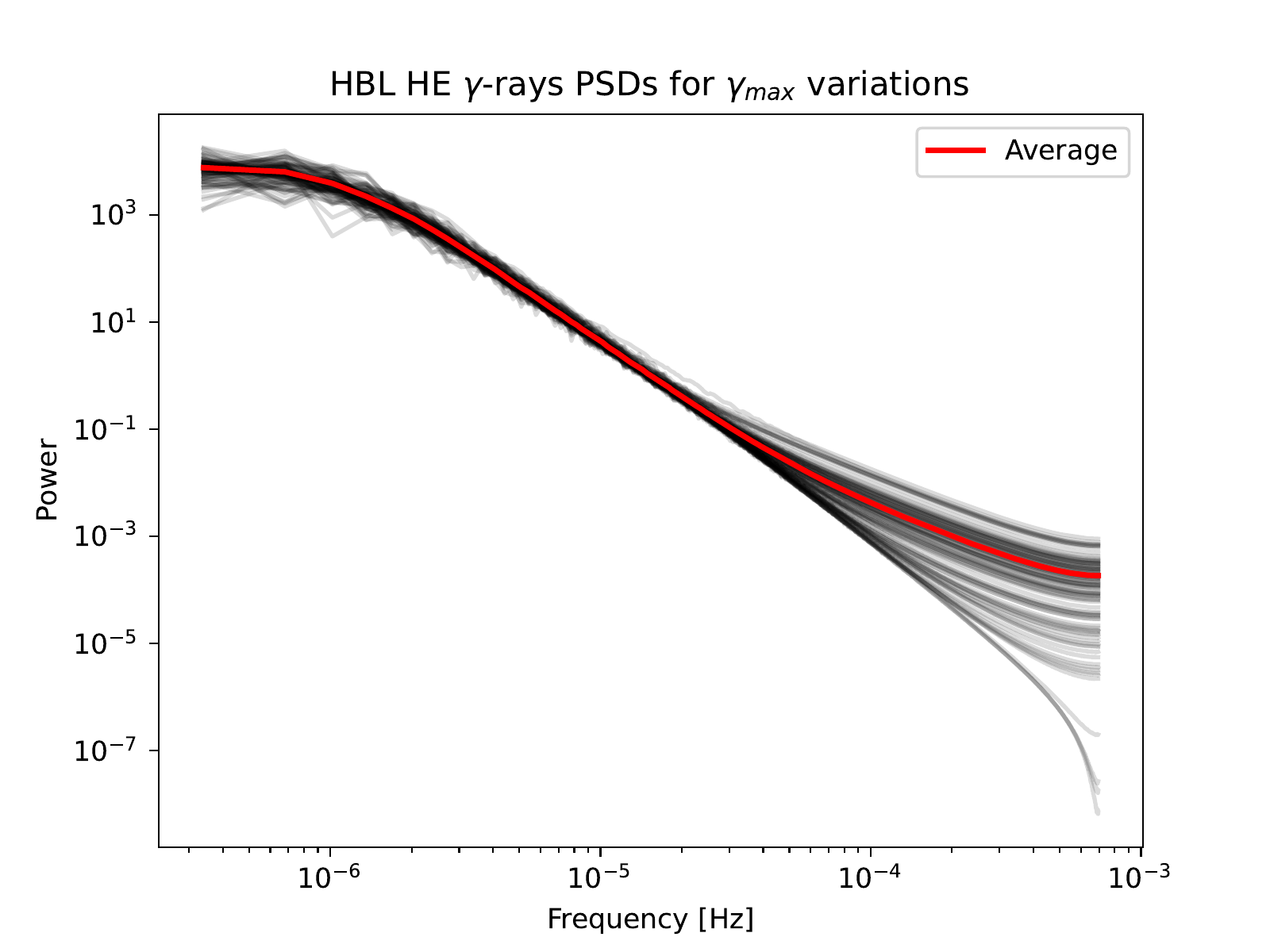}{.25\textwidth}{}
        \fig{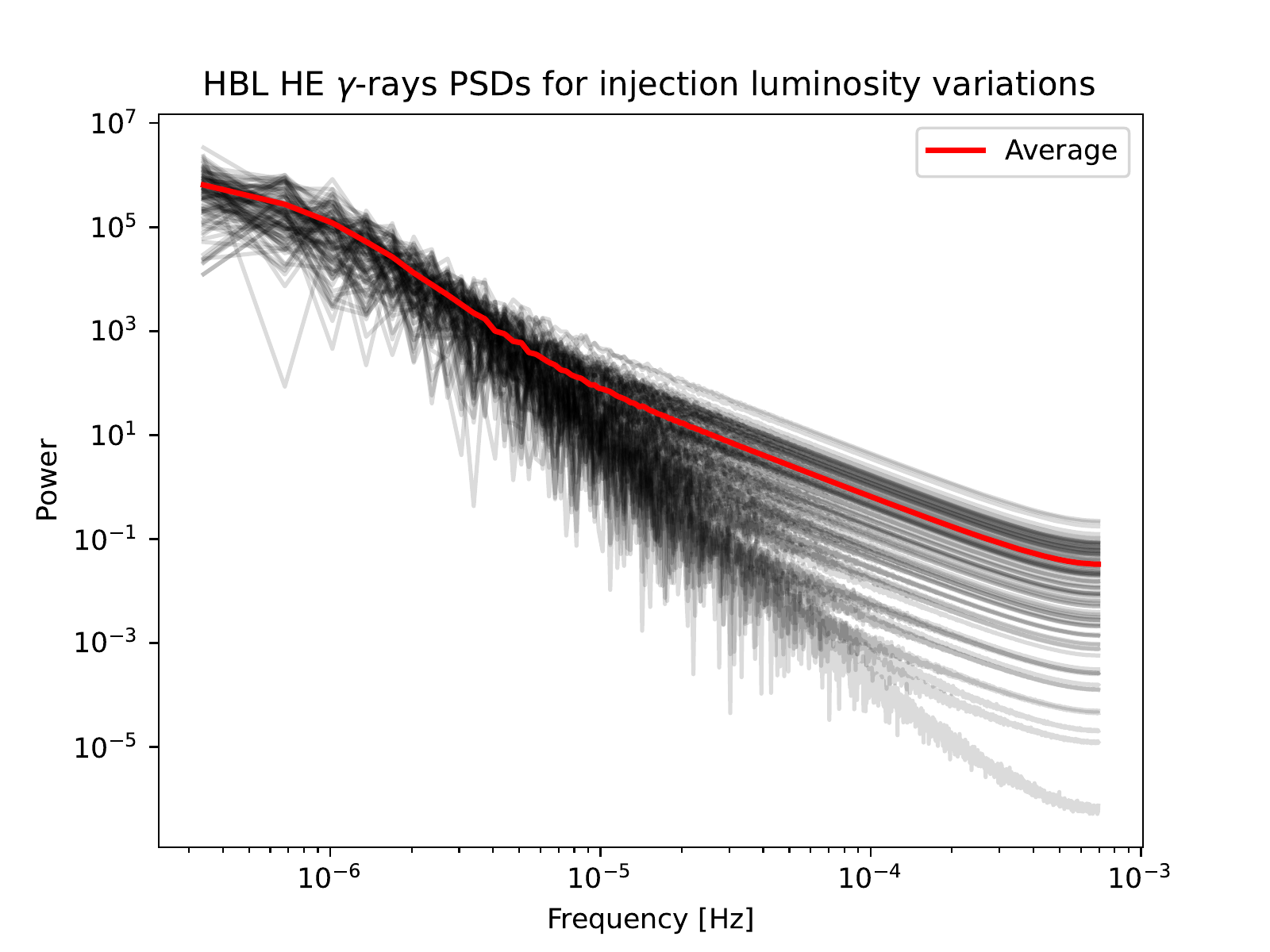}{.25\textwidth}{}
        \fig{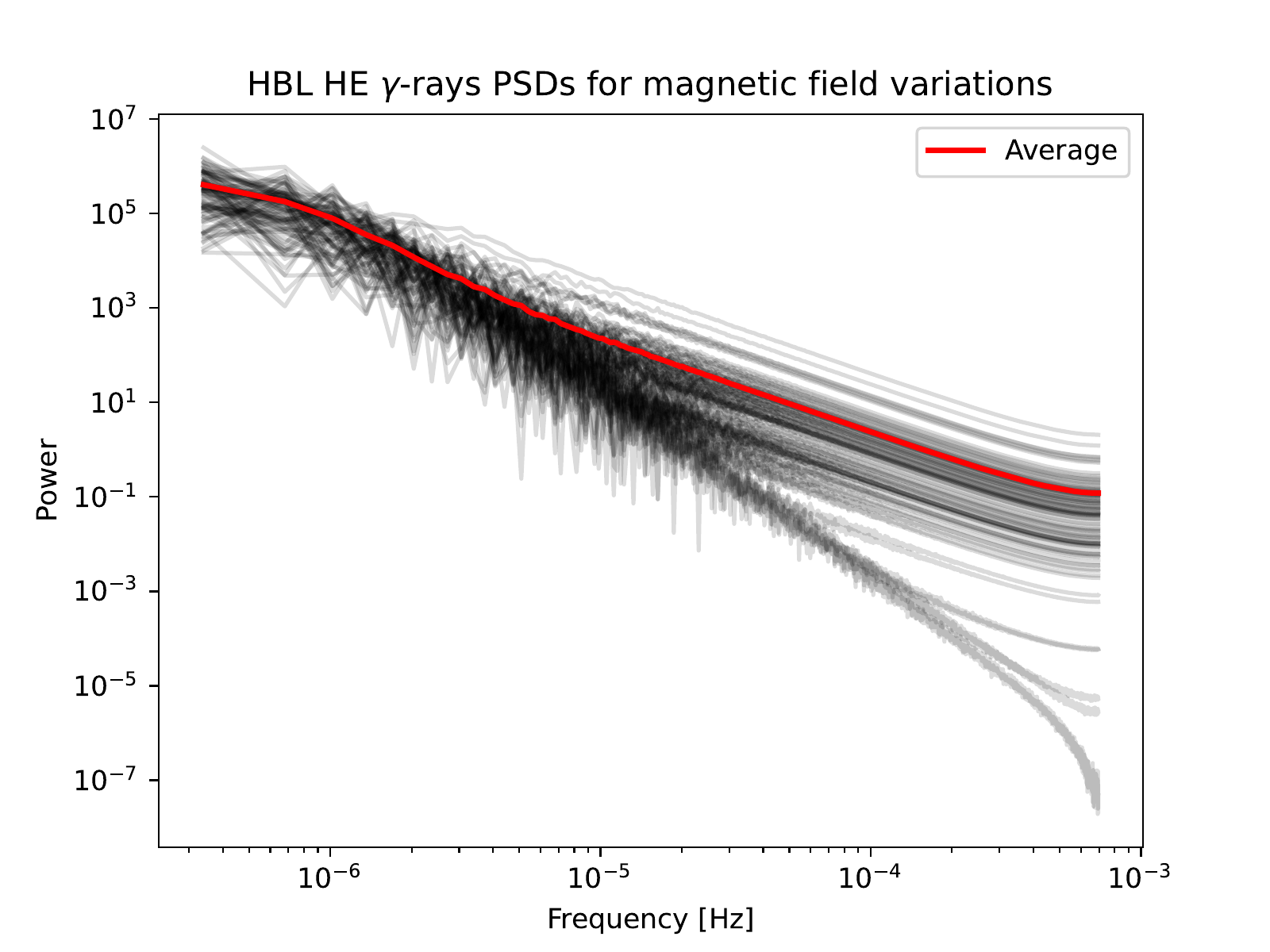}{.25\textwidth}{}
        \fig{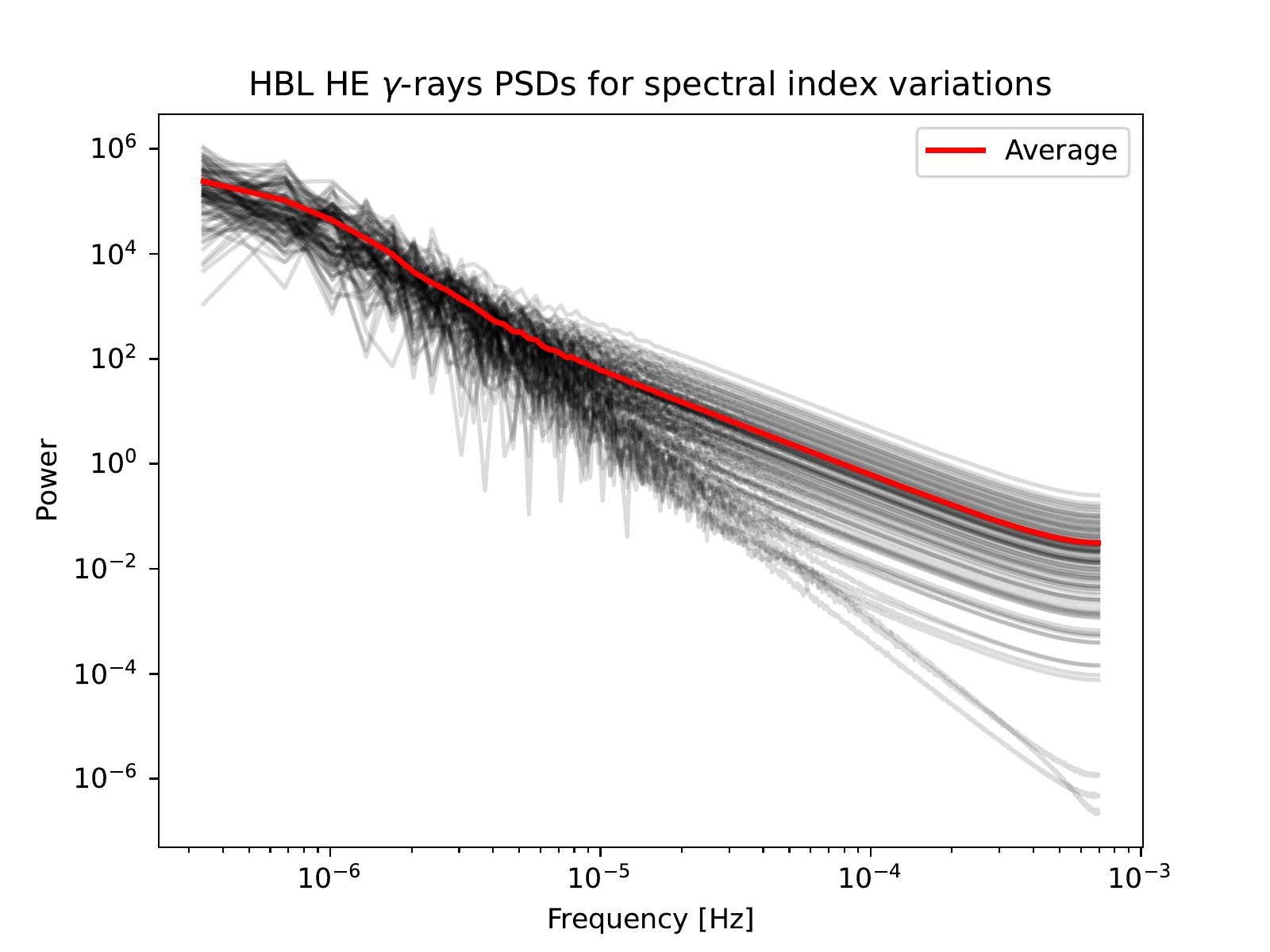}{.25\textwidth}{}
    }
    \gridline{
        \fig{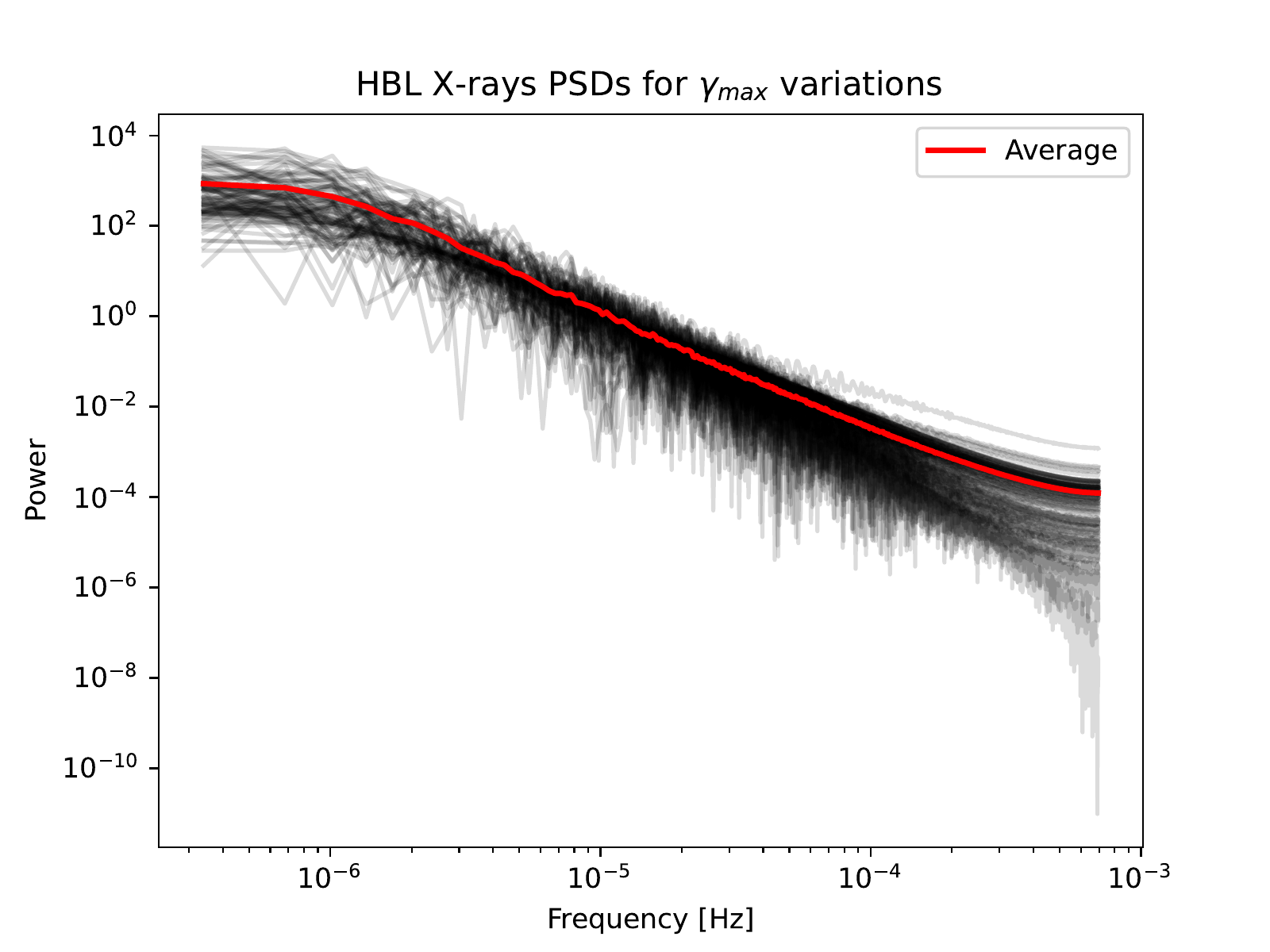}{.25\textwidth}{}
        \fig{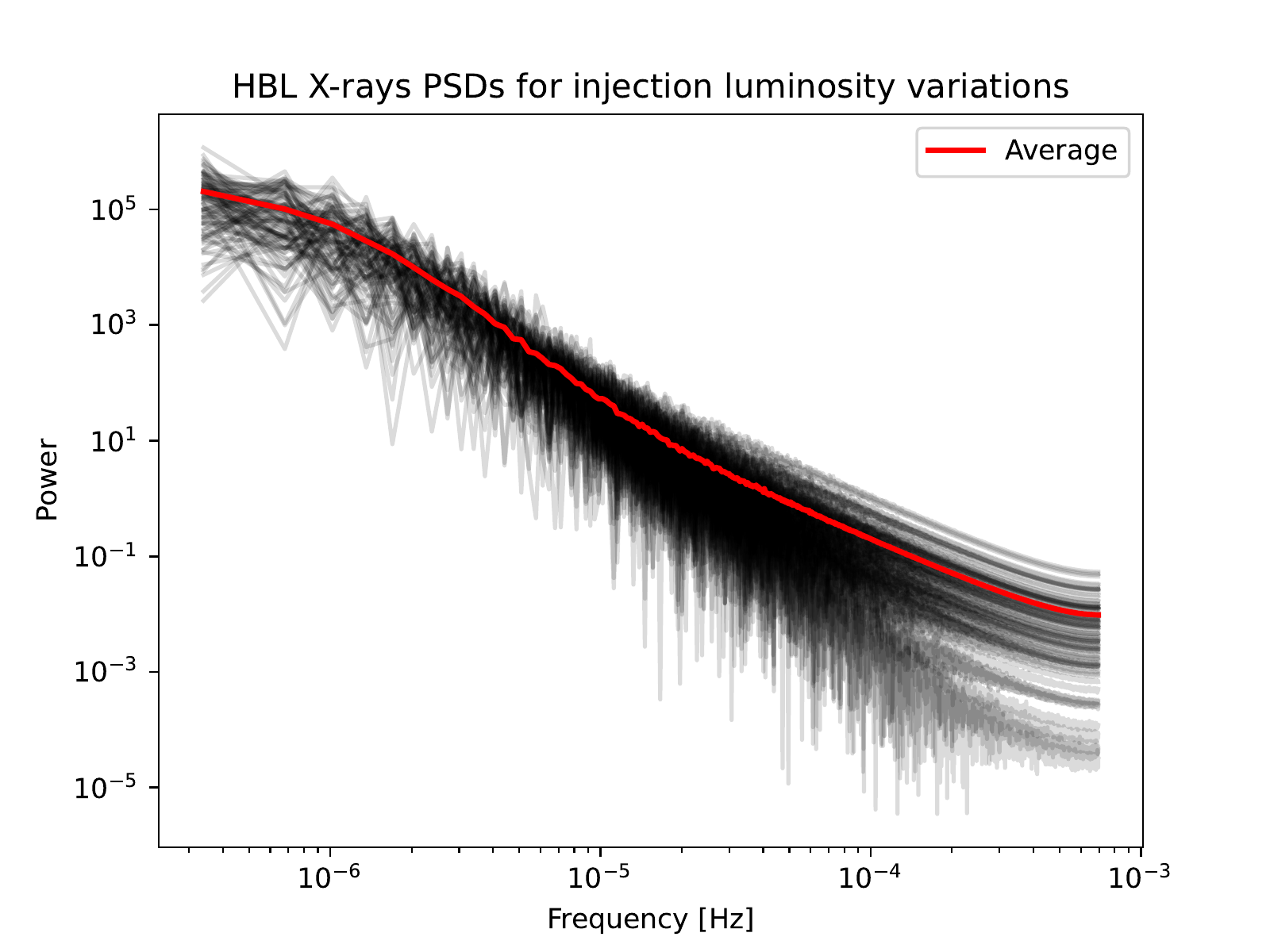}{.25\textwidth}{}
        \fig{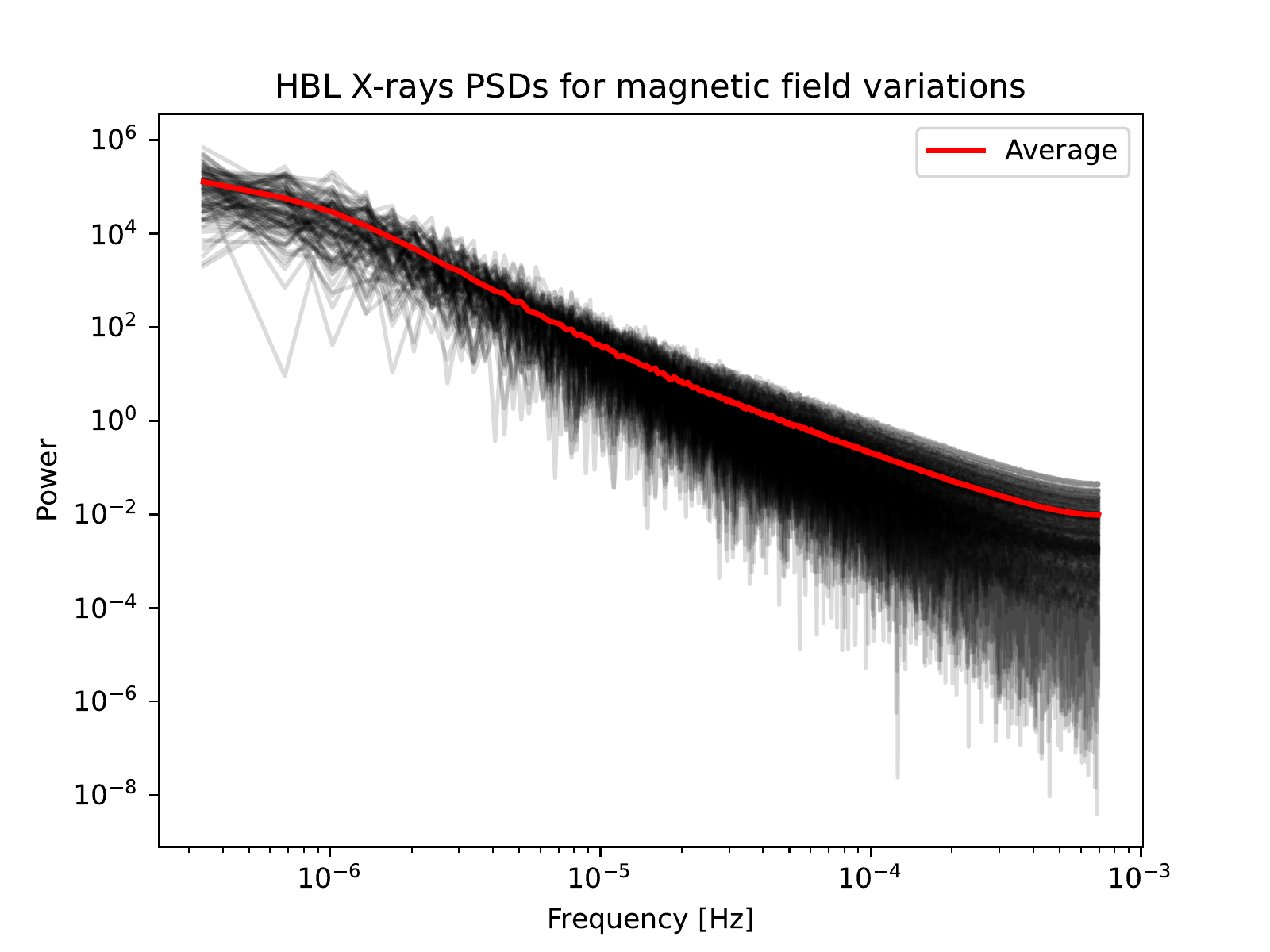}{.25\textwidth}{}
        \fig{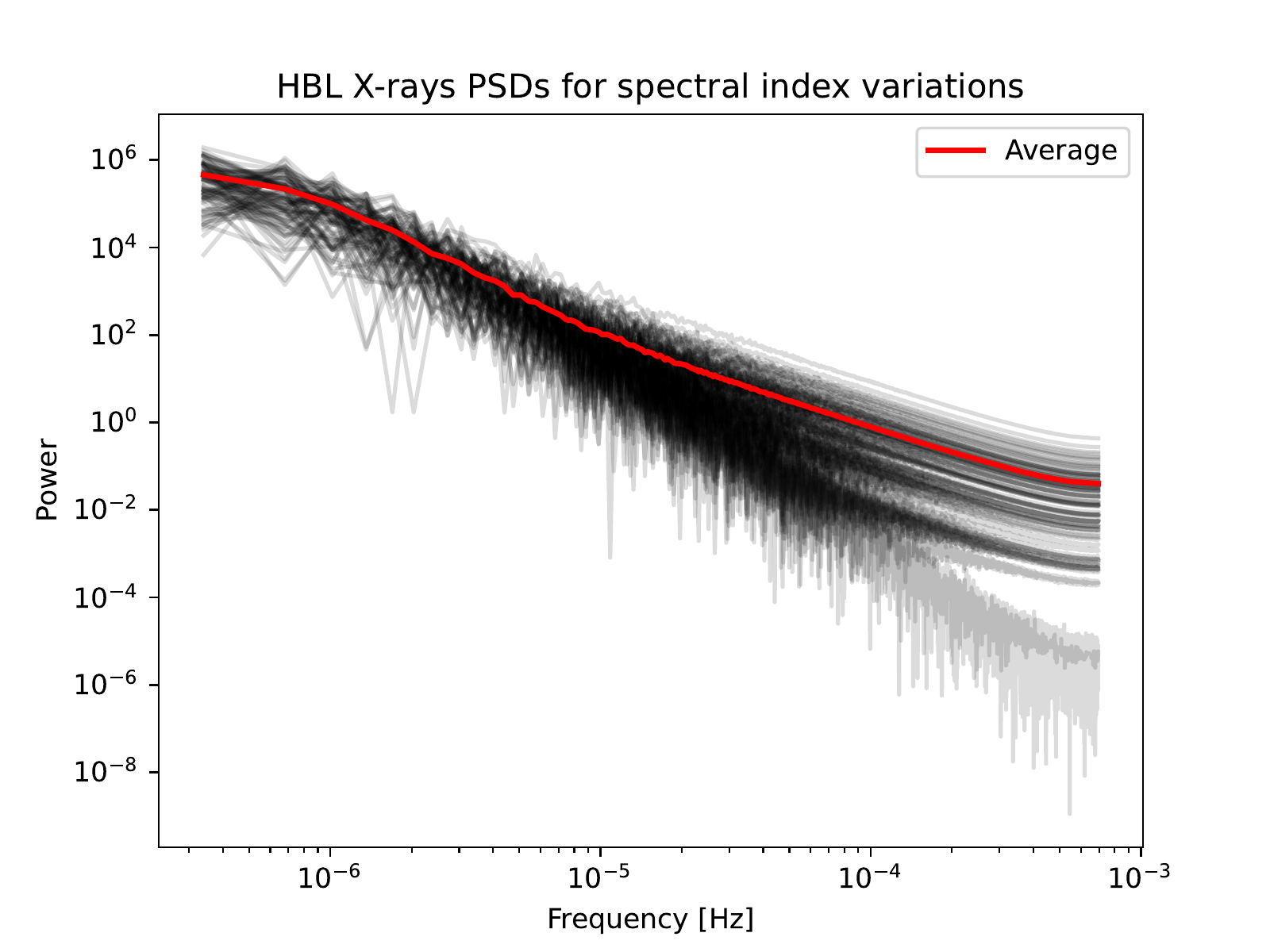}{.25\textwidth}{}
    }
    \gridline{
        \fig{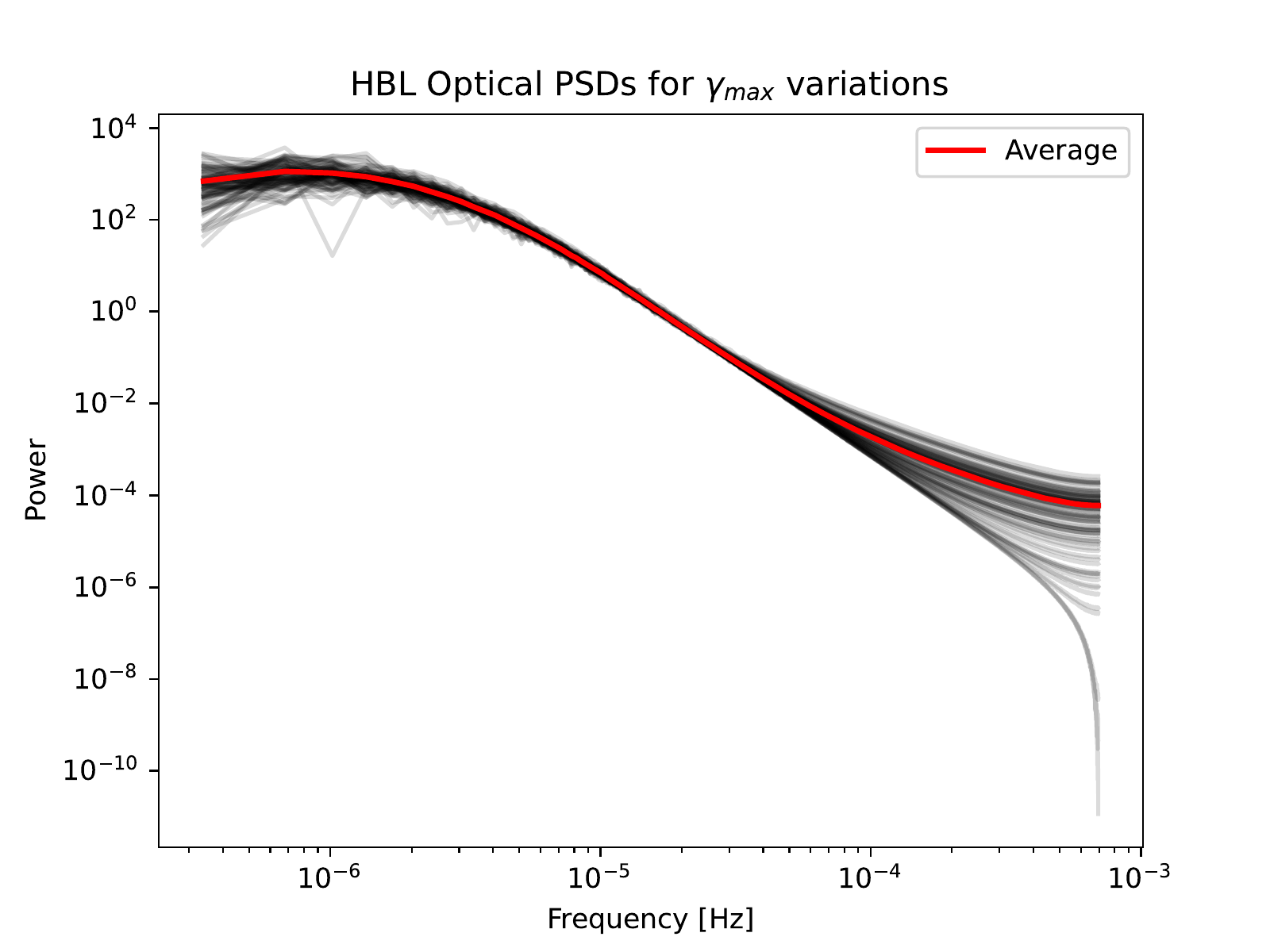}{.25\textwidth}{}
        \fig{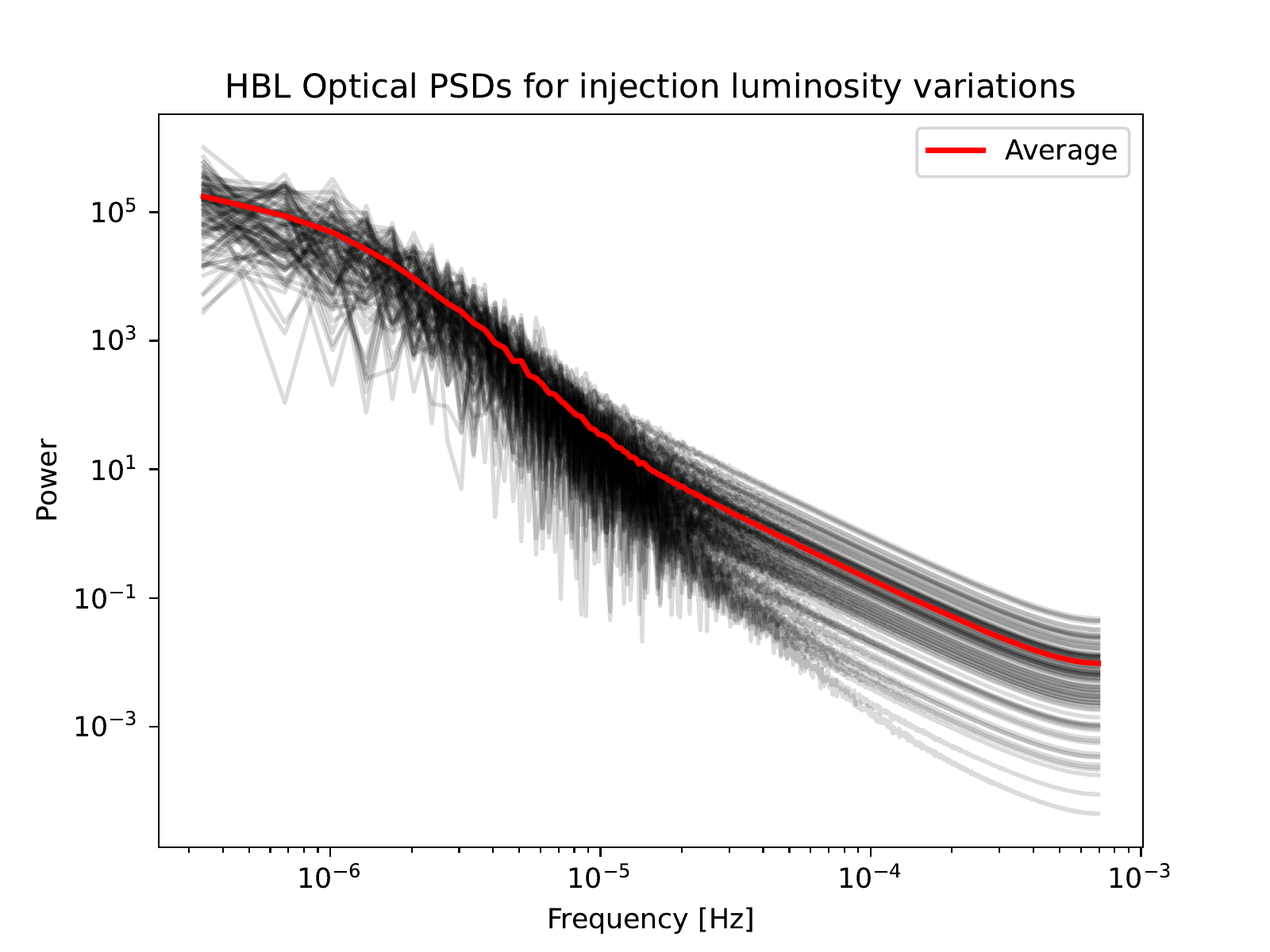}{.25\textwidth}{}
        \fig{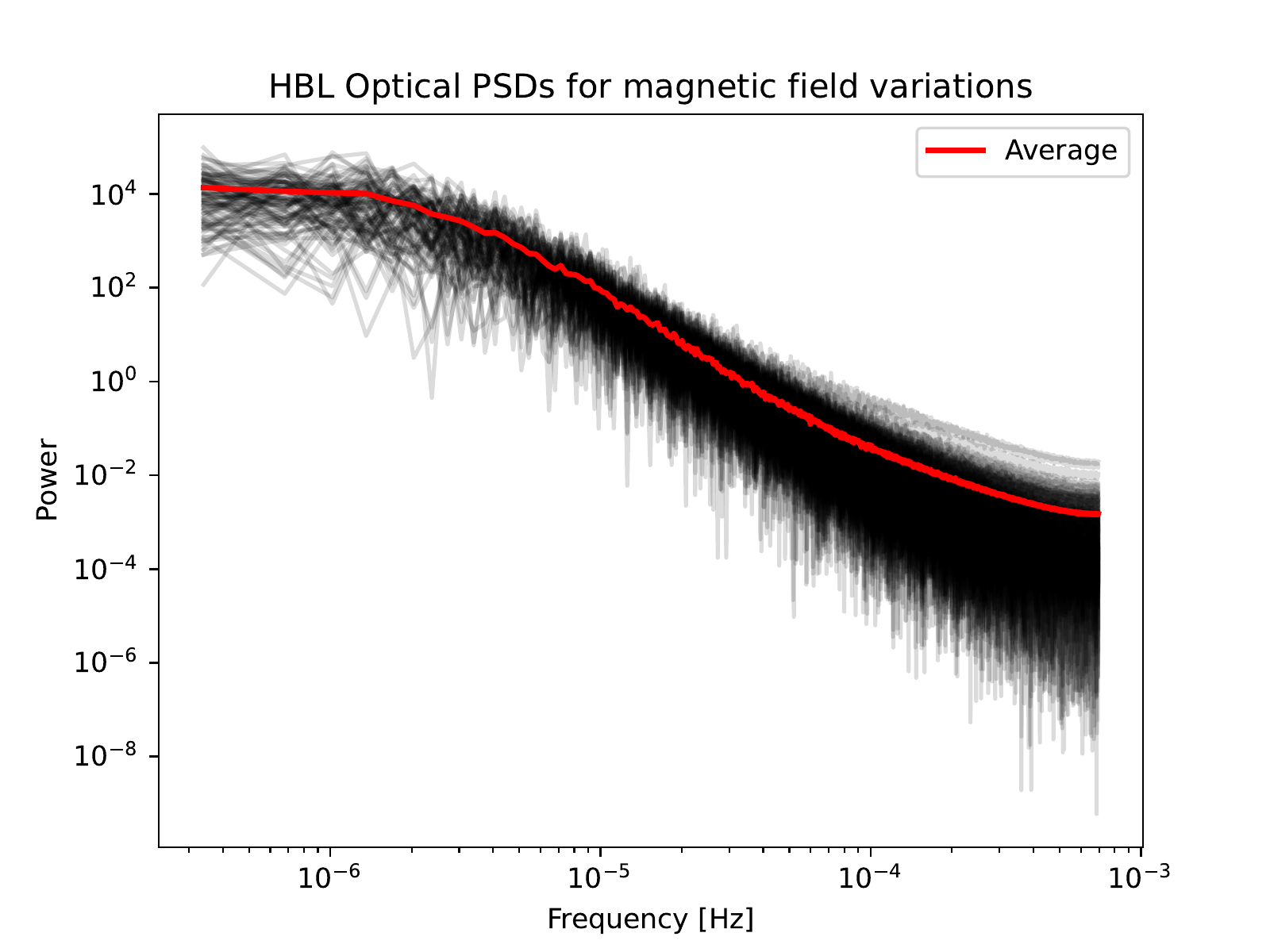}{.25\textwidth}{}
        \fig{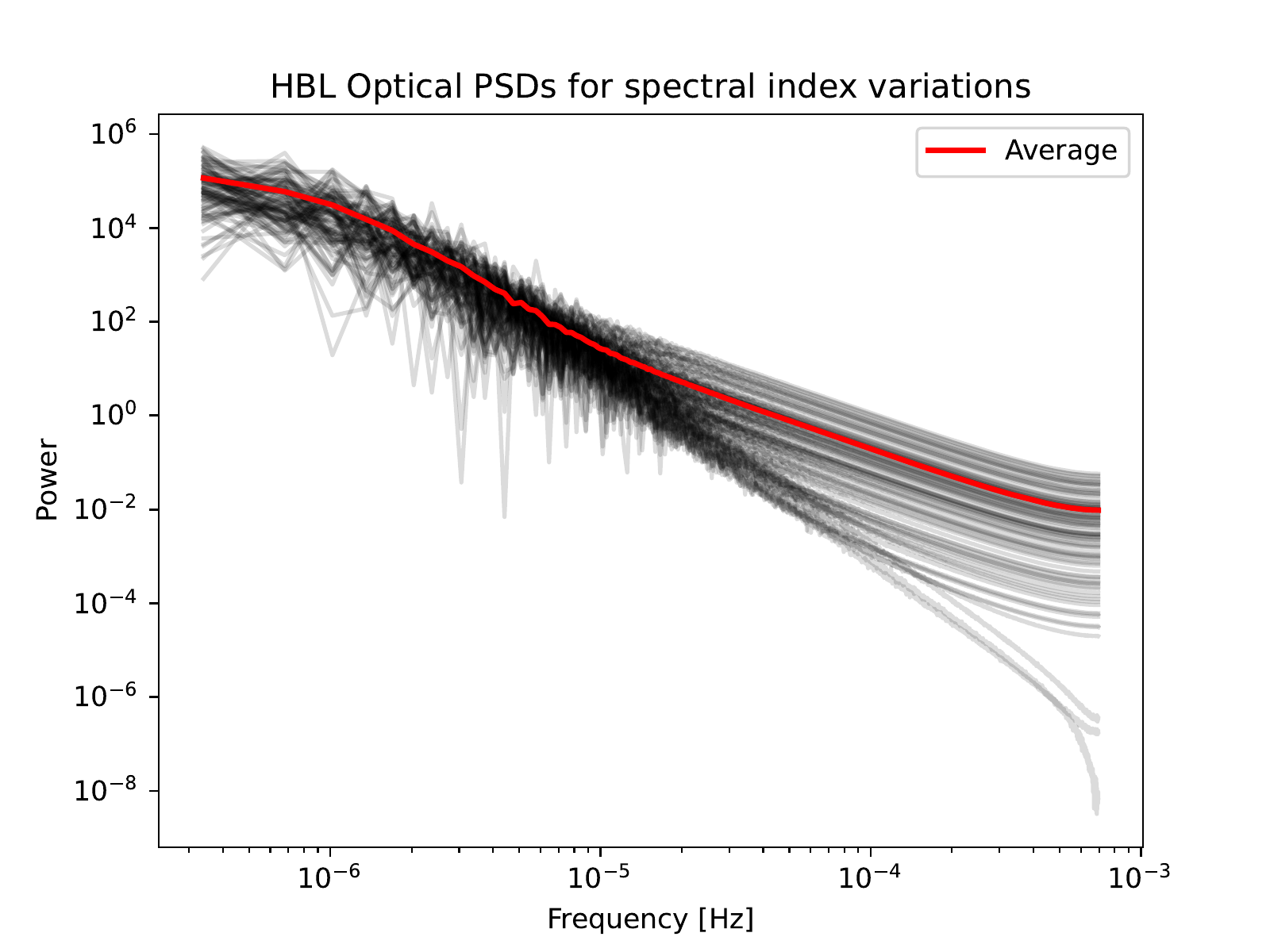}{.25\textwidth}{}
    }
    \caption{
        Average and individual PSD comparisons for HBL simulation realizations.}
    \label{fig:SSC-PSD-ave_vs_indiv}
\end{figure}

\begin{figure}[h]
    \gridline{
        \fig{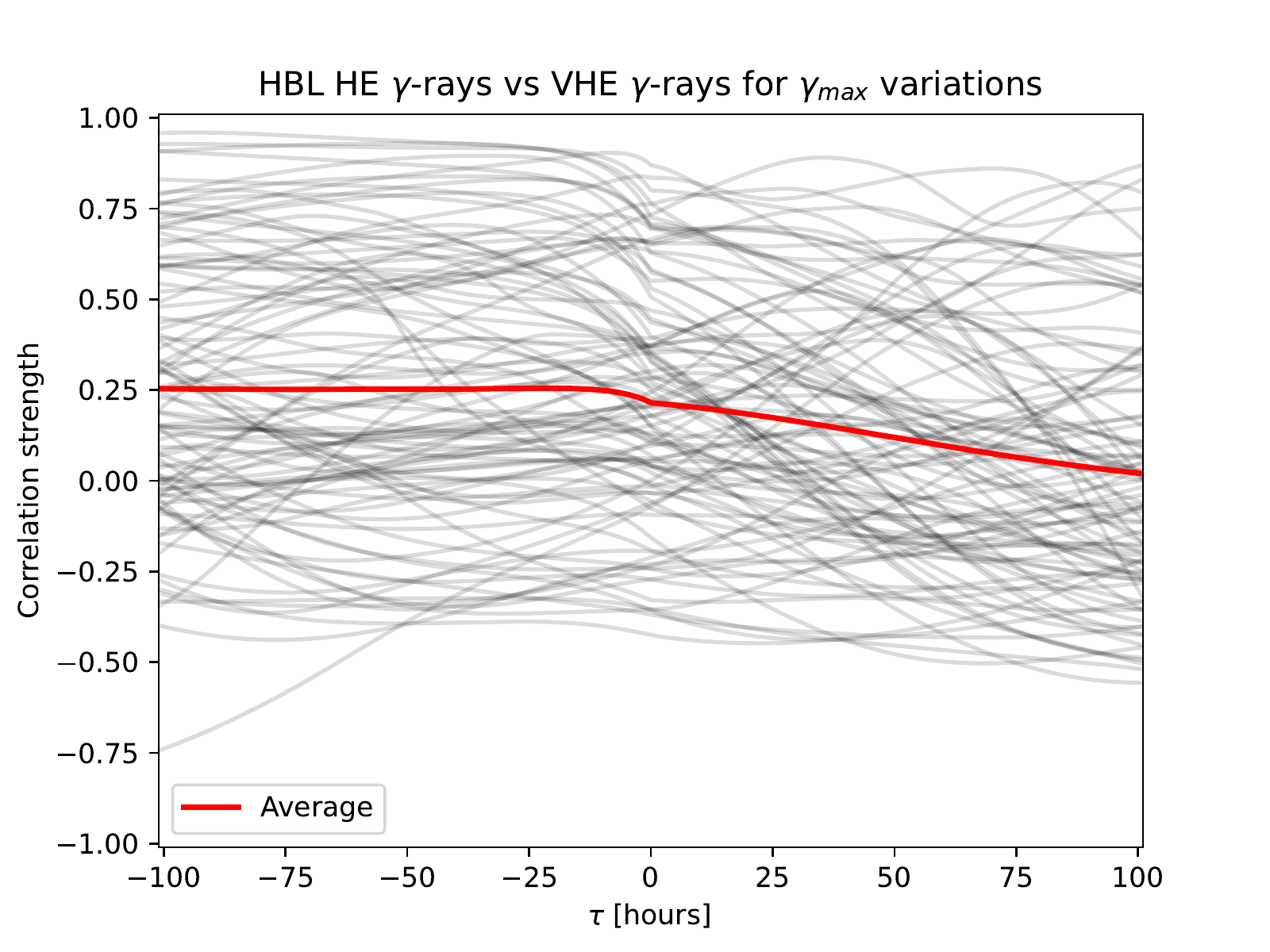}{.3\textwidth}{}
        \fig{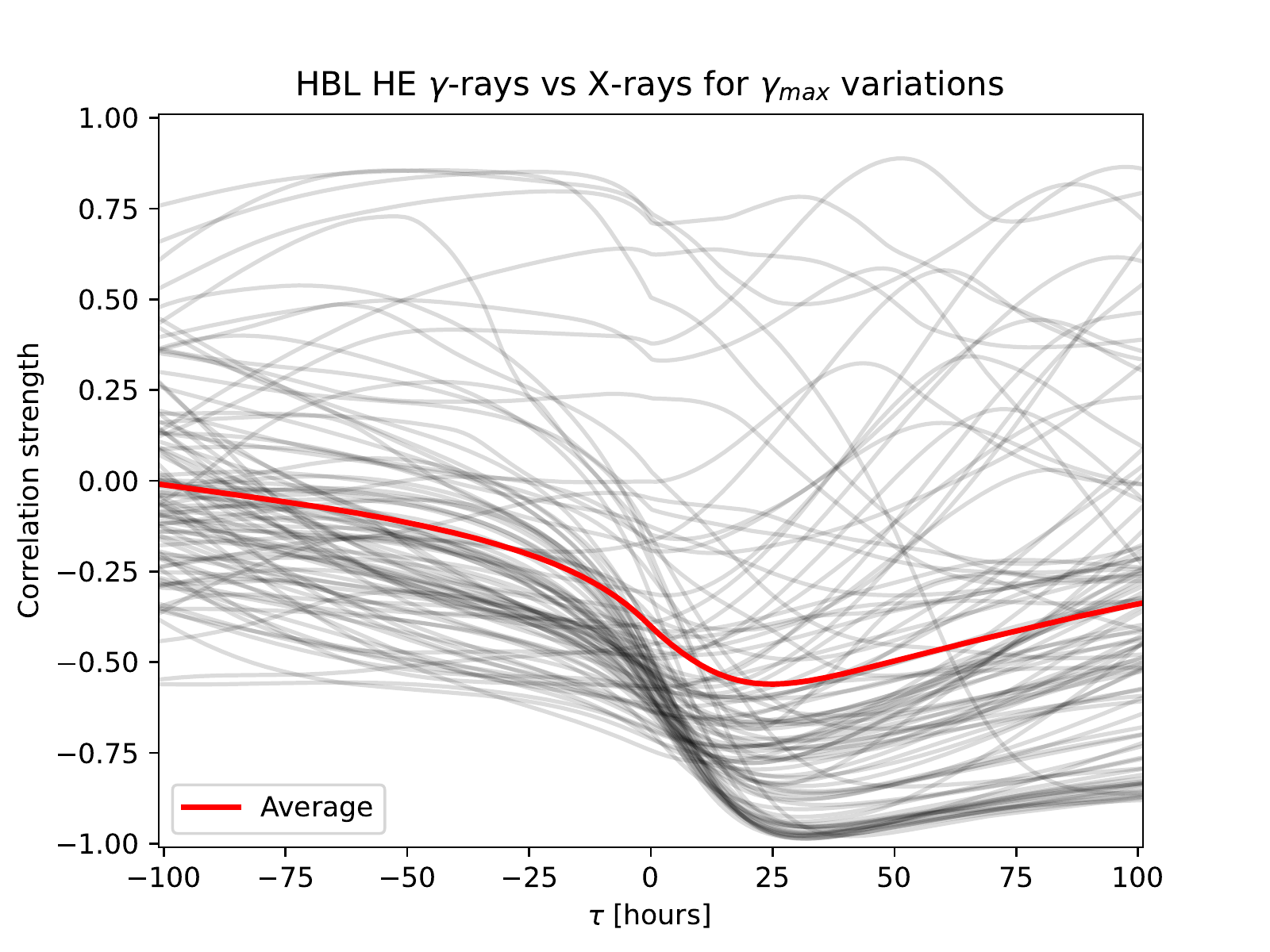}{.3\textwidth}{}
        \fig{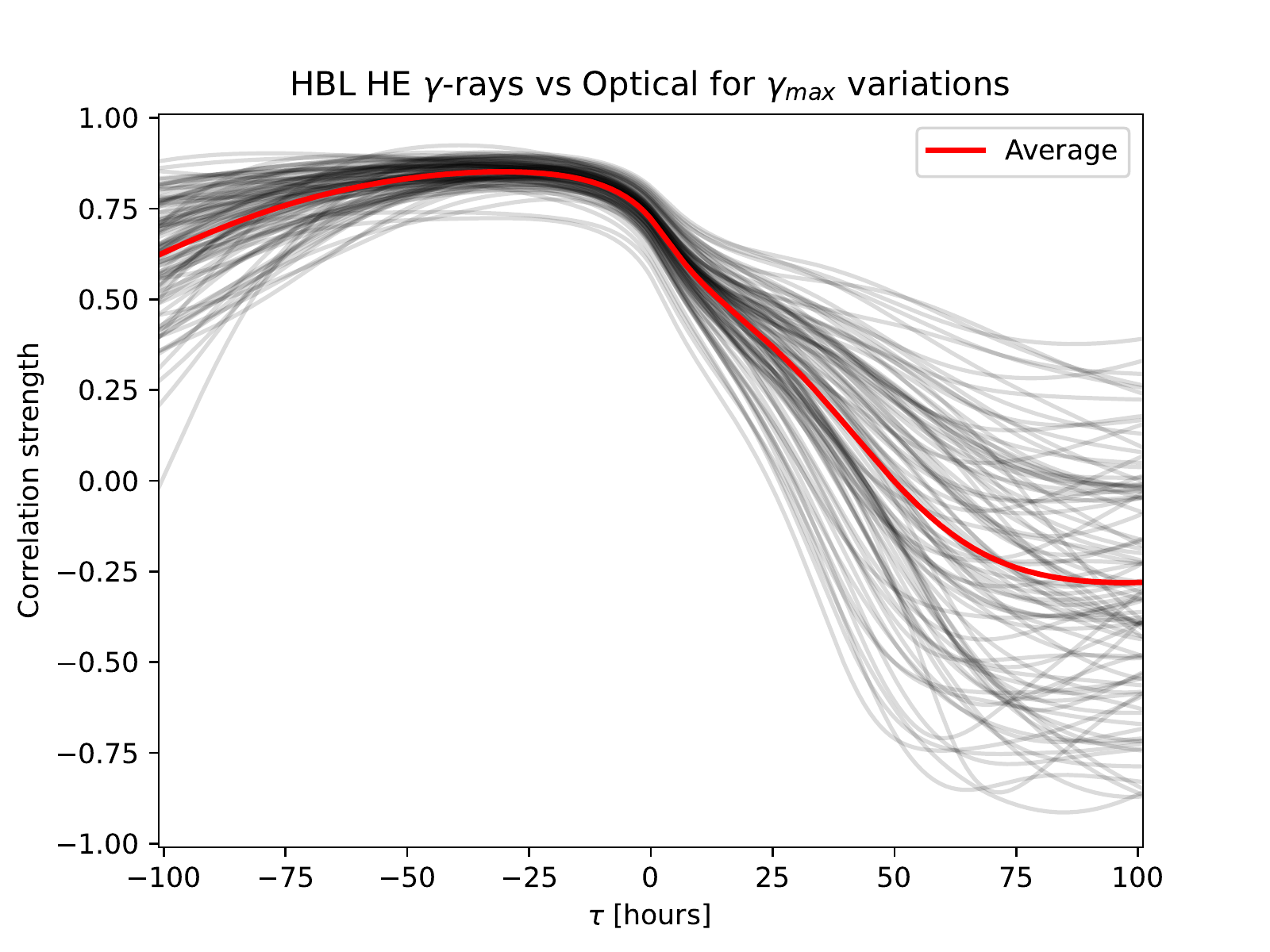}{.3\textwidth}{}
    }
    \gridline{
        \fig{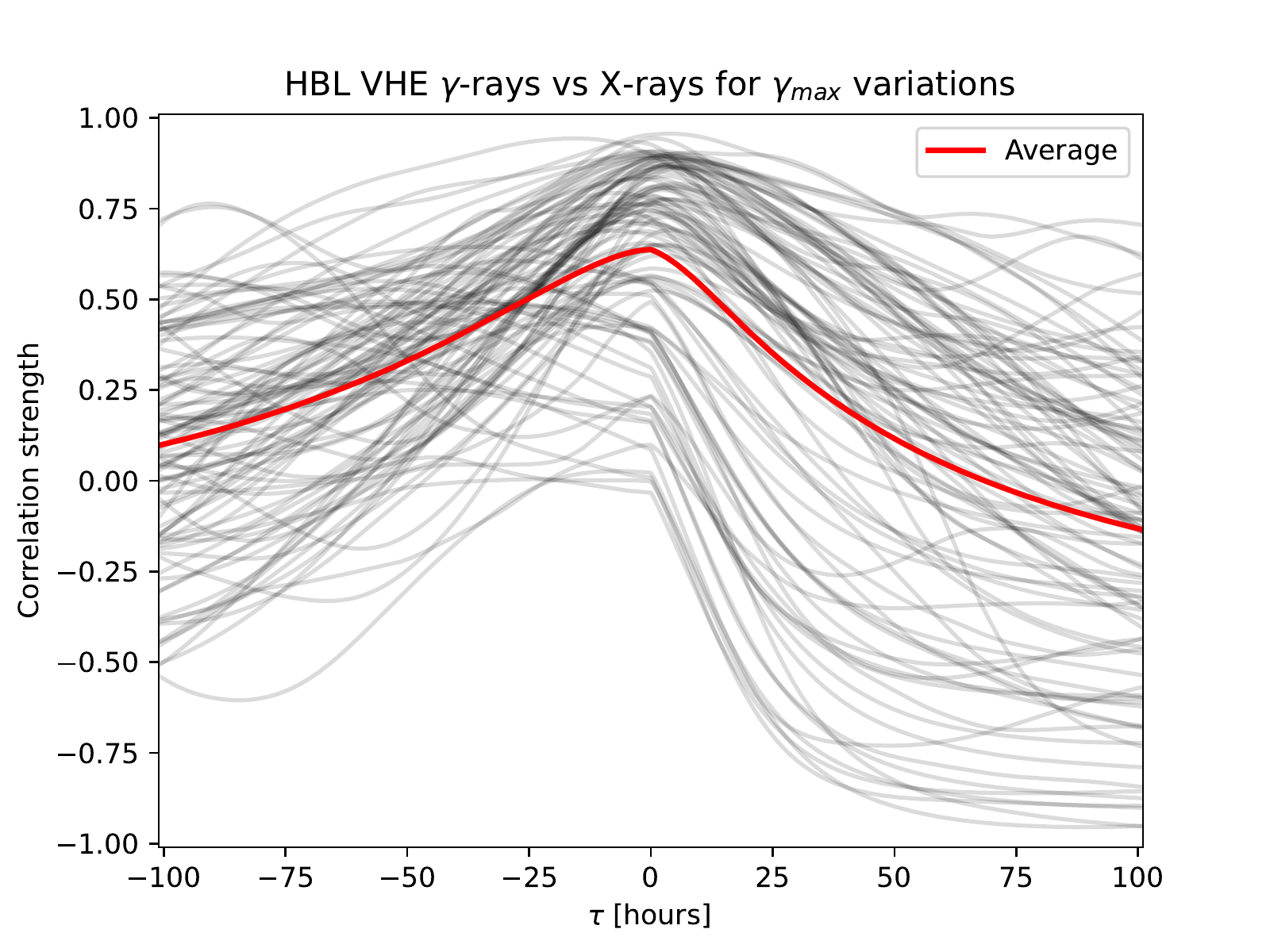}{.3\textwidth}{}
        \fig{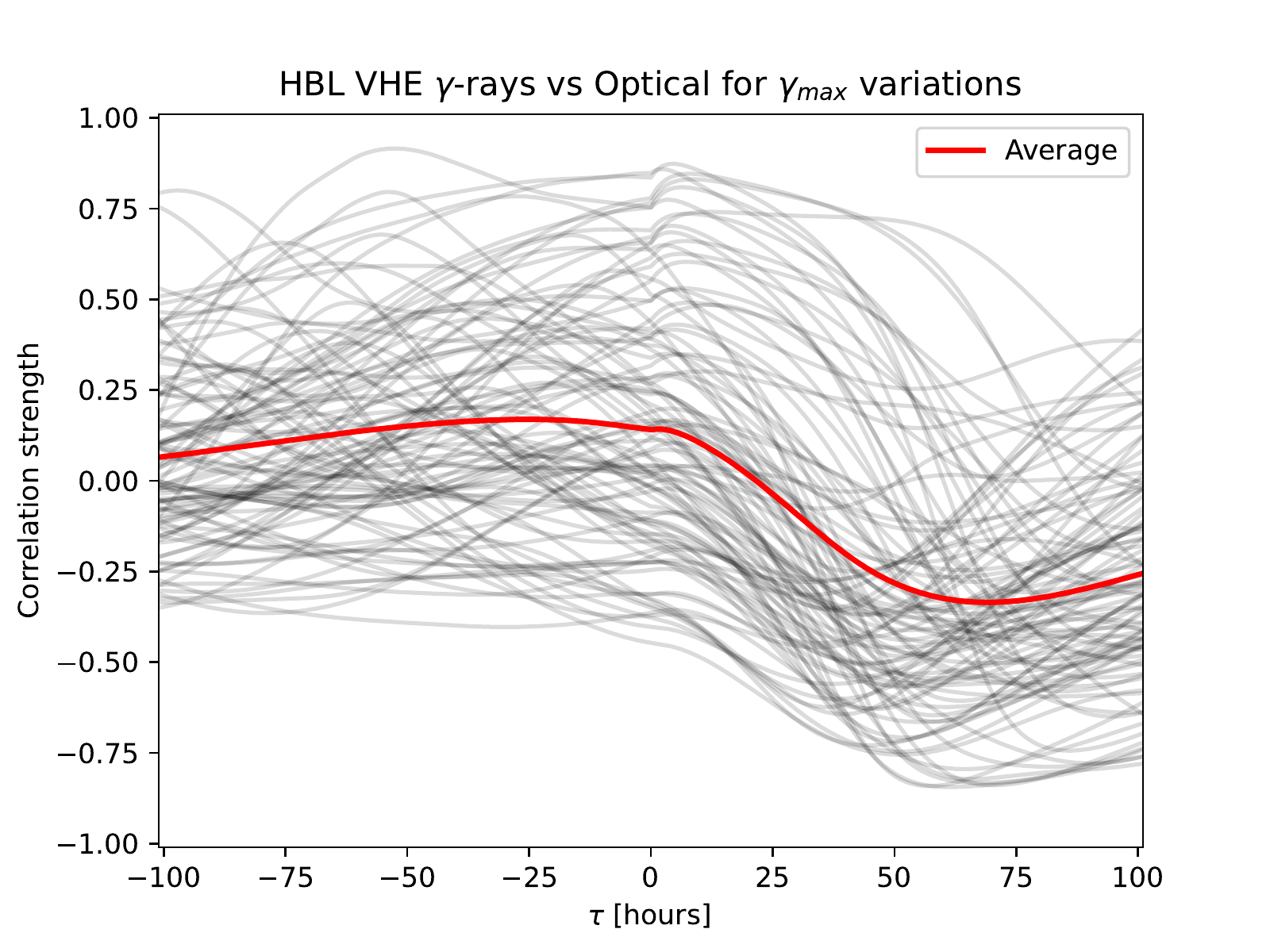}{.3\textwidth}{}
        \fig{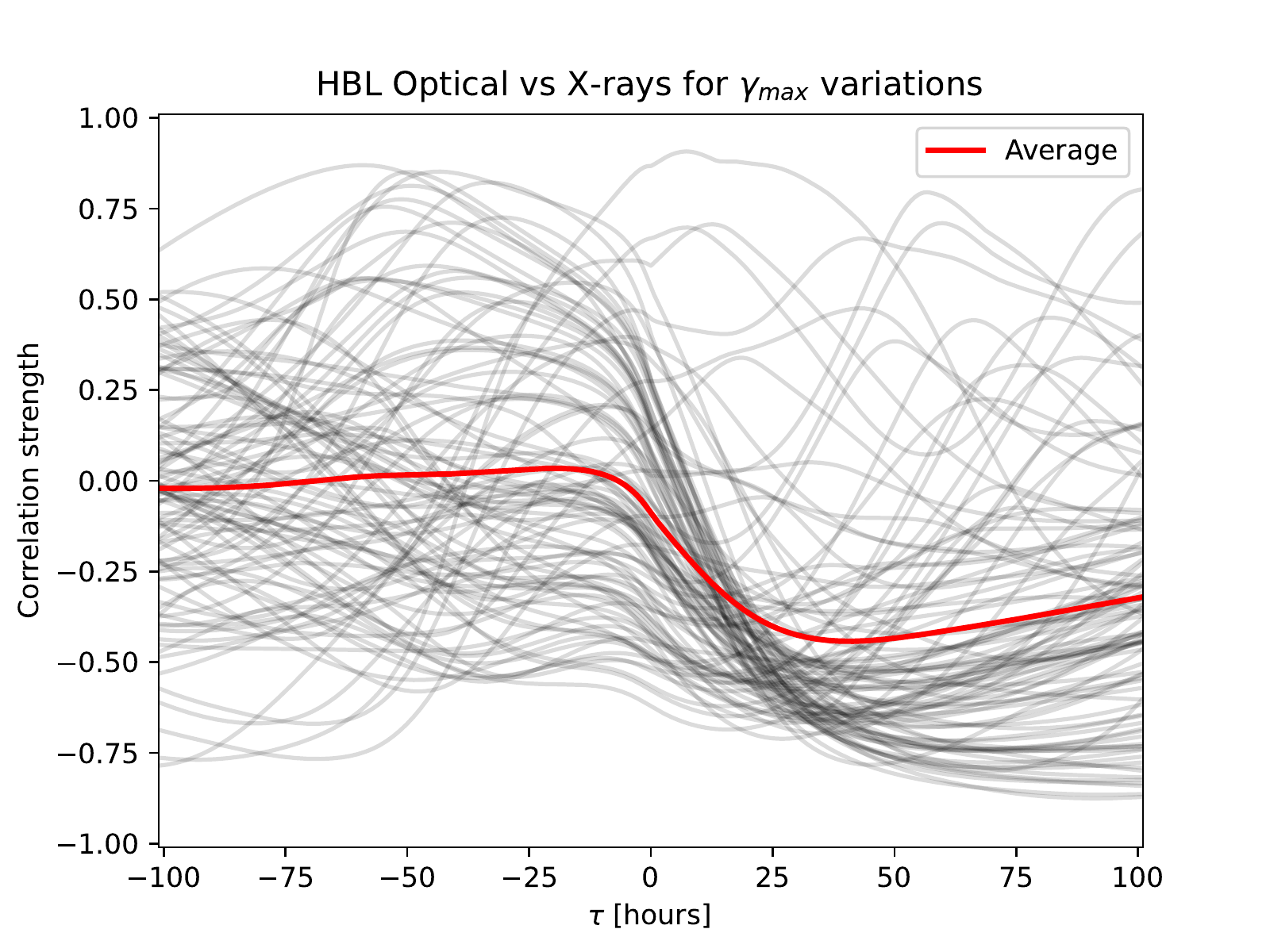}{.3\textwidth}{}
    }
    \caption{
        Average and individual cross-correlation comparisons for HBL simulation
        realizations with maximum  Lorentz factor variations.}
    \label{fig:SSC-Corr-inj-ave_vs_indiv}
\end{figure}

\begin{figure}[h]
    \gridline{
        \fig{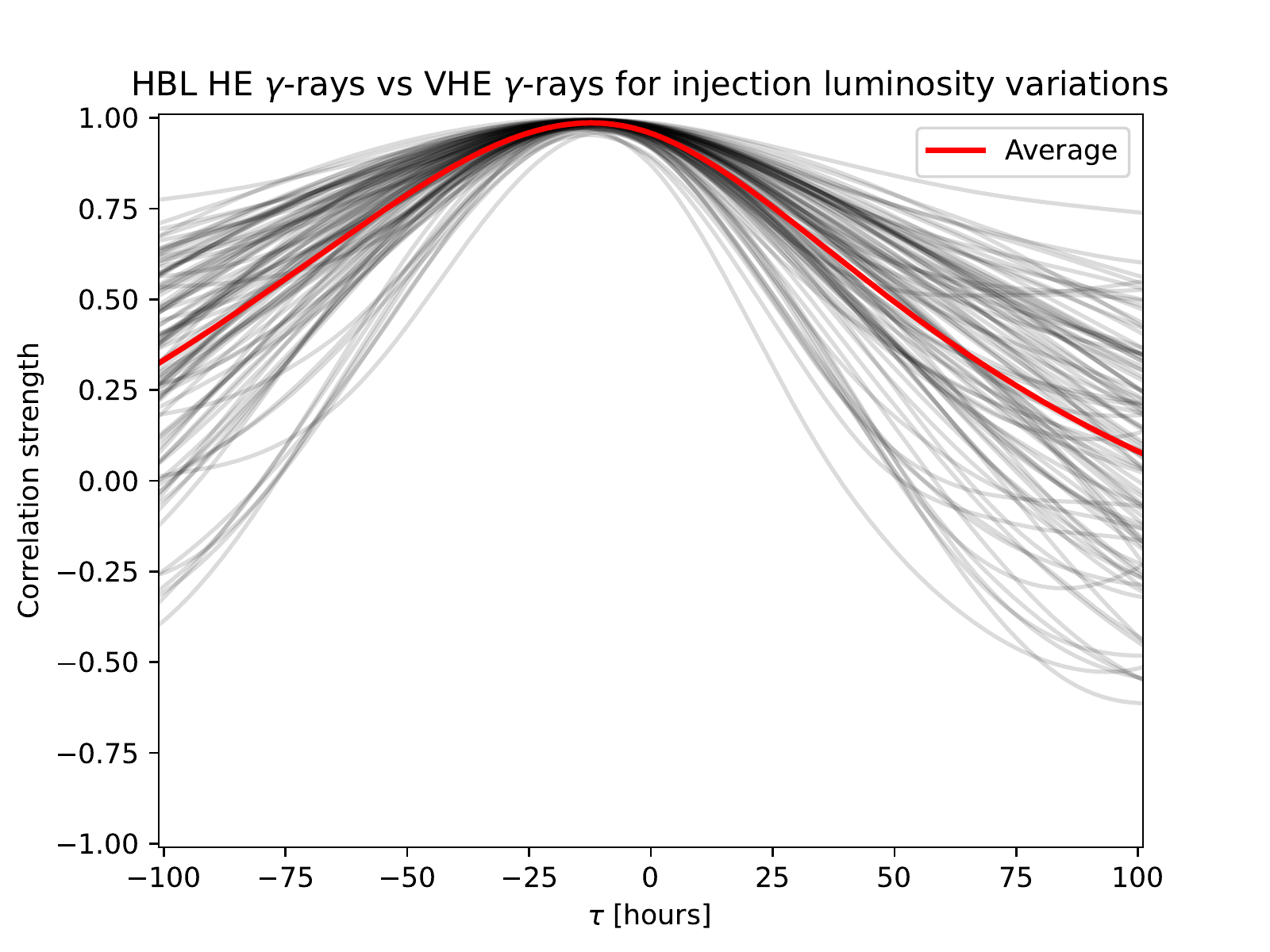}{.3\textwidth}{}
        \fig{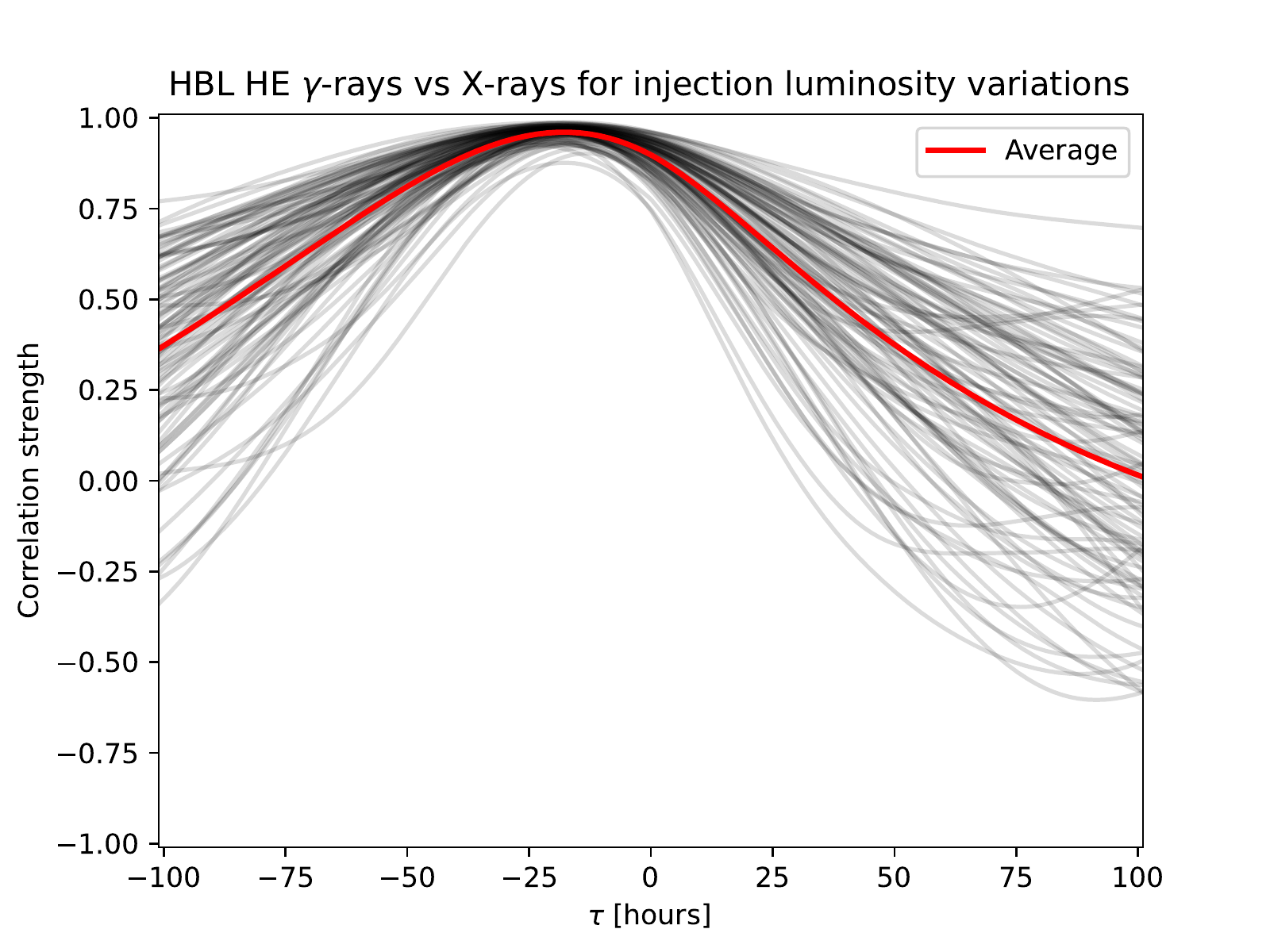}{.3\textwidth}{}
        \fig{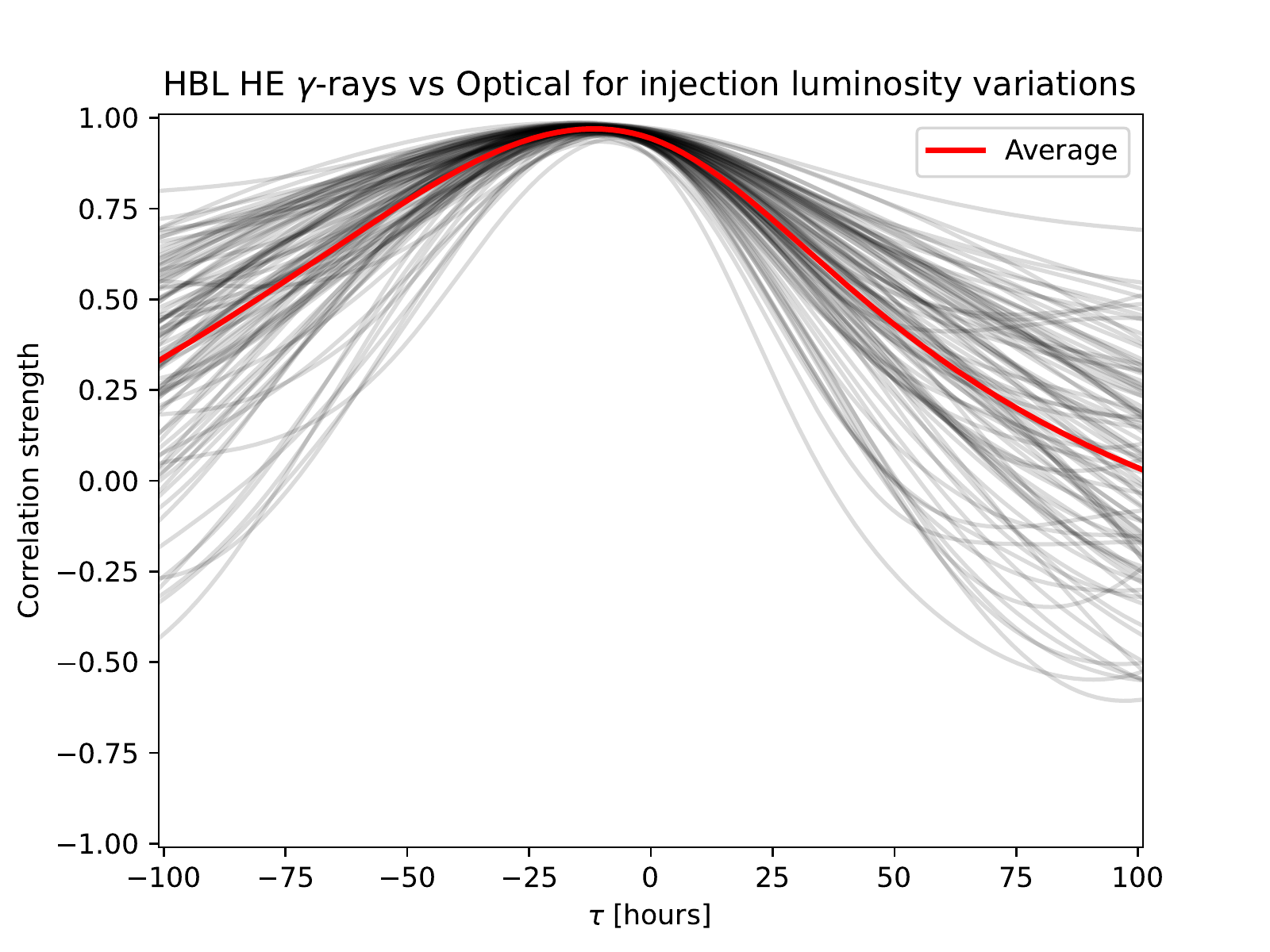}{.3\textwidth}{}
    }
    \gridline{
        \fig{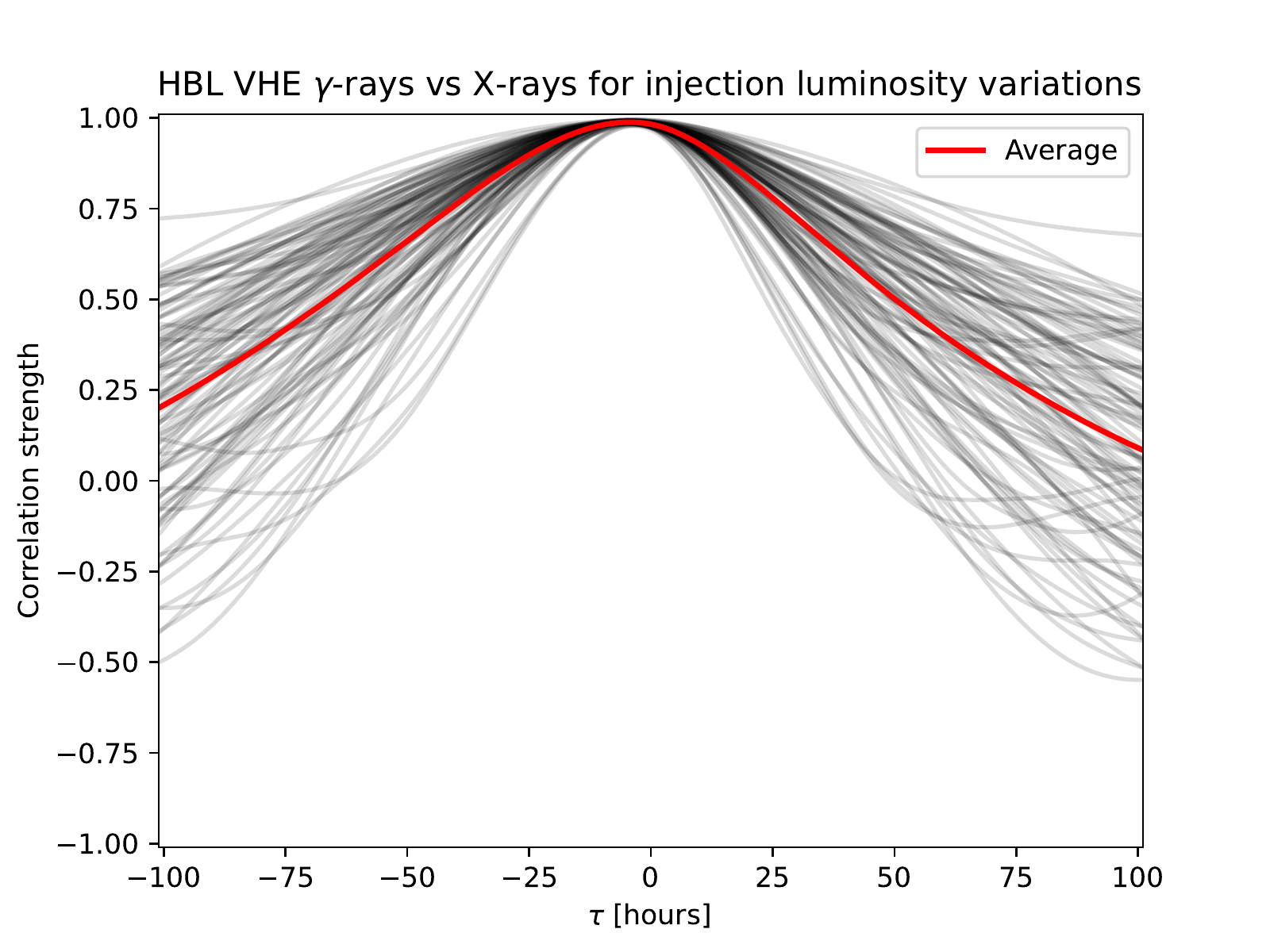}{.3\textwidth}{}
        \fig{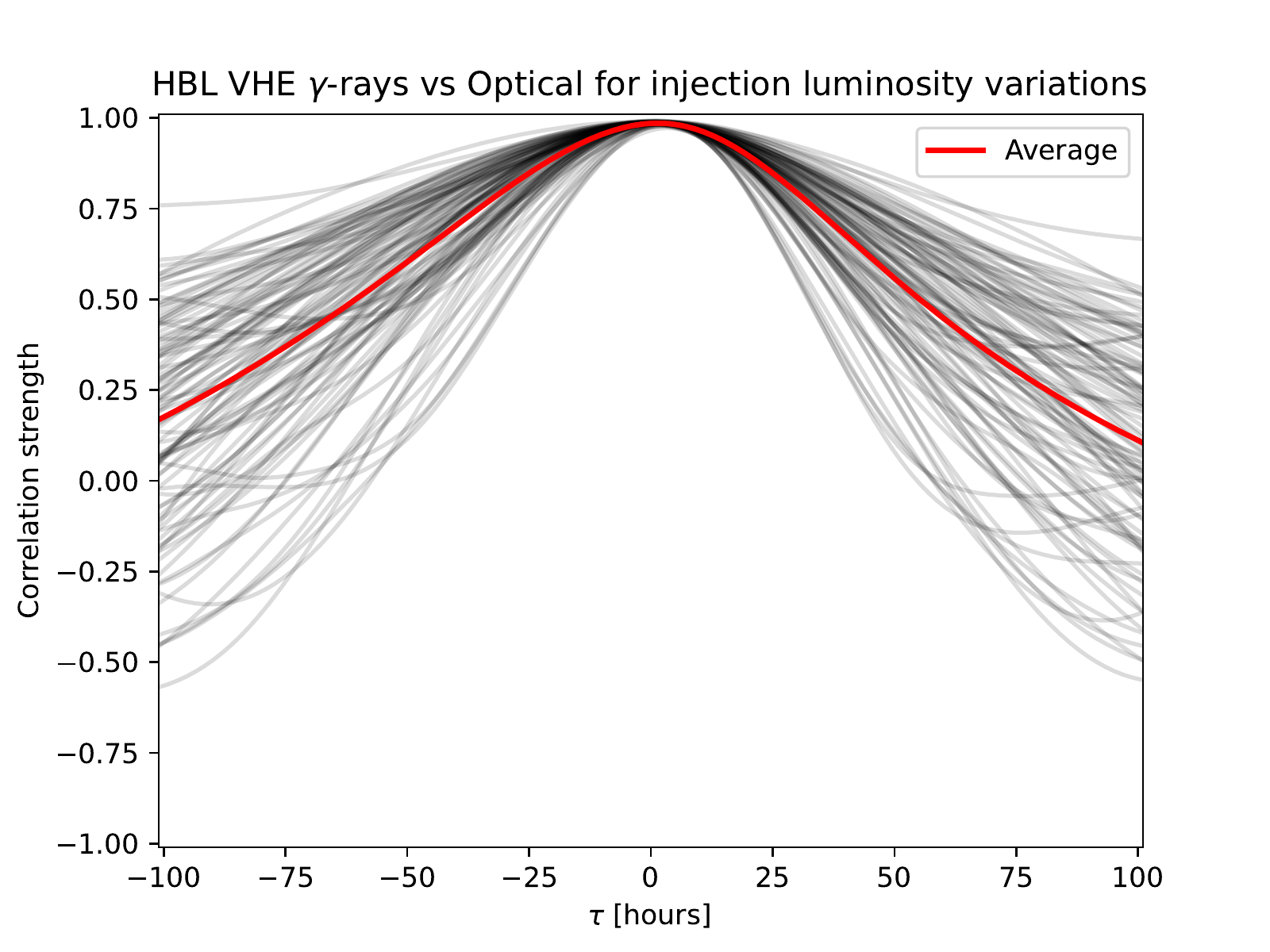}{.3\textwidth}{}
        \fig{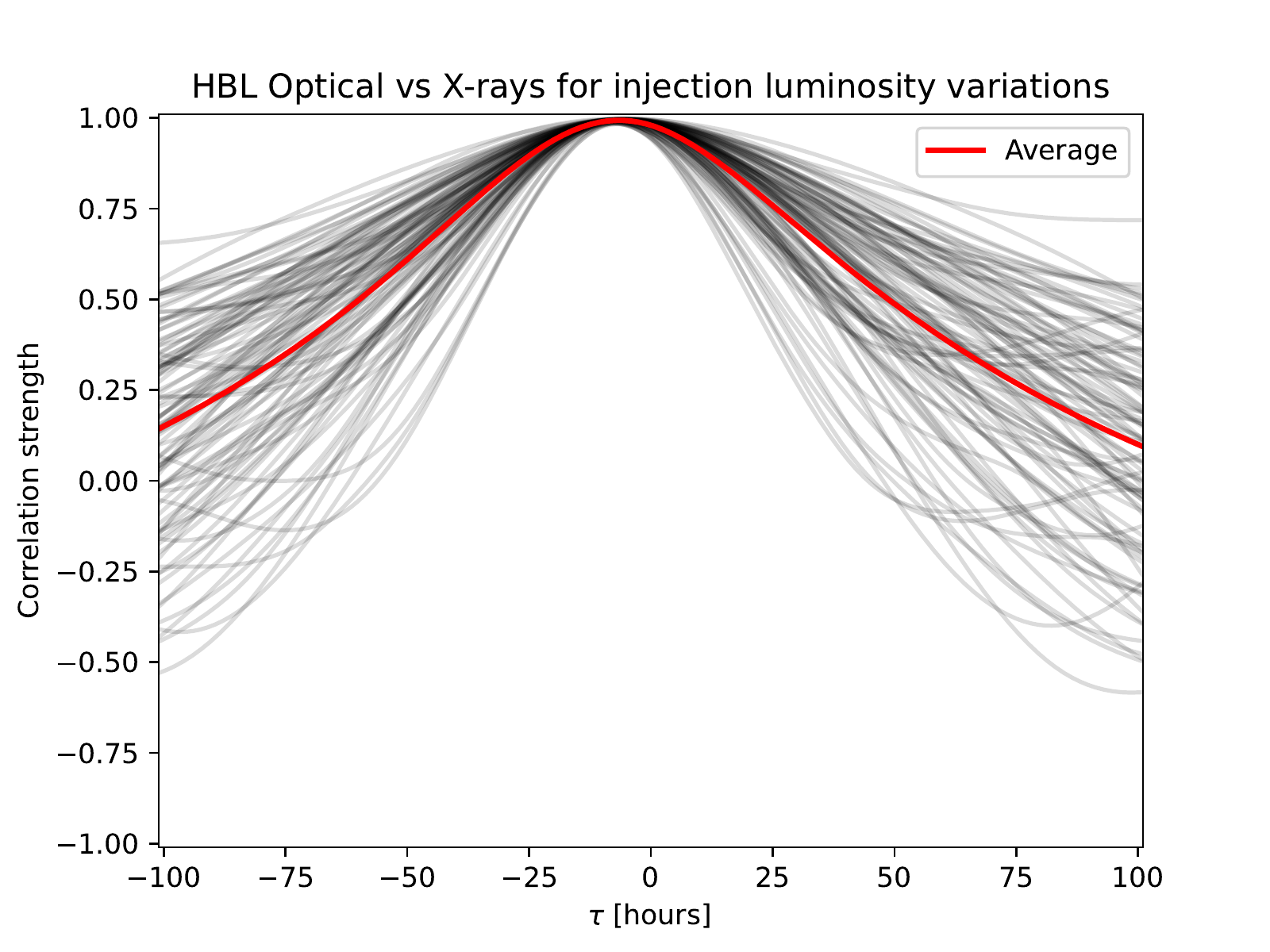}{.3\textwidth}{}
    }
    \caption{
        Average and individual cross-correlation comparisons for HBL simulation
        realizations with injection luminosity variations.}
    \label{fig:SSC-Corr-inj-ave_vs_indiv}
\end{figure}

\begin{figure}[h]
    \gridline{
        \fig{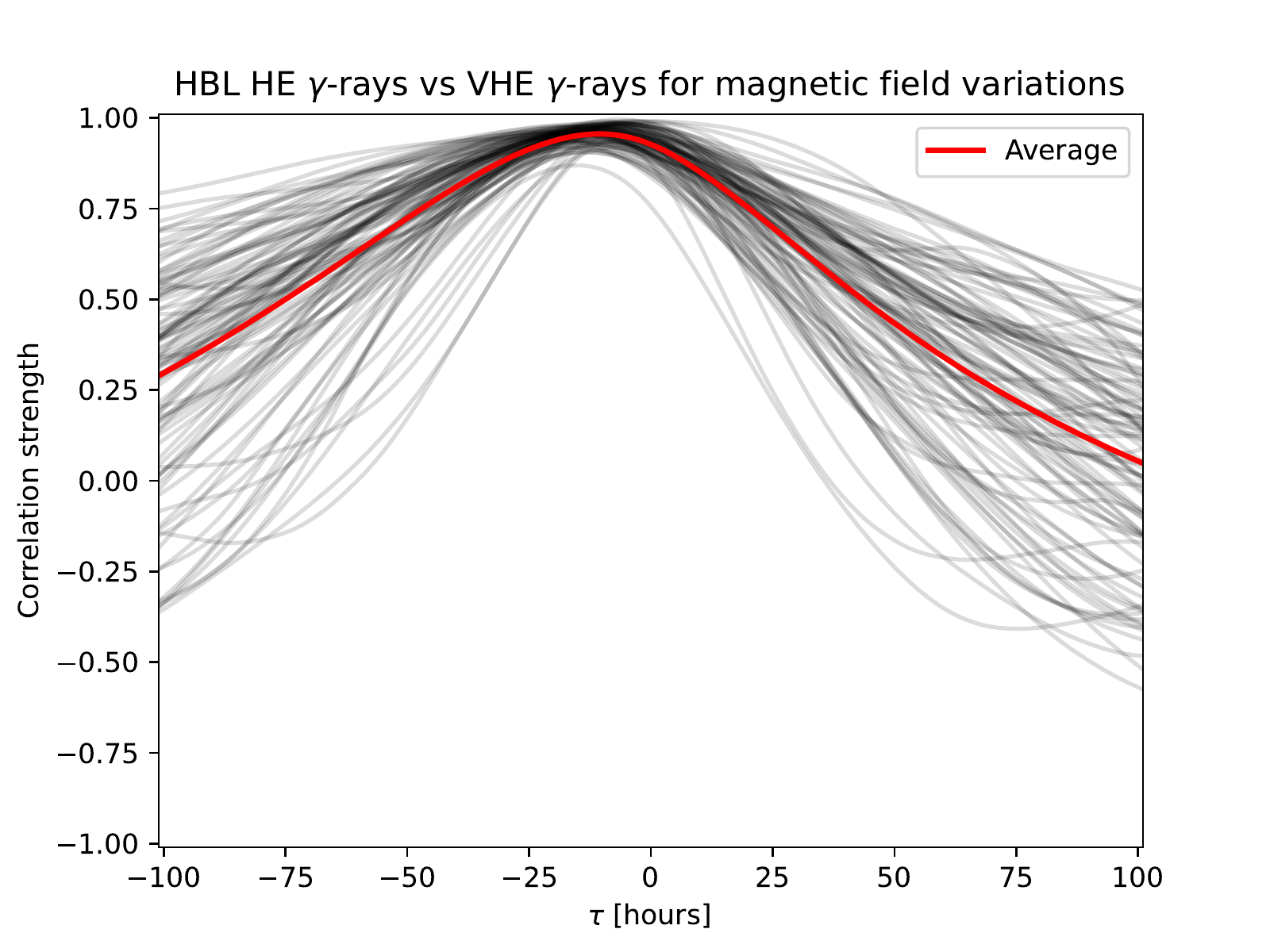}{.3\textwidth}{}
        \fig{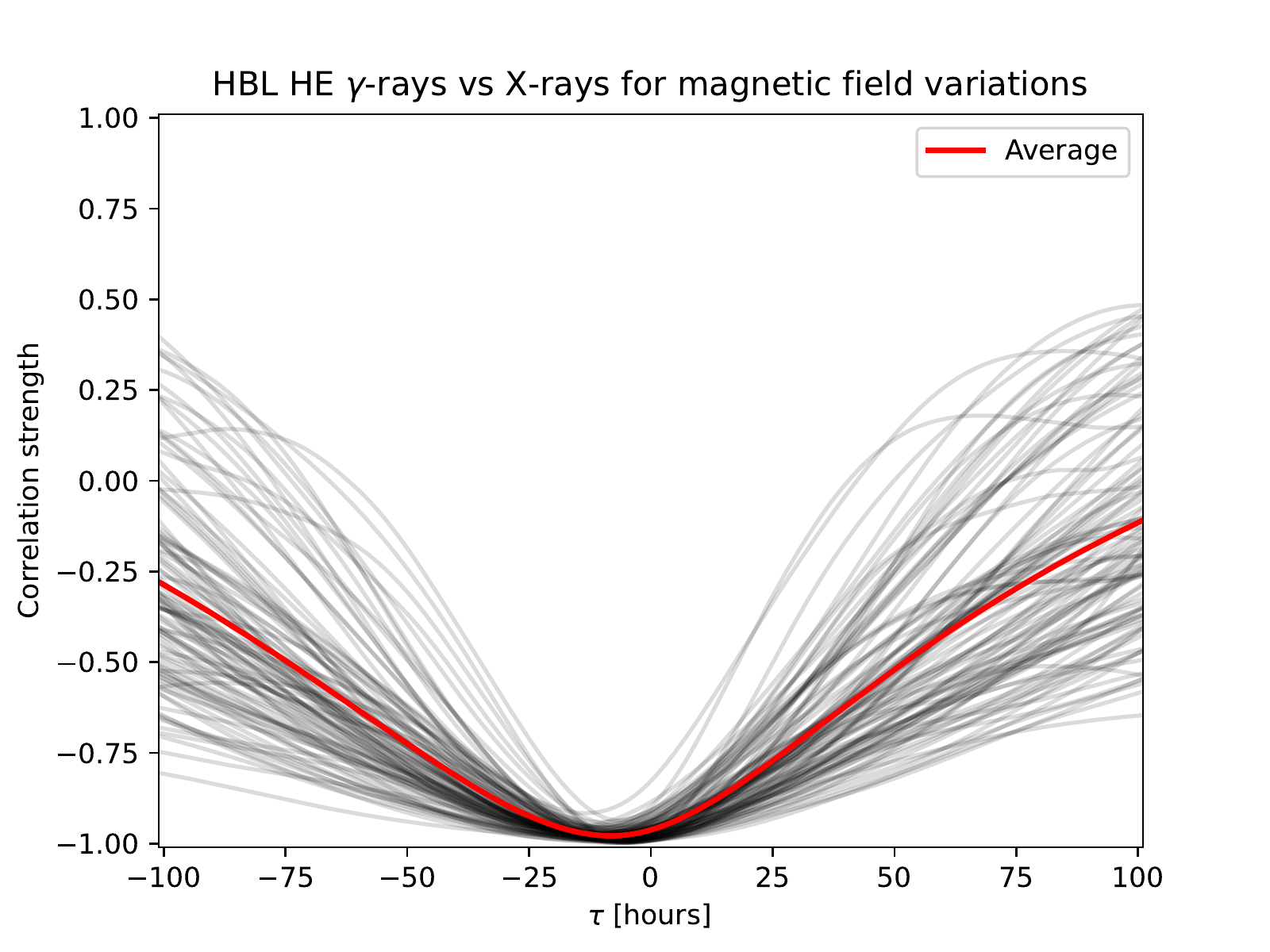}{.3\textwidth}{}
        \fig{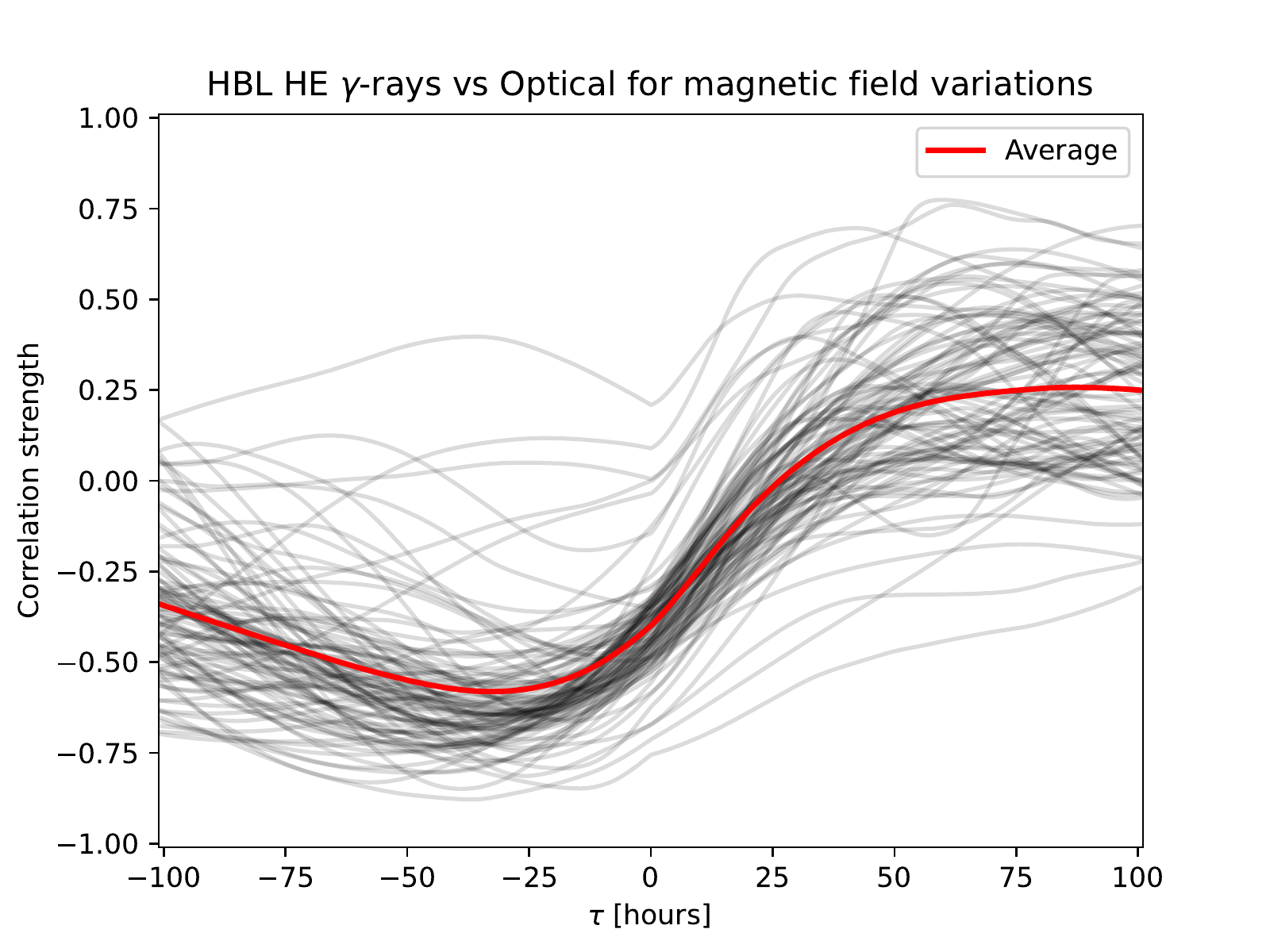}{.3\textwidth}{}
    }
    \gridline{
        \fig{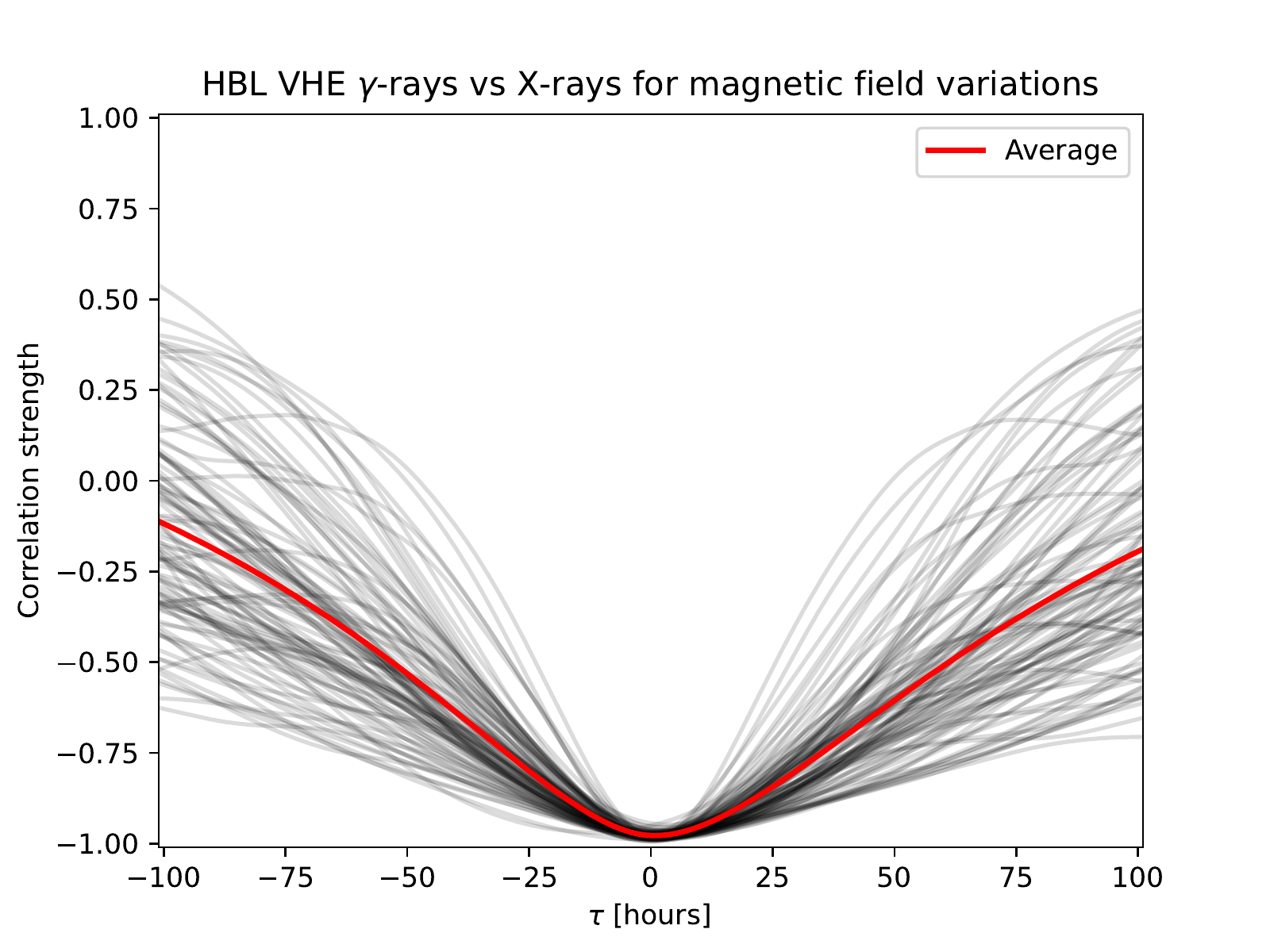}{.3\textwidth}{}
        \fig{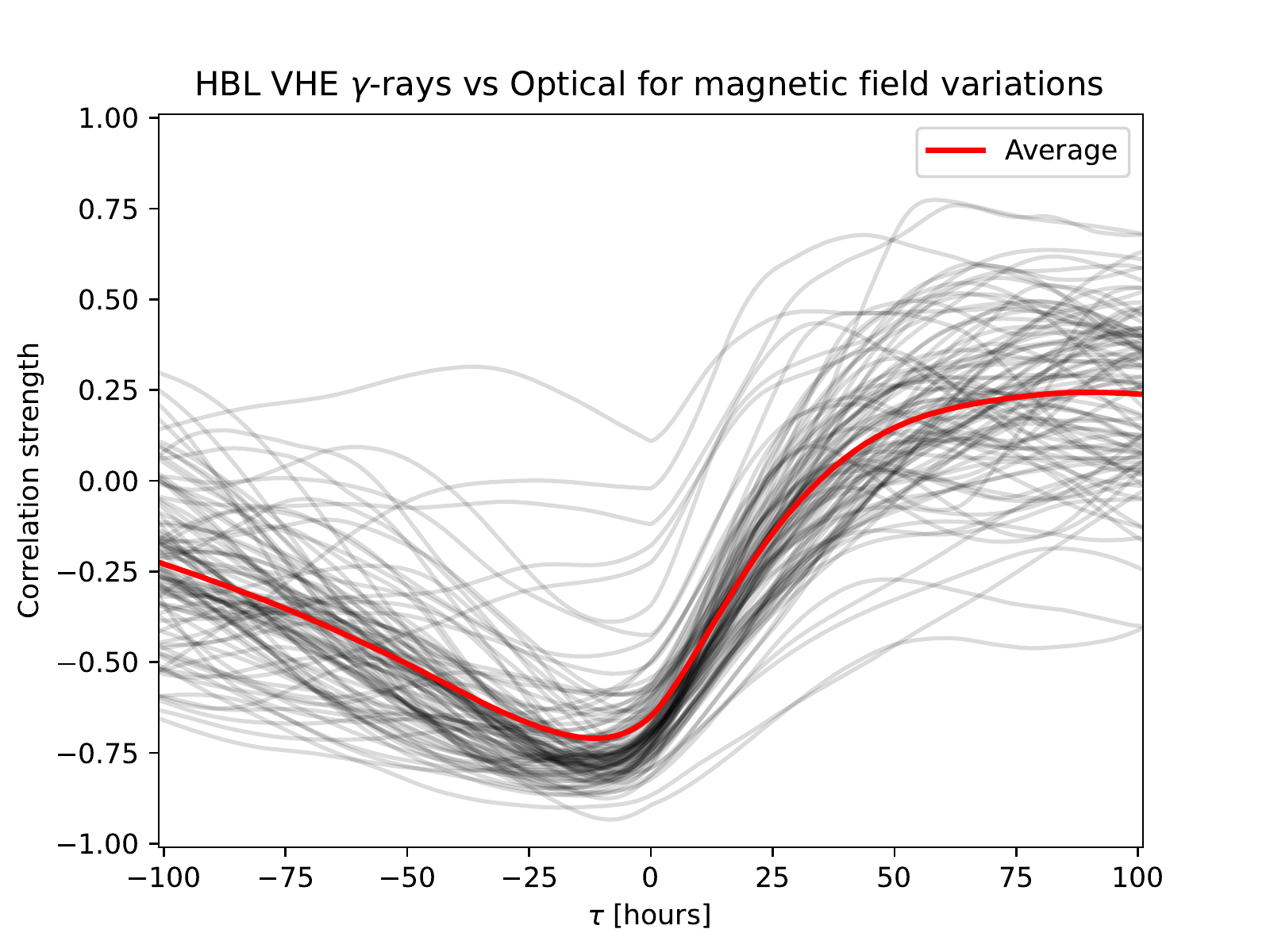}{.3\textwidth}{}
        \fig{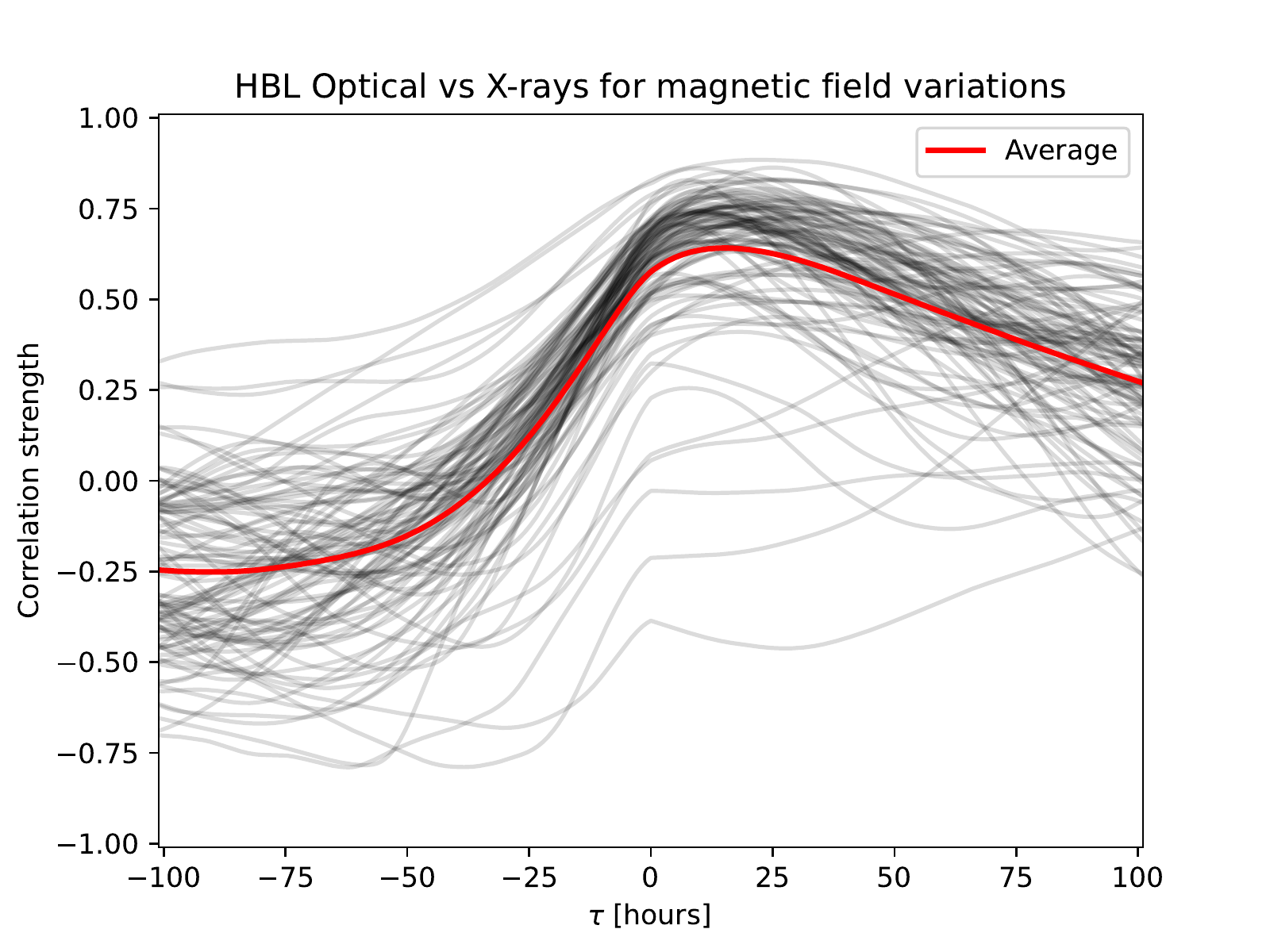}{.3\textwidth}{}
    }
    \caption{
        Average and individual cross-correlation comparisons for HBL simulation
        realizations with magnetic field variations.}
    \label{fig:SSC-Corr-mag-ave_vs_indiv}
\end{figure}

\begin{figure}[h]
    \gridline{
        \fig{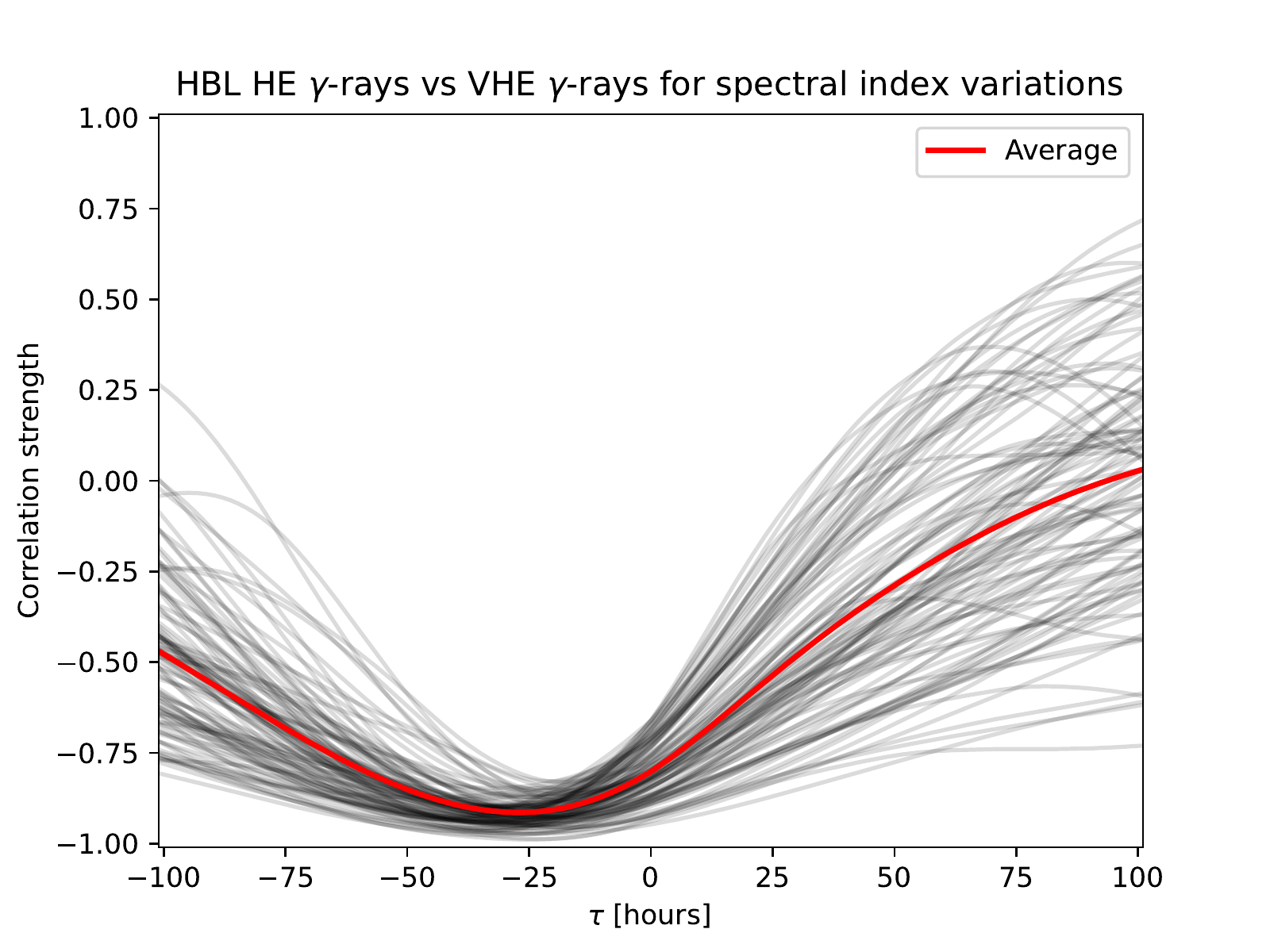}{.3\textwidth}{}
        \fig{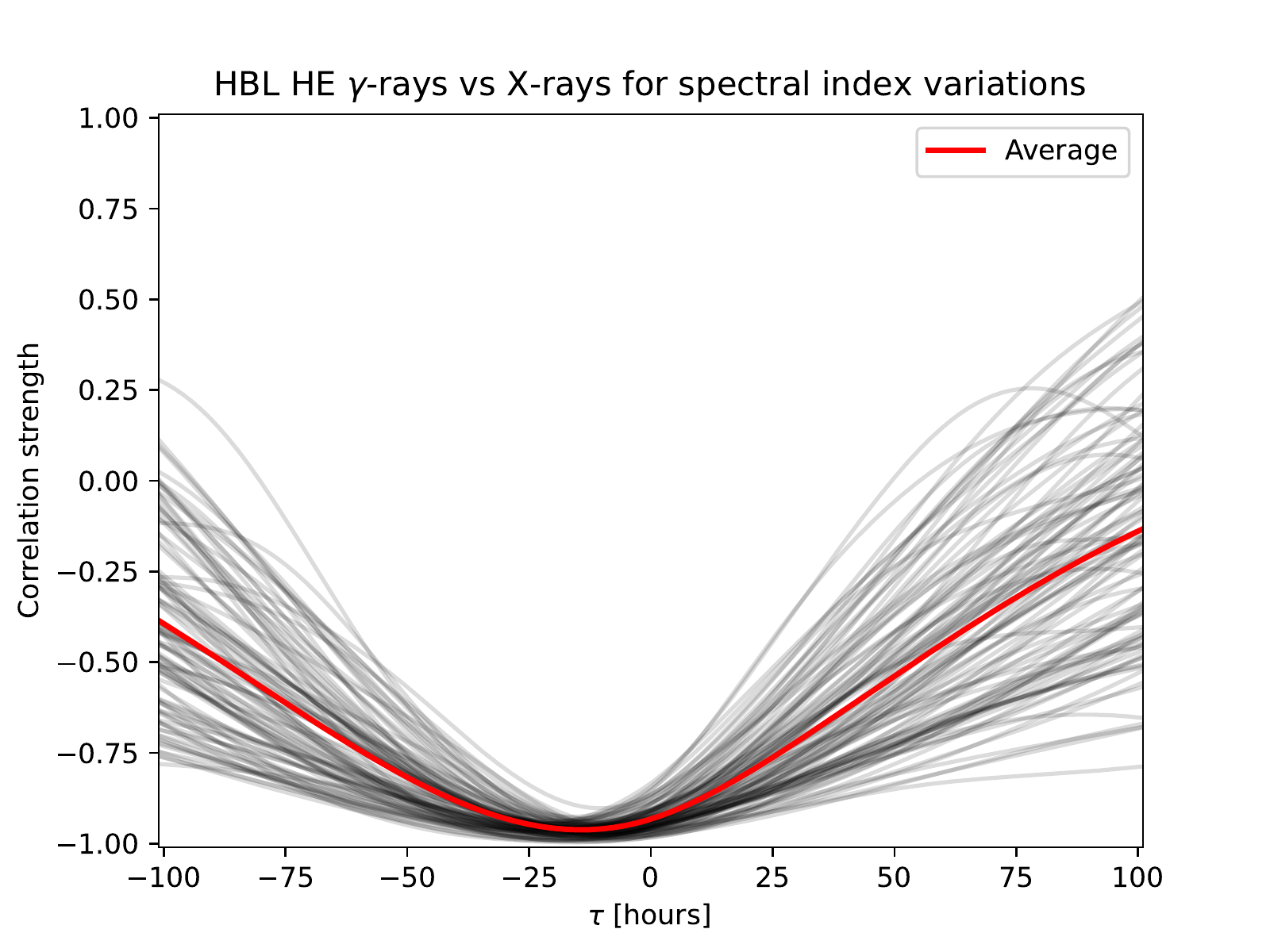}{.3\textwidth}{}
        \fig{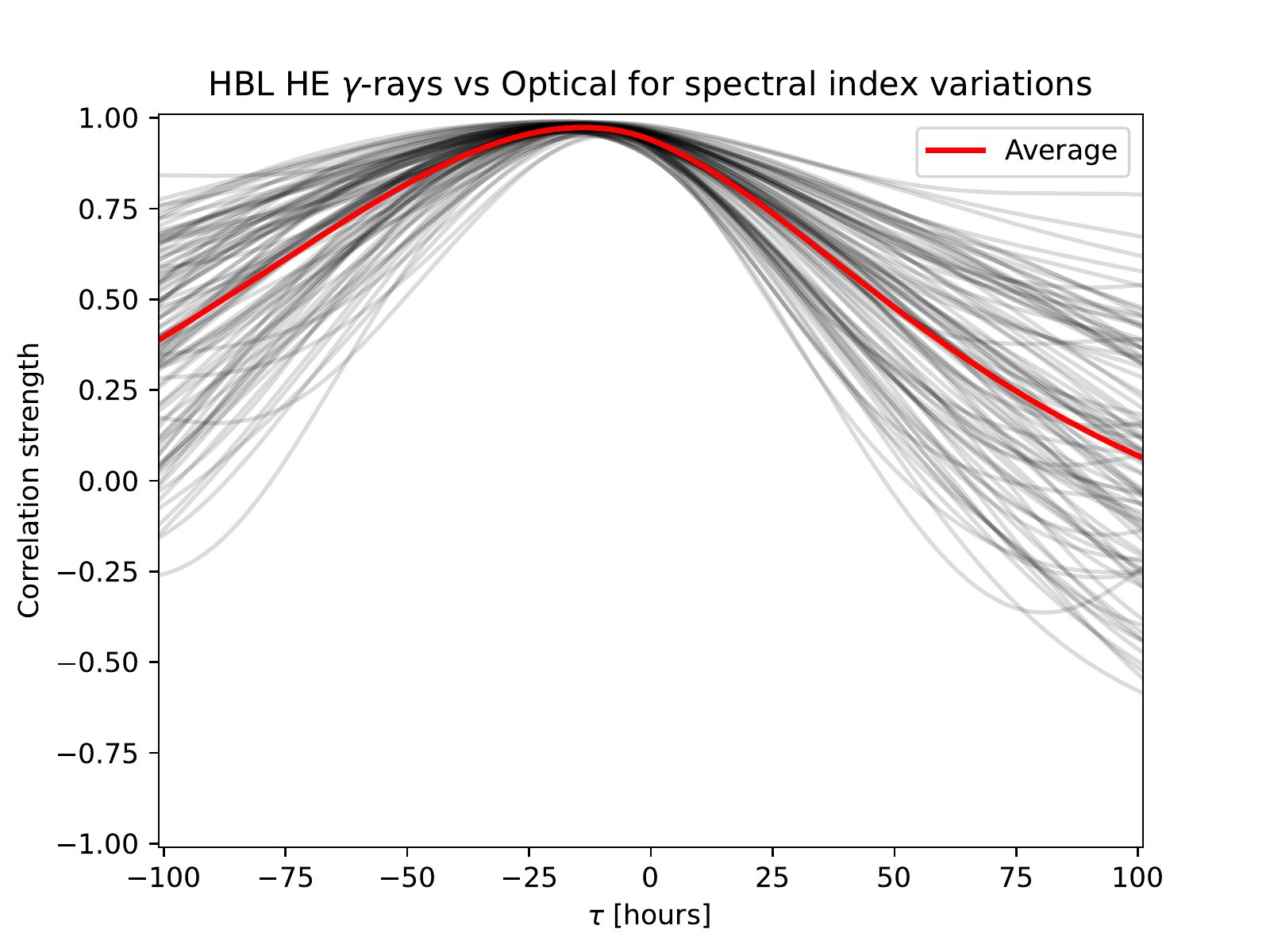}{.3\textwidth}{}
    }
    \gridline{
        \fig{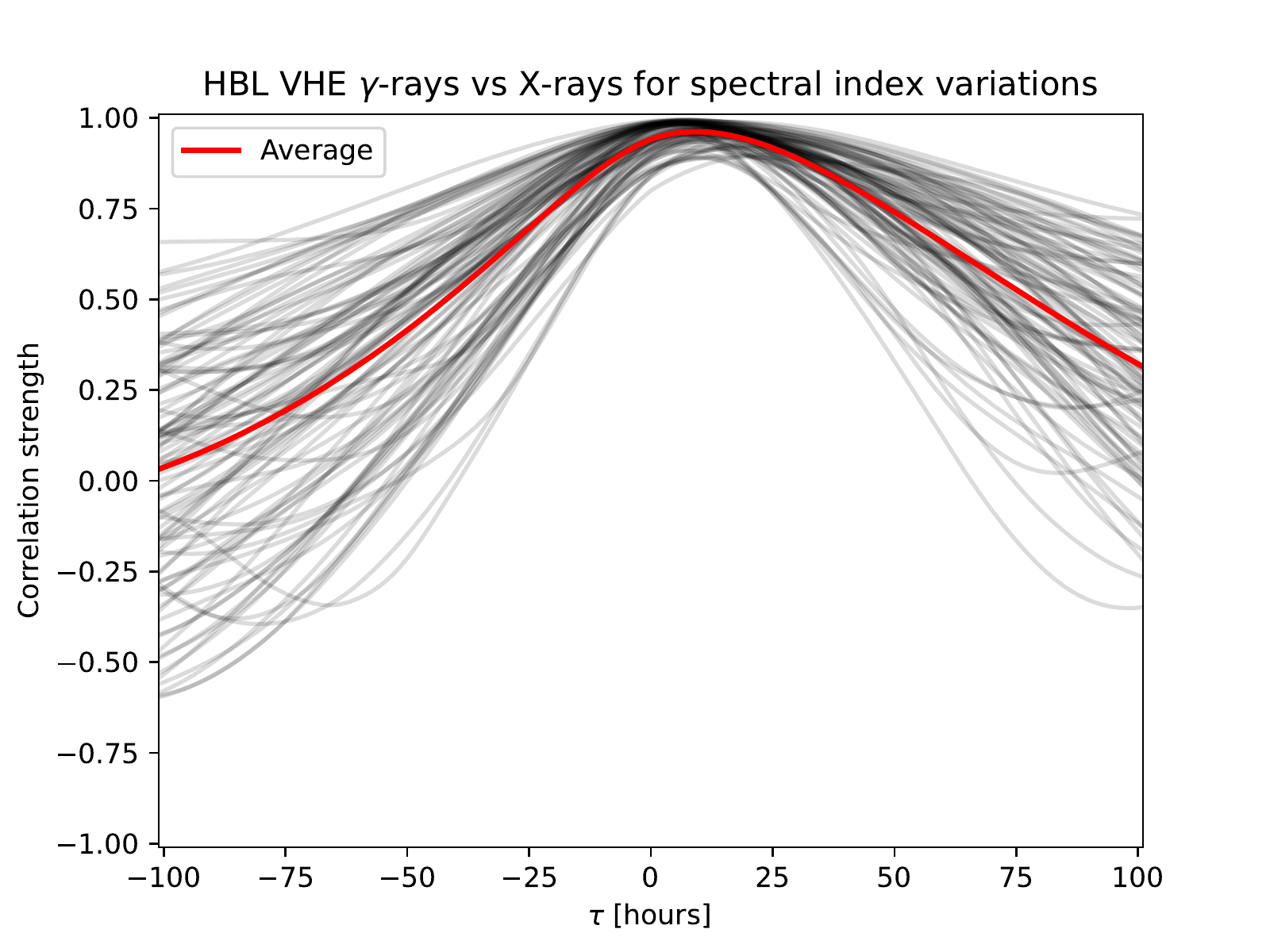}{.3\textwidth}{}
        \fig{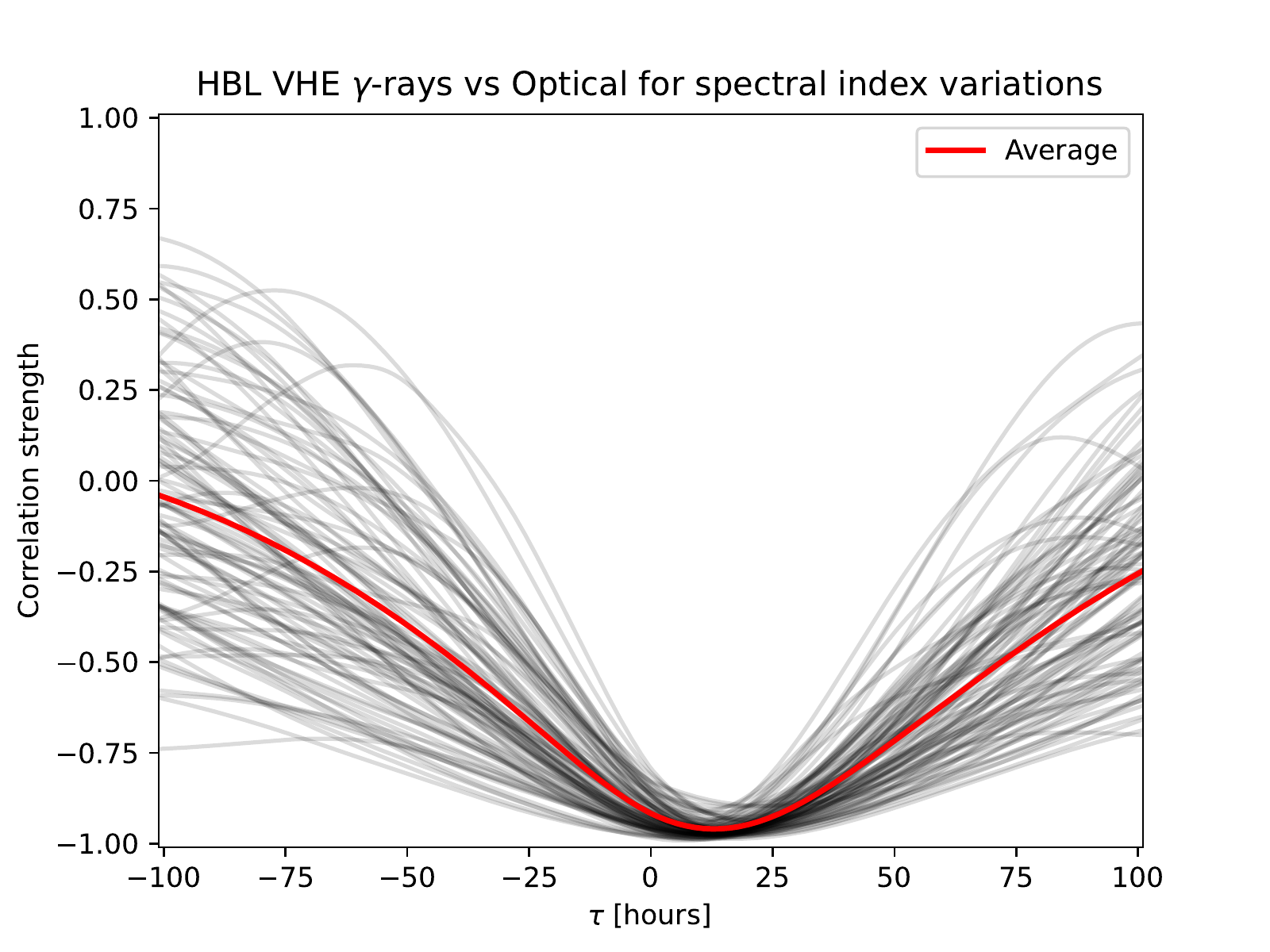}{.3\textwidth}{}
        \fig{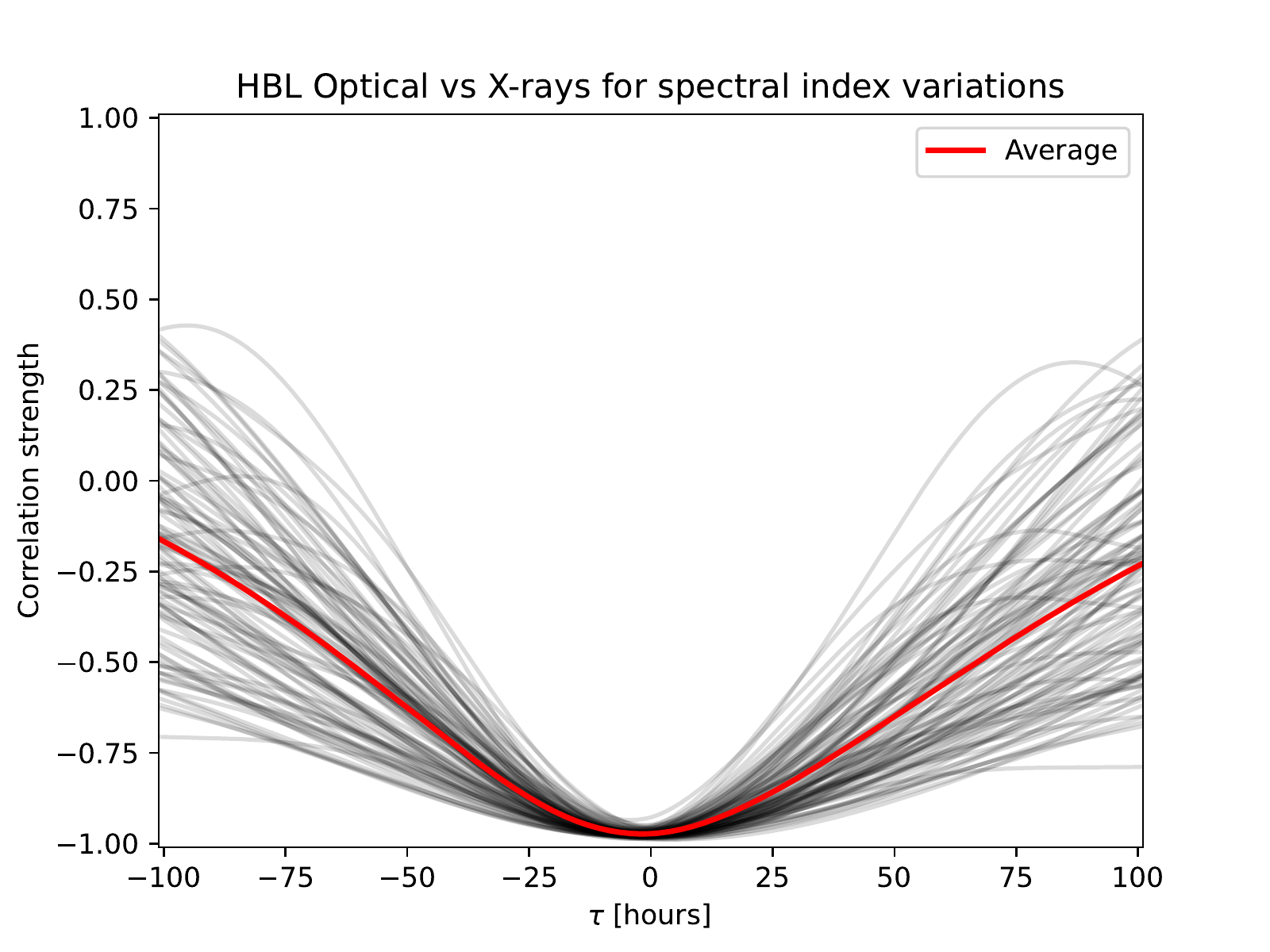}{.3\textwidth}{}
    }
    \caption{
        Average and individual cross-correlation comparisons for HBL simulation
        realizations with spectral index variations.}
    \label{fig:SSC-Corr-spc-ave_vs_indiv}
\end{figure}

\section{Cross-correlation peak analyses} \label{app:cc_peaks}

Figure~\ref{fig:EC-PSD-ave_vs_indiv} shows the peak in each cross-correlation function within a time delay range of $[-100, 100]$ hours for each of the simulations from which average cross-correlation peaks (crosses) are calculated.

\begin{figure}[h]
    \centering
    \gridline{
        \fig{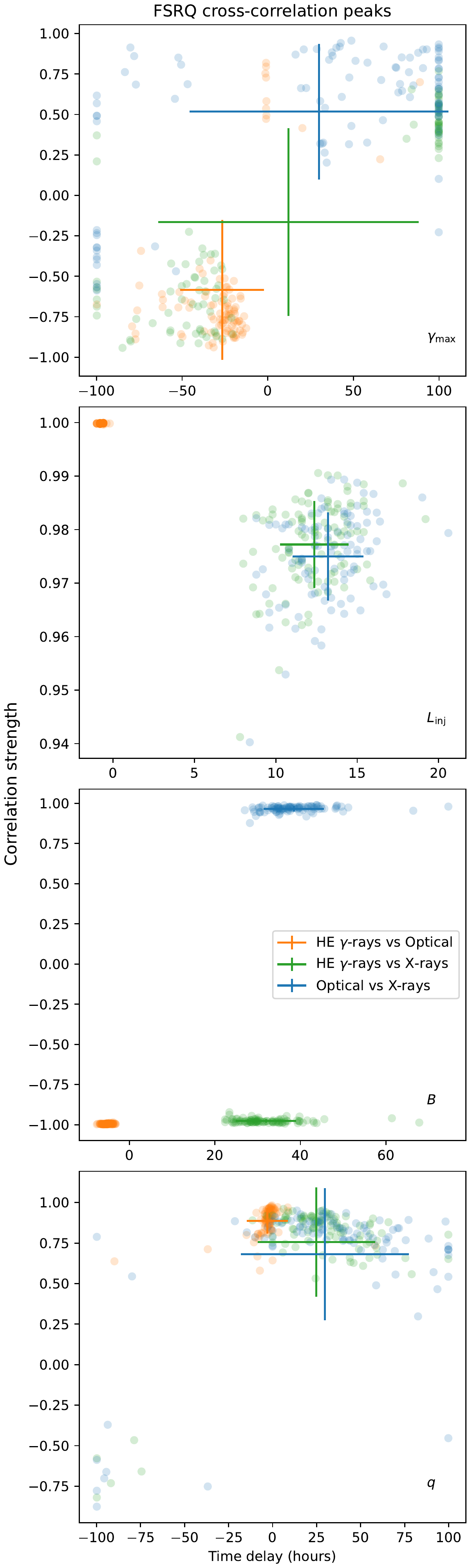}{.35\textwidth}{}
        \fig{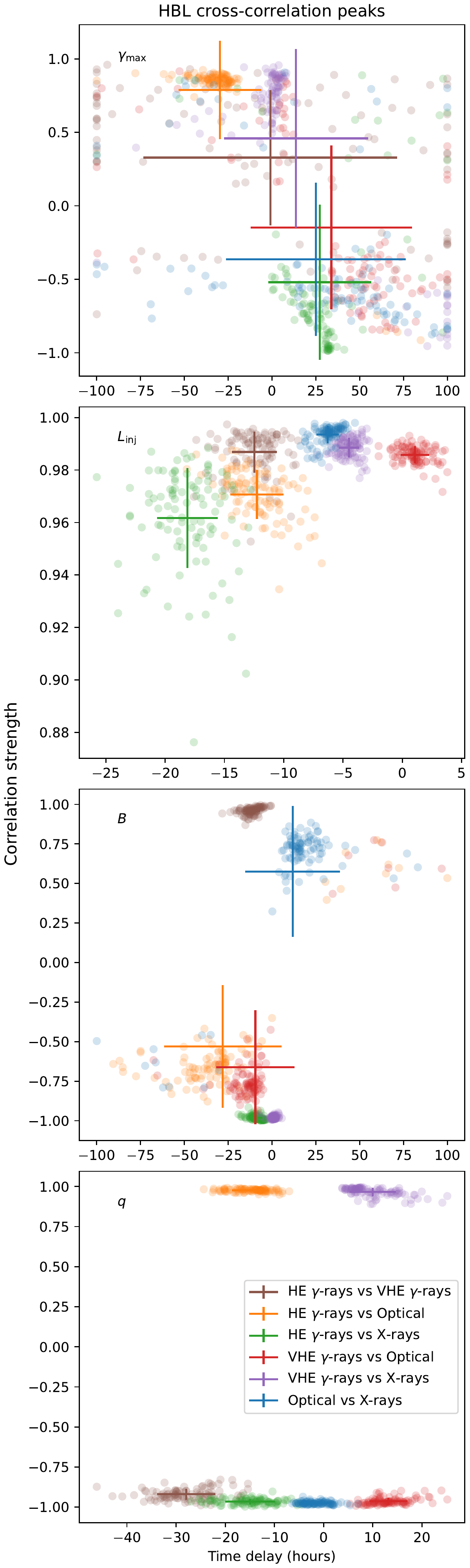}{.35\textwidth}{}
    }
\caption{
    Individual cross-correlation peaks of all simulations as well as the
    average cross-correlation peaks.}
\label{fig:EC-PSD-ave_vs_indiv}
\end{figure}

\bibliography{sample63}{}
\bibliographystyle{aasjournal}


\end{document}